\global\def\draftcontrol{0}

   \def\versionno{ stress tensor -- draft   }

\catcode`\@=11

\expandafter\ifx\csname draftcontrol\endcsname\relax\global\def\draftcontrol{0}
\fi

{\count255=\time\divide\count255 by 60
\xdef\hourmin{\number\count255}
\multiply\count255 by-60\advance\count255 by\time
\xdef\hourmin{\hourmin:\ifnum\count255<10 0\fi\the\count255}}
\def\draftdate{\number\month/\number\day/\number\year\ \ \ \hourmin }

\newcommand\makepapertitle{\par
  \begingroup
    \renewcommand\thefootnote{\@fnsymbol\c@footnote}%
    \def\@makefnmark{\rlap{\@textsuperscript{\normalfont\@thefnmark}}}%
    \long\def\@makefntext##1{\parindent 1em\noindent
            \hb@xt@1.8em{%
                \hss\@textsuperscript{\normalfont\@thefnmark}}##1}%
     \newpage
     \global\@topnum\z@   
     \@makepapertitle
     \thispagestyle{empty}\@thanks
  \endgroup
  \setcounter{footnote}{0}%
  \global\let\thanks\relax
  \global\let\makepapertitle\relax
  \global\let\@makepapertitle\relax
  \global\let\@thanks\@empty
  \global\let\@author\@empty
  \global\let\@date\@empty
  \global\let\@title\@empty
  \global\let\title\relax
  \global\let\author\relax
  \global\let\date\relax
  \global\let\and\relax
  \def\version{\let\version\@version\@gobble}
}
\def\@makepapertitle{%
  \newpage
   \ifnum\draftcontrol=1 {}
   \version\versionno
   \vskip 3em%
   \else
   \hfill\hbox to 4cm {\parbox{5cm}{\@pubnum}\hss}%
   \vskip 3em%
   \fi
   \begin{center}%
   \let \footnote \thanks
     {\LARGE {\@title}}%
     \vskip 1.5em%
     {\normalsize
       \lineskip .5em%
       \begin{tabular}[t]{c}%
         \@author
       \end{tabular}\par}%
     \vskip 1.5em%
     {\@bstract}%
     \end{center}%
     \vskip 1.5em 
     \@date%
   \par
}

\gdef\@pubnum{}
\def\pubnum#1{%
  \gdef\@pubnum{#1}}

\gdef\@bstract{}
\def\Abstract#1{%
  \gdef\@bstract{%
   \parbox{\textwidth-0pc}{%
   \centerline{\bf Abstract}\penalty1000
   \noindent
   \renewcommand\baselinestretch{1.0}
   {#1}}}
}

\def\ps@paper{\let\@mkboth\@gobbletwo%
     \ifnum\draftcontrol=1
        \def\@oddfoot{\hbox to \textwidth{\tiny \versionno \hfil\tiny\draftdate}%
        \hskip -\textwidth \hbox to \textwidth{\hfil\rm\thepage\hfil}}%
     \else\def\@oddfoot{\hbox to \textwidth{\hfil\rm\thepage\hfil}}
     \fi
     \let\@evenfoot\@oddfoot
}

\def\body{\clearpage
          \pagestyle{paper}
        }

\def\@version#1{\ifnum\draftcontrol=1
\typeout{}\typeout{#1}\typeout{}
\vskip3mm\centerline{\hbox{\fbox{\normalsize{\tt DRAFT -- #1 -- }
                   {\draftdate}}}}\vskip3mm
\fi}
\let\version\@version
\long\def\eqlabel#1{\ifnum\draftcontrol=1
                    \tag@false  
                    \tag*{(\theequation) \hbox to -0.2cm{\hspace{0cm}\small{#1}\hss}}
                    \refstepcounter{equation} 
                    \edef\@currentlabel{\theequation}
                    \ltx@label{#1}          
                    \else
                    \label{#1}
                    \fi
                    }
\let\st@bibitem\@bibitem
\let\st@lbibitem\@lbibitem
\ifnum\draftcontrol=1
  \def\@bibitem#1{%
    \st@bibitem{#1}\a@@label{#1}\ignorespaces}
  \def\@lbibitem[#1]#2{%
    \st@lbibitem[#1]{#2}\a@@label{#2}\ignorespaces}
  \def\a@@label#1{%
    \gdef\a@lab{\smash{\normalfont\small#1}}
    \ifvmode
      \if@inlabel
        \global\setbox\@labels\hbox{%
          \llap{\a@lab\let\a@lab\relax
                \kern\@totalleftmargin\kern\marginparsep}%
          \box\@labels}%
      \fi
    \fi}
\fi

\documentclass[12pt,letterpaper]{article}

\usepackage{amsmath,amssymb,array,calc,rotating,epsfig,psfrag,cite}

\ifnum\draftcontrol=1
\fi

\renewcommand\baselinestretch{1.25}
\renewcommand\section{\@startsection {section}{1}{\z@}%
                                   {-3.5ex \@plus -1ex \@minus -.2ex}%
                                   {2.3ex \@plus.2ex}%
                                   {\normalfont\large\bfseries}}
\renewcommand\subsection{\@startsection{subsection}{2}{\z@}%
                                   {-3.25ex\@plus -1ex \@minus -.2ex}%
                                   {1.5ex \@plus .2ex}%
                                   {\normalfont\normalsize\bfseries}}
\renewcommand\subsubsection{\@startsection{subsubsection}{3}{\z@}%
                                   {-3.25ex\@plus -1ex \@minus -.2ex}%
                                   {1.5ex \@plus .2ex}%
                                   {\normalfont\normalsize\it}}
\renewcommand\paragraph{\@startsection{paragraph}{4}{\z@}%
                                   {-3.25ex\@plus -1ex \@minus -.2ex}%
                                   {1.5ex \@plus .2ex}%
                                   {\normalfont\normalsize\bf}}

\numberwithin{equation}{section}

\def\ie{{\it i.e.}}

\def\revise#1       {\raisebox{-0em}{\rule{3pt}{1em}}%
                     \marginpar{\raisebox{.5em}{\vrule width3pt\
                     \vrule width0pt height 0pt depth0.5em
                     \hbox to 0cm{\hspace{0cm}{%
                     \parbox[t]{4em}{\raggedright\footnotesize{#1}}}\hss}}}}

\newcommand\nxt[1]  {\\\fnxt#1}

\def\cala         {{\cal A}}

\def\calb         {{\cal B}}

\def\calf         {{\cal F}}

\def\call         {{\cal L}}
\def\calm         {{\cal M}}
\def\caln         {{\cal N}}
\def\calo         {{\cal O}}
\def\calp         {{\cal P}}

\def\calr         {{\cal R}}

\def\reals        {{\mathbb R}}
\def\zet          {{\mathbb Z}}

\def\del          {\partial}

\def\sqr#1#2{{\vcenter{\vbox{\hrule height.#2pt  
 \hbox{\vrule width.#2pt height#1pt \kern#1pt
 \vrule width.#2pt}\hrule height.#2pt}}}}
\def\square{%
  \mathop{\mathchoice{\sqr{12}{15}}{\sqr{9}{12}}{\sqr{6.3}{9}}{\sqr{4.5}{9}}}}

\def\om{\Omega}
\def\r{\rho}
\def\a{\alpha}
\def\dd{\delta}

\def\la{\lambda}

\def\hr{\hat{\rho}}
\def\ga{\gamma}
\def\pz{p_0}
\def\kz{K_0}
\def\vev#1{\langle #1 \rangle}

\catcode`\@=12

\begin{document}


\title{Holographic renormalization of cascading gauge theories}

\pubnum{
WIS/14/05-JUN-DPP\\
{\tt hep-th/0506002}}
\date{June 2005}

\author{Ofer Aharony$ ^1$, Alex Buchel$ ^{2,3}$ and 
Amos Yarom$ ^{4}$\\[0.4cm]
\it $^1$Department of Particle Physics, Weizmann Institute of Science,\\ 
\it Rehovot 76100, Israel\\[0.2cm]
\it $^2$Perimeter Institute for Theoretical Physics\\
  \it Waterloo, Ontario N2J 2W9, Canada\\[0.2cm]
  \it $^3$Department of Applied Mathematics, University of Western Ontario\\
  \it London, Ontario N6A 5B7, Canada\\[0.2cm]
\it $^4$Department of Physics, Ben-Gurion University, Be'er Sheva 84105,
Israel\\
 }

\Abstract{
We perform a holographic renormalization of cascading gauge theories.
Specifically, we find the counter-terms that need to be
added to the gravitational action of the backgrounds dual to the
cascading theory of
Klebanov and Tseytlin, compactified on an arbitrary four-manifold,
in order to obtain finite correlation functions (with a limited set of 
sources).  We
show that it is possible to truncate the action for deformations of
this background to a five dimensional system coupling
together the metric and four scalar fields. Somewhat surprisingly,
despite the fact that these theories involve an infinite number of
high-energy degrees of freedom, we find 
finite answers for all one-point functions
(including the conformal anomaly). We compute explicitly the
renormalized stress tensor for the cascading gauge theories at high
temperature and show how our finite answers are consistent with the
infinite number of degrees of freedom. 
Finally, we discuss ambiguities
appearing in the holographic renormalization we propose for the
cascading gauge theories;
our finite results for the one-point functions have some
ambiguities in curved space (including the conformal anomaly) but not
in flat space. }


\makepapertitle

\body

\version\versionno

\section{Introduction}

The traditional definition of quantum field theories starts from a
fixed point of the renormalization group (free or interacting) at
high energies, which is some local conformal field theory, and defines
the theory by a renormalization group flow starting from that 
fixed point. One of the interesting ``side-effects'' of the progress
in our understanding of string theory in the last few years is the
realization that there exist consistent quantum field theories which
cannot be defined in this way. In some cases these theories may be
described by a decoupling limit of some sector of string theory, and
more generally they can always be defined by a background of type IIB
string theory
which is holographically dual to these theories, in the same sense that the
$AdS_5\times S^5$ background of string theory is dual to the $d=4$ $\caln=4$
supersymmetric Yang-Mills theory \cite{mal}.

The new types of theories which were discovered seem to fall into two
classes. One class of theories is ``little string theories'' --
non-local theories which share some features with string theories
(though they do not include gravity). In this paper we focus on the
second class, of ``cascading theories'' \footnote{An interesting relation
between these two types of theories was recently discovered in 
\cite{Butti:2004pk}.}.
The prototypical example of such theories (which we will focus on
here, though we expect to be able to generalize our methods also to other
cascading theories, including theories in other dimensions) is the
theory related to fractional $3$-branes at a conifold singularity, first
studied in \cite{kn,kt,ks} (see \cite{Herzog:2002ih,Strassler:2005qs} 
for reviews). 
These theories do not have a direct
definition in field theory terms (since they do not seem to have a UV
fixed point), so their only known direct definition is via the holographic
duality; in this paper we will attempt to understand this definition
better and verify that it can be used for well-defined computations in
these theories. When one introduces a finite high-energy cutoff, 
these theories at the cutoff scale resemble $\caln=1$
supersymmetric $SU(K)\times SU(K+P)$ gauge theories with two
bifundamental and two anti-bifundamental chiral superfields and some
superpotential; when one flows down in energy from this cutoff one of
the gauge theories becomes strongly coupled and the theory seems to
undergo a series of Seiberg duality \cite{Seiberg:1994pq}
``cascades'', ending with a confining theory in the IR (which is in the
same universality class as the $\caln=1$ supersymmetric pure $SU(P)$
Yang-Mills theory when $K$ is a multiple of $P$). However, these
gauge theories are never asymptotically free, so they cannot be used
to define the theory -- rather it seems that more and more degrees of
freedom are needed to define the theory at higher energies, and that
the ultimate definition of the cascading theories requires a theory with an
infinite number of fields.  It may be possible to define the cascading
gauge theories by a limiting procedure, starting from a theory with a
finite number of degrees of freedom which can flow to the cascade (as
in \cite{Hollowood:2004ek}) and taking the limit in which the number
of degrees of freedom goes to infinity -- it would be interesting to
make such a definition precise. In this paper we will not discuss the
interpretation of the cascade as a gauge theory, but rather we will
take the holographic dual to {\it define} the cascading theory, and attempt
to see if such a definition makes sense.

In order to use a holographic dual as a definition of a field theory we
need to have a prescription for the computation of all correlation
functions in the field theory. For the AdS/CFT correspondence such a
prescription was given in \cite{Gubser:1998bc,Witten:1998qj}, 
and it can be generalized to other
holographic dualities as well. In principle one should be able to perform
the computations of correlation functions directly in string theory on the
holographic dual background, but in practice, since this string theory is
very complicated, computations can only be done in a low-energy gravitational
approximation, and we will use this approximation in this paper. In this
approximation the correlation functions may be defined as derivatives of
the (super)gravity action on the holographic dual background with respect
to sources at the boundary of space-time. When one tries to perform such
computations one encounters divergences, which from the gravity point of
view are IR divergences related to the infinite distance to the boundary.
These divergences may be dealt with by the process of ``holographic
renormalization'' \cite{Henningson:1998gx,bk,Emparan:1999pm,Kraus:1999di,
Nojiri:1999jj,Nojiri:2000kh,deHaro:2000xn,Taylor-Robinson:2001pp,
Muck:2001cy,bfs1,bfs2,Skenderis:2002wp,Berg:2002hy,Papadimitriou:2004rz}.
First, one regularizes the theory by imposing
a cutoff on the radial direction (the precise meaning of this cutoff in
the field theory language is not clear, but it certainly provides a good
regularization). Next, one adds counter-terms to the gravitational action,
which are local functions of the fields at the cutoff, in such a way that
the action remains finite when the cutoff is taken to infinity. This
process is very analogous to renormalization in field theory, and it seems
that it should be mapped to this by the holographic duality; in both cases
in a well-defined theory there is just a finite number of divergences which
need to be canceled, after which correlation functions may be computed,
depending on a finite number of parameters (some of which are coupling 
constants of the theory, while others are related to vacuum expectation
values of fields or to ambiguities in the
definitions of operators). 

The process of holographic renormalization is by now conceptually
well-understood in asymptotically anti-de Sitter spaces, where several
examples have been analyzed in detail \cite{bfs1,bfs2,Berg:2002hy}, 
though the general
renormalization for theories with many fields in the bulk is quite
complicated and has not yet been performed. In principle one would
expect a similar renormalization process to apply in other holographic
dualities such as those of cascading gauge theories\footnote{
Some general properties of the stress tensor in 
holographic renormalization, which apply
also to cascading backgrounds, were derived 
in \cite{Hollands:2005ya}. It is not
obvious whether holographic renormalization 
should be possible also in the duals to non-local
theories such as ``little string theories''.}, and our goal in
this paper is to understand how this works. One interesting question
which immediately arises is the following. In standard field theories
there are some correlation functions, such as the one-point function
of the trace of the stress-energy tensor (related to the conformal
anomaly), which are proportional to the number of degrees of freedom
in the theory. As discussed above, cascading theories do not seem to
have a finite number of degrees of freedom, but rather more and more
degrees of freedom as one goes to shorter and shorter distance
scales. So, should the correlation functions of cascading theories be
finite as one takes the cutoff to infinity or should some of them
diverge ?

We have attempted to perform a holographic renormalization of the cascading
theories both under the assumption that all correlation functions must be
finite, and under the assumption that correlation functions are allowed
to diverge as the cutoff is taken to infinity, in a way which depends on
the effective number of degrees of freedom. Somewhat surprisingly, we found
that it is possible to renormalize the theory with finite correlation
functions, but we were not able to renormalize the theory using the other
assumption. Thus, we claim that cascading gauge theories should be
renormalized just like standard theories, with all correlation functions
finite. At first sight this seems to contradict the fact that these
theories have an infinite number of high-energy degrees of freedom. We
claim that this is not the case, and that these theories have an infinite
number of high-energy degrees of freedom even though all correlation functions
(including the conformal anomaly) are finite. We illustrate this by analyzing
the thermodynamics of the cascading theories, showing that the effective
number of degrees of freedom diverges at high temperatures even though all
correlation functions are finite (at any fixed temperature).

At this point we should describe the precise assumptions under which
we perform the holographic renormalization of the cascading gauge
theories.  Usually holographic renormalization is performed using a
consistent truncation of the theory to a small number of fields
\footnote{As far as we know no more than two fields (coupled to gravity)
were analyzed
until now.}, for which any sources are allowed. In our case we also
truncate the full spectrum of fields in the holographic dual
background to a finite number of fields.  One important truncation we
make is that we only include fields which preserve the $SU(2)\times SU(2)
\times \zet_{2P}\times \zet_2$
isometry which is present at high energies in the cascading background
of \cite{kt}, related to the global symmetry in the field theory (the
metric actually has a $U(1)$ isometry but this is broken to a
$\zet_{2P}$ subgroup by the fluxes \cite{Klebanov:2002gr}). In
infinite space the cascading theories spontaneously break the
$\zet_{2P}$ symmetry, as found in \cite{ks}, so our analysis cannot be
directly used to analyze the infinite volume cascading
theories. However, at finite (high enough) temperature or at finite
(small enough) volume (for instance on $S^3\times \reals$, $S^4$ or
$dS_4$) this symmetry is expected to be preserved
\cite{bhkt1,bhkt2,bhkt3,bt}, 
so our analysis may be directly used for such backgrounds. We
expect that it should be straightforward (though technically
difficult) to extend our analysis to include also fields which break
the $SU(2)\times SU(2)\times \zet_{2P}\times \zet_2$ symmetry.

Even with the truncation to the $(SU(2)\times SU(2)\times
\zet_{2P}\times \zet_2)$-invariant sector, 
we are left with a large number of five
dimensional fields in the bulk -- the metric and four scalar
fields. These fields all mix together so we were not able to truncate
the theory further. Moreover, some of these fields are dual to
irrelevant operators \footnote{Since the cascading theories are close
to conformal field theories at high energies, with the characteristic
power law behavior of conformal field theories replaced by powers multiplying
logs, we will use the standard terminology of conformal field
theories.}, so it is not known how to introduce arbitrary sources for
these fields, as is usually done in the holographic renormalization
process in order to systematically compute the counter-terms. Thus, we
do not consider the most general sources; we allow a generic source
for the five dimensional metric (this source is identified with the 
four dimensional metric of the space-time
on which the cascading theory lives), but only constant sources for
the other scalar fields (and, in particular, vanishing sources for the
two scalar fields corresponding to irrelevant operators). This
simplifies the analysis considerably, but there are three
disadvantages. First, usually in holographic renormalization the
finiteness of the action (for arbitrary sources) guarantees the
finiteness of all correlation functions, but we do not allow arbitrary
sources so we have to separately check that the correlation functions
are finite in addition to the finiteness of the action (we check this
only for one-point functions; additional counter-terms may be needed
to ensure the finiteness of all correlation functions).  Second, we
can no longer translate the divergent terms in the action directly to
counter-terms, as usually done in holographic
renormalization. Therefore, we are forced to use a different
procedure, of guessing the counter-terms and verifying that they lead
to finite correlation functions (with a finite number of
ambiguities). Again, we expect that it should be possible to
generalize our analysis to include arbitrary sources (at least for all
the marginal and relevant operators), though this
will be technically complicated. We hope to return to this problem in
the future. The counter-terms that we find in this method are far from
being unique, and are certainly not the precise counter-terms that
lead to finiteness of all correlation functions. However, we expect
that the difference between the counter-terms we find and the correct
counter-terms will not affect the unambiguous results which we obtain.
Third, since we do not have arbitrary sources we cannot compute
arbitrary correlation functions, but only the derivatives of the
action with respect to the sources we include. Thus, our procedure
allows us to compute any correlation functions of the stress-energy
tensor (dual to the bulk metric), but in the scalar sector we can only
compute one-point functions.

In this paper we show that, in the truncation described above, it is
possible to holographically renormalize the cascading gauge theory
background and to obtain finite one-point functions. We begin in
section 2 by describing the background, the ansatz we use for the
solutions with the sources described above, and the solutions we find.
In section 3 we describe in detail the holographic renormalization
process and the form of the counter-terms we find. In section 4 we
discuss the example of cascading theories at finite (high)
temperature, following \cite{bhkt1,bhkt2,bhkt3}, and compute their
thermodynamic properties. We end in section 5 with our conclusions
and a discussion of future directions.
Various technical results are relegated to the appendix.

\section{The action and asymptotic behavior of cascading backgrounds}

In this section we construct the asymptotic (near the boundary)
solutions corresponding to cascading gauge theories compactified on
arbitrary manifolds, generalizing the flat space asymptotic solution
found in \cite{kt}.

\subsection{The gravitational action and its KK reduction and truncation}

We will work in the gravitational approximation to type IIB string theory,
using the type IIB supergravity action. This action takes the form (in 
the Einstein frame)
\begin{equation}
\begin{split}
S_{10}=\frac{1}{16\pi G_{10}}\int_{\calm_{10}}&\ \biggl(
R_{10}\wedge \star 1 -\frac 12 d\Phi\wedge \star d\Phi
-\frac 12 e^{-\Phi} H_3\wedge\star H_3  -\frac 12 
e^{\Phi} F_3\wedge\star F_3\\
& \qquad -\frac 14 F_5\wedge \star F_5-\frac 12 C_4\wedge H_3\wedge F_3\biggr),
\end{split}
\eqlabel{10action}
\end{equation} 
where $\calm_{10}$ is the ten dimensional bulk space-time, $G_{10}$ is
the ten dimensional gravitational constant, and we have consistently
set the axion $C_0$ to zero (it vanishes in all the solutions we
are interested in). In this action
\begin{equation}
F_3=dC_2,\qquad F_5=dC_4 -C_2\wedge H_3,
\eqlabel{fluxes}
\end{equation}
where $C_2$ and $C_4$ are the Ramond-Ramond (RR) potentials.
The equations of motion following from the 
action \eqref{10action} have to be supplemented by
the self-duality condition
\begin{equation}
\star F_5=F_5.
\eqlabel{5self}
\end{equation}
It is important to remember that the self-duality 
condition \eqref{5self} 
can not be imposed at the level of the action,
as this would lead to wrong equations of motion.

Next, we perform a Kaluza-Klein (KK) reduction of this action to five
dimensions, using a
specific ansatz for the metric and for the various forms. This ansatz
includes in particular the solution of \cite{kt}, and it is the most
general ansatz describing a deformation of this solution which preserves 
the $SU(2)\times SU(2)\times \zet_{2P}\times \zet_2$ symmetry of this 
solution\footnote{Except for two modes of the RR fields 
which we consistently set to zero as mentioned in the text.}
(the discrete $\zet_2$ 
symmetry acts by exchanging the two global $SU(2)$ factors).
We take $\calm_{10}$ to be  a direct warped product of 
$\calm_5$ with metric $g_{\mu\nu}(y)$ and the `squashed' $T^{1,1}$ 
coset appearing in the solution of \cite{kt}. 
So, the Einstein-frame metric ansatz is
\begin{equation}
ds_{10}^2 =g_{\mu\nu}(y) dy^{\mu}dy^{\nu}+\om_1^2(y) e_{\psi}^2 
+\om_2^2(y) \sum_{a=1}^2\left(e_{\theta_a}^2+e_{\phi_a}^2\right),
\eqlabel{10met}
\end{equation}
where $y$ denotes the coordinates of $\calm_5$ (greek indices $\mu,\nu$
will run from $0$ to $4$) and
the one-forms $e_{\psi},\ e_{\theta_a},\ e_{\phi_a}$ ($a=1,2$) are given by
(see also \cite{bt}) :
\begin{equation}
e_{\psi}=\frac 13 \left(d\psi+\sum_{a=1}^2 \cos\theta_a\ 
d\phi_a\right),\qquad
e_{\theta_a}=\frac{1}{\sqrt{6}} d\theta_a\,,\qquad 
e_{\phi_a}=\frac{1}{\sqrt{6}} \sin\theta_a\ d\phi_a.
\eqlabel{1forms}
\end{equation}
Additionally, we assume the following ansatz for the 
fluxes $H_3\equiv d B_2$, $F_3$ and the 
dilaton $\Phi$ :
\begin{equation}
F_3=P\ e_{\psi}\wedge \left(e_{\theta_1}\wedge e_{\phi_1}-
e_{\theta_2}\wedge e_{\phi_2}\right),\quad B_2={\tilde k}(y) 
\left(e_{\theta_1}\wedge e_{\phi_1}-
e_{\theta_2}\wedge e_{\phi_2}\right),\quad
\Phi= \Phi(y),
\eqlabel{fdil}
\end{equation} 
where 
$P$ is an integer corresponding to the RR 3-form flux on the compact 3-cycle
(and to the number of fractional branes on the conifold).
Special care should be taken with the RR 5-form. 
From \eqref{fluxes} we get the Bianchi identity
\begin{equation}
dF_5=-F_3\wedge H_3,
\end{equation}
which for the background fluxes \eqref{fdil}
is solved by
\begin{equation}
F_5=dC_4-\bigg({\tilde K}_0+2 P {\tilde k}(y)\bigg)\ 
e_{\psi} \wedge
e_{\theta_1}\wedge e_{\phi_1}\wedge 
e_{\theta_2}\wedge e_{\phi_2}
\eqlabel{f5sol}
\end{equation}
with some constant ${\tilde K}_0$.
In our ansatz the RR four-form does not 
depend on the compact coordinates, that is $C_4\equiv C_4(y)$ (note that
$C_4\wedge F_3\wedge H_3\ne 0$), and the RR five-form is proportional to
the volume form of $\calm_5$ (plus its dual). We define $K(y)$ by
\begin{equation}
dC_4=\frac{K(y)}{\Omega_1 \Omega_2^4}
\ {\rm vol}_{\calm_5}\equiv \frac{K(y)}{\Omega_1 \Omega_2^4}
\sqrt{-\det(g_{\mu\nu})}\ 
dy^{1}\wedge\cdots
\wedge dy^5,
\eqlabel{dc4}
\end{equation} 
and then the self-duality condition \eqref{5self}  implies 
\begin{equation}
K(y)= {\tilde K}_0+2 P {\tilde k}(y)
\eqlabel{kdef}
\end{equation}
(again, in deriving the effective action we should keep $C_4$  
unconstrained and impose this equation later). 
Altogether, from the five-dimensional perspective we allow fluctuations 
in the metric $g_{\mu\nu}(y)$, in the scalar fields
$\Omega_1(y)\,, \Omega_2(y)\,, {\tilde k}(y)\,,  \Phi(y)$
and in the four-form $C_4(y)$ (which is determined in terms of the others
by the self-duality condition). We have set to zero various fluctuations
of the form fields which are $p$-forms on $\calm_5$, and also fluctuations
of $C_2$ of the same form as the fluctuation of $B_2$ in \eqref{fdil},
even though they are
allowed by the symmetries. This is a consistent truncation of the
full ten dimensional supergravity action.

We now perform the KK reduction of \eqref{10action} by plugging into
it the ansatz described above. Recall that 
\begin{equation}
{\rm vol}_{T^{1,1}}\equiv \int e_{\psi} \wedge
e_{\theta_1}\wedge e_{\phi_1}\wedge 
e_{\theta_2}\wedge e_{\phi_2}=\frac {16 \pi^3}{27}.
\end{equation}
First, we have
\begin{equation}
\int_{\calm_{10}} 1\wedge \star 1= 
{\rm vol}_{T^{1,1}}\int_{\calm_5} \Omega_1\Omega_2^4\ {\rm vol}_{\calm_5}.
\eqlabel{int5}
\end{equation}
With a straightforward but somewhat tedious computation 
we find that in the background \eqref{10met}
\begin{equation}
\begin{split}
R_{10}=R_5&-2\om_1^{-1}g^{\lambda\nu}\biggl(\nabla_{\lambda}\nabla_{\nu}\om_1
\biggr)-8\om_2^{-1}g^{\lambda\nu}\biggl(\nabla_{\lambda}\nabla_{\nu}\om_2
\biggr)\\
&-4 g^{\lambda\nu}\biggl(2\ \om_1^{-1}\om_2^{-1}\ 
\nabla_\lambda\om_1\nabla_\nu\om_2
+3\ \om_2^{-2}\ \nabla_\lambda\om_2\nabla_{\nu}\om_2\biggr)\\
&+24\ \om_2^{-2}-4\ \om_1^2\ \om_2^{-4},
\end{split}
\eqlabel{ric5}
\end{equation}
where $R_5$ is the five dimensional Ricci scalar of the metric 
\begin{equation}
ds_{5}^2 =g_{\mu\nu}(y) dy^{\mu}dy^{\nu}.
\eqlabel{5met}
\end{equation}
In \eqref{ric5}, $\nabla_\lambda$ denotes the covariant derivative 
with respect to the metric \eqref{5met}, explicitly given by
\begin{equation}
\begin{split}
\nabla_\lambda \Omega_i &= \del_\lambda \Omega_i,\\
\nabla_\lambda\nabla_\nu \Omega_i &= \del_\lambda\del_\nu \Omega_i
-\Gamma^{\rho} _{\lambda\nu}\ \del_\rho \Omega_i.
\end{split}
\eqlabel{remcov}
\end{equation}
Now, by plugging our ansatz into \eqref{10action} we find that it
reduces to the following 
effective action : 
\begin{equation}
\begin{split}
S_5= \frac{1}{16\pi G_5} \int_{\calm_5} {\rm vol}_{\calm_5}\
\biggl\lbrace &
\Omega_1 \Omega_2^4 \biggl(R_{10}-\frac 12 \nabla_\mu \Phi \nabla^\mu
\Phi\biggr)-\Omega_1 e^{-\Phi}\biggl(
\nabla_\mu {\tilde k}\nabla^\mu {\tilde k}+\frac{P^2 e^{2\Phi}}{\Omega_1^2}
\biggr)\\
&-\frac 14\biggl(\frac{({\tilde K}_0+2 P {\tilde k})^2}{\Omega_1\Omega_2^4}
+\frac{5}{24}\ \Omega_1\Omega_2^4\ \calf_{\mu_1\cdots\mu_5}
\calf^{\mu_1\cdots\mu_5}\biggr)
\biggr\rbrace\\
+\frac{1}{16\pi G_5}\ P\int_{\calm_5}\ & d{\tilde k}\wedge C_4,
\end{split}
\eqlabel{5actionzero}
\end{equation}
where 
\begin{equation}
\calf_{\mu_1\cdots\mu_5}\equiv \del\ _{[\mu_1} C_4\ _{\mu_2\cdots \mu_5]}
=\frac 15 \frac{K}{\Omega_1 \Omega_2^4} 
\sqrt{-\det(g_{\mu\nu})}\ \epsilon_{\mu_1\cdots\mu_5}
\eqlabel{defff}
\end{equation}
($[\cdots]$ denotes anti-symmetrization with 
weight one) 
and $G_5$ is the five dimensional effective gravitational constant  
\begin{equation}
G_5\equiv \frac{G_{10}}{{\rm vol}_{T^{1,1}}}.
\end{equation}
Note that our gravitational action is not the standard five dimensional
action because of the factor of $\Omega_1 \Omega_2^4$ in front of
the five dimensional Einstein-Hilbert term.

In the five dimensional action it turns out to be possible to ``integrate
out'' the field $C_4$ using the self-duality equation \eqref{kdef} and
to obtain an action involving only the other fields. This leads to the
action we will be using in this paper
\begin{equation}
\begin{split}
S_5= \frac{1}{16\pi G_5} \int_{\calm_5} {\rm vol}_{\calm_5}\
 \biggl\lbrace &
\Omega_1 \Omega_2^4 \biggl(R_{10}-\frac 12 \nabla_\mu \Phi \nabla^\mu
\Phi\biggr)-P^2 \Omega_1 e^{-\Phi}\biggl(
\frac{\nabla_\mu K\nabla^\mu K}{4P^4}+\frac{e^{2\Phi}}{\Omega_1^2}
\biggr)\\
&-\frac 12\frac{K^2}{\Omega_1\Omega_2^4}\biggr\rbrace,
\end{split}
\eqlabel{5action}
\end{equation}
where $R_{10}$ is given by \eqref{ric5} and $K(y)$ is related to
${\tilde k}(y)$ by \eqref{kdef}.
 
\subsection{The equations of motion and the ansatz for the solution}\label{eom}

From the effective action \eqref{5action} 
we obtain the following equations of motion :
\begin{equation}
0=\frac{1}{\sqrt{-g}}\ \del_\mu\left[\frac{e^{-\Phi}\om_1}
{2P^2}\sqrt{-g} g^{\mu\nu}\del_\nu K\right]-\frac{K}{\om_1\om_2^4},
\eqlabel{keq}
\end{equation} 
\begin{equation}
0=\frac{1}{\sqrt{-g}}\del_\mu\left[\om_1\om_2^4\sqrt{-g}
g^{\mu\nu}\del_\nu\Phi\right]+\frac{\om_1e^{-\Phi}(\del K)^2}{4P^2}
-\frac{P^2e^\Phi}{\om_1},
\eqlabel{peq}
\end{equation}
\begin{equation}
\begin{split}
0=&\ \om_2^4
R_5-12\om_2^2(\del\om_2)^2+24\om_2^2-12\om_1^2-8\om_2^3\square_5\om_2\\
&\ -\frac 12
\om_2^4(\del\Phi)^2+\frac{P^2e^\Phi}{\om_1^2}-\frac{e^{-\Phi}(\del K)^2}
{4P^2}+\frac {K^2}{2\om_1^2\om_2^4},
\end{split}
\eqlabel{om1}
\end{equation}
\begin{equation}
\begin{split}
0=&\ 4\om_1\om_2^3R_5-8\om_2^3\square_5\om_1-24\om_1\om_2^2\square_5\om_2
-24\om_2^2\del\om_1\del\om_2-24\om_1\om_2(\del\om_2)^2+48\om_1\om_2\\
&-2\om_1\om_2^3(\del\Phi)^2+\frac{2K^2}{\om_1\om_2^5},
\end{split}
\eqlabel{om2}
\end{equation}
\begin{equation}
\begin{split}
\om_1\om_2^4\ R_{5\mu\nu}=&\frac{g_{\mu\nu}}{3}
\left\{\frac{P^2e^\Phi}{\om_1}+\frac{K^2}{2\om_1\om_2^4}
+\square_5\left(\om_1\om_2^4\right)-24\om_1\om_2^2+4\om_1^3\right\}\\
&+\nabla_\mu\nabla_\nu\left(\om_1\om_2^4\right)-4\om_2^3
\left(\del_\mu\om_1\del_\nu\om_2+\del_\nu\om_1\del_\mu\om_2\right)
-12\om_1\om_2^2\del_\mu\om_2\del_\nu\om_2\\
&+\frac{\om_1e^{-\Phi}}{4P^2}\ \del_\mu K\del_\nu K
+\frac 12 \om_1\om_2^4\ \del_\mu \Phi\del_\nu \Phi,
\end{split}
\eqlabel{r5}
\end{equation}
where $(\del F)^2$ denotes $g^{\mu \nu} \del_{\mu} F \del_{\nu} F$ and
$\square_5$ is the Laplacian in the metric \eqref{5met}.

In order to proceed further we make a convenient gauge choice for the
five dimensional metric which separates the radial direction, which
we will call $\rho$, from the four space-time dimensions of the
cascading theory which we will denote by $x^i$ :
\begin{equation}
ds_5^2=h^{-1/2}(x,\r)\r^{-2}\biggl(G_{ij}(x,\r) dx^idx^j\biggr)+
h^{1/2}(x,\r) \r^{-2}(d\r)^2.
\eqlabel{5metric}
\end{equation}
The boundary of the space will be taken to be at $\rho \to 0$.
In this gauge choice some of the off-diagonal components of the metric vanish,
partly fixing the diffeomorphism symmetry.
We also define new scalar fields $f_2$ and $f_3$ related to the $\Omega_i$
fields by
\begin{equation}
\om_1^2=h^{1/2} f_2,\qquad \om_2^2=h^{1/2} f_3.
\end{equation}
The motivation for this parameterization is that in the solution of
\cite{kt} the function $h$ diverges logarithmically near the boundary
$\rho \to 0$, but $G_{ij}$, $f_2$ and $f_3$ approach constant values.
It is not difficult to rewrite the equations of motion \eqref{keq}-\eqref{r5}
using the new variables $G_{ij}$, $h$, $f_2$, $f_3$, $K$ and $\Phi$, and
using the parameterization \eqref{5metric} of the metric, and we
present the results in appendix \ref{neweom}.

We now wish to find the solutions for the cascading theories on
arbitrary space-time manifolds, in an expansion near the boundary at
$\rho\to 0$. In the case of asymptotically anti-de Sitter spaces,
fields in the bulk are dual to operators in the field theory of some
dimension $\Delta$, and they may be expanded in a power series in the
radial $\rho$ coordinate, with a leading term of order
$\rho^{4-\Delta}$ corresponding to the source of the operator,
multiplied by a power series in $\rho^2$, and then a subleading term
of order $\rho^{\Delta}$ corresponding to the one-point function of
the operator (again multiplied by a power series in $\rho^2$)\footnote{
For integer values of $\Delta$ there are also some logarithmic terms.}. In the
cascading gauge theory we expect a similar picture to arise, but with
logarithmic corrections to all the terms 
corresponding to the logarithmic deviation
from conformal invariance, and we will see below that this is indeed
the case.

Let us first analyze the dimensions of the fields in the action
\eqref{5action} in the conformal case of $P=0$ \cite{kw}; note that
naively the action \eqref{5action}
we wrote is singular as $P\to 0$, but if we change
variables from $K(y)$ to ${\tilde k}(y)$ (using \eqref{kdef}) the action
becomes non-singular, so our analysis everywhere in this paper should
be valid also in the $P=0$ case. Obviously, the metric $G_{ij}$ is
dual to a dimension four operator which is just the stress-energy
tensor. The dilaton $\Phi$ and $\tilde k$ both correspond to dimension four
scalar operators (which are the real parts of the two 
exactly marginal single-trace
deformations of the SCFT of \cite{kw} which preserve the $SU(2)\times
SU(2)$ global symmetry), and one can show that combinations of the
scalars $\Omega_1$ and $\Omega_2$ (or $f_2$ and $f_3$) correspond to
operators of dimensions six and eight\footnote{Note that $h$ does not
correspond to an operator in the dual field theory, since it can be
gauged away by reparametrizations of the radial coordinate.}. For
$P=0$ all these fields are decoupled (at leading order in the deformation
from the solution of \cite{kw}), but for non-zero $P$ the
equations of motion couple them all together, and we need to analyze
all of them at the same time.

The usual procedure of holographic renormalization starts by finding
the solution for arbitrary sources, continues by computing the
divergences of the action as a function of the sources, and then
introduces counter-terms to cancel these divergences by expressing
them as a function of the local fields (the transformation from the
sources to the fields is invertible). In our case we have a problem
with implementing this procedure because some of the operators
involved in our action are irrelevant, meaning that we cannot find the
solution with arbitrary sources for these operators as a power series
with bounded powers of the radial coordinate as usual.  In order to
find the sources we need to expand the equations of motion around some
solution to linear order and look at the solution to the linearized
equations of motion which is larger near the boundary. When expanding
the equations of motion of appendix \ref{neweom} around the solution
of
\cite{kt}, which in our parametrization is given (to leading order in
$\rho$) by
\begin{equation}
\begin{split}
G_{ij}(x,\rho) &= \eta_{ij},\\
\Phi(x,\rho) &= \ln(\pz),\\ 
h(x,\rho) &= \frac{1}{8} P^2 \pz + \frac{1}{4} \kz - 
\frac{1}{2} P^2 \pz \ln \rho,\\
K(x,\rho) &= \kz - 2 P^2 \pz \ln \rho,\\
f_2(x,\rho) &= 1,\\
f_3(x,\rho) &= 1,\\
\end{split}
\eqlabel{ktsol}
\end{equation}
(with some constants $\pz$ and $\kz$ which are the parameters of the solution),
we find the following independent solutions to the linearized equations of
motion which we identify with the 
sources for the various operators :
\begin{equation}
\begin{split}
(i) & \ \delta G_{ij} = {\tilde G}_{ij}(x);\\
(ii) & \ \delta K = {\tilde K}(x),\ \delta h = \frac{1}{4} {\tilde K}(x);\\
(iii) & \ \delta \Phi = {\tilde p}(x)/\pz,\ \delta h = \frac{1}{8} P^2
{\tilde p}(x) - \frac{1}{2} P^2 {\tilde p}(x) \ln \rho,\  
\delta K = - 2 P^2 {\tilde p}(x) \ln \rho;\\
(iv) & \ \delta f_3 = \alpha_6(x) \rho^{-2},\ \delta f_2 = - 4 \alpha_6(x)
\rho^{-2},\ \delta h = \alpha_6(x) P^2 \pz \rho^{-2},\ 
\delta K = 2 \alpha_6(x) P^2 \pz \rho^{-2};\\
(v) & \ \delta h = \alpha_8(x) \rho^{-4}.\\
\end{split}
\eqlabel{sources}
\end{equation}
The first three sources are those of the operators which are marginal
for $P=0$ -- the stress-energy tensor and the two scalar operators of
dimension four -- and the last two are the sources of the two irrelevant
scalar operators\footnote{The fifth source $\alpha_8(x)$ naively couples
only to $h$ which, as we mentioned, is not a physical field, but in fact
by a reparametrization of the $\rho$ coordinate one can rewrite the
corresponding solution in a way which involves $f_2$ and $f_3$. We chose
to write the solution in the form above for convenience.}. In all cases
we wrote down only the leading $\rho$-dependence of the solutions -- in
general there are corrections to the expressions above which involve
powers of $\rho^2 \ln^n(\rho)$ (for some integer $n$) multiplying the 
$\rho$-dependence of the terms we wrote,
and which generally involve all the fields (not just the ones which are
turned on at the leading order). 

Finding a solution to the full non-linear equations involving all the sources
above is an ill-defined question since some of these sources are irrelevant.
In order to have a well-defined solution we need to set the sources
$\alpha_6(x) = \alpha_8(x) = 0$ (later we will take these sources to be
infinitesimal in order to compute the correlation functions of the 
corresponding operators, but we cannot take them to be more than
infinitesimal). Once we do this there are no negative powers of $\rho$ in
any of the fields, so all fields have a well-defined expansion in powers
of $\rho$ (and $\ln \rho$). We will also make another simplifying assumption
and we will not introduce any sources for the other two scalar operators,
leaving the corresponding fields to take (as $\rho \to 0$)
the ($x$-independent) values they take
in \eqref{ktsol}. It would be interesting to analyze the solutions with
arbitrary sources for these fields, but we postpone this to future work.
Thus, the only field that we allow an arbitrary source for is the metric,
which we take to be of the form $G_{ij}(x,\rho) = G_{ij}^{(0)}(x) +
{\cal O}(\rho^2 \ln^n(\rho))$. This means that the solutions we construct will
describe the cascading gauge theory compactified on an arbitrary manifold
(with metric $G_{ij}^{(0)}$, since the source for the stress-energy tensor
is just the metric of the cascading field theory), but without any deformations
to its Lagrangian. Of course, the fact that we do not allow arbitrary sources
means that even though we will be able to perform the second step of 
holographic renormalization -- expressing the divergences of the action in
terms of the sources in our equations -- we will not be able to uniquely
translate these divergences into functions of the local fields (since we
have many fields but just one arbitrary source), and we will be forced to
use other methods to determine the counter-terms. We will discuss this
further in the next section.

Next, we would like to find the solution with the source described above.
We do this in a perturbative expansion in $\rho^2$, as usual in holographic
renormalization. The difference from the usual case is that already our
leading order solution contains logarithms of $\rho$, and when we solve the
equations we find that we need even higher powers of logarithms at the higher
orders in $\rho$. We use the following parametrization for the solution :
\begin{equation}
\begin{split}
&G_{ij}(x,\r)=G_{ij}^{(0)}(x)+\r^2\biggl[G_{ij}^{(2,0)}(x)
+\ln\r\ G_{ij}^{(2,1)}(x)\biggr]\\
&\qquad+\r^4\biggl[G_{ij}^{(4,0)}(x)
+\ln\r\ G_{ij}^{(4,1)}(x)+\ln^2\r\ G_{ij}^{(4,2)}(x)+\ln^3\r\ G_{ij}^{(4,3)}(x)\biggr]\\
&\qquad+\calo(\r^6\ln^5\r)
\end{split}
\eqlabel{help1}
\end{equation} 
\begin{equation}
\begin{split}
&h(x,\r)=\frac 18P^2 \pz+\frac 14 \kz 
-\frac 12 P^2 \pz \ln\r+\r^2\biggl[h^{(2,0)}(x)
+\ln\r\ h^{(2,1)}(x)+\ln^2\r\ h^{(2,2)}(x)\biggr]\\
&\qquad+\r^4\biggl[h^{(4,0)}(x)
+\ln\r\ h^{(4,1)}(x)+\ln^2\r\ h^{(4,2)}(x)+\ln^3\r\ h^{(4,3)}(x)
+\ln^4\r\ h^{(4,4)}(x)\biggr]\\&\qquad+\calo(\r^6\ln^6\r)
\end{split}
\eqlabel{help2}
\end{equation} 
\begin{equation}
\begin{split}
&K(x,\r)=\kz  -2P^2\pz\ln\r+\r^2\biggl[K^{(2,0)}(x)
+\ln\r\ K^{(2,1)}(x)\biggr]\\
&\qquad+\r^4\biggl[K^{(4,0)}(x)
+\ln\r\ K^{(4,1)}(x)+\ln^2\r\ K^{(4,2)}(x)+\ln^3\r\ K^{(4,3)}(x)\biggr]\\
&\qquad+\calo(\r^6\ln^5\r)
\end{split}
\eqlabel{help3}
\end{equation} 
\begin{equation}
\begin{split}
&\Phi(x,\r)=\ln \pz+\r^2\biggl[p^{(2,0)}(x)
+\ln\r\ p^{(2,1)}(x)\biggr]\\
&\qquad+\r^4\biggl[p^{(4,0)}(x)
+\ln\r\ p^{(4,1)}(x)+\ln^2\r\ p^{(4,2)}(x)+\ln^3\r\ p^{(4,3)}(x)\biggr]\\
&\qquad+\calo(\r^6\ln^5\r)
\end{split}
\eqlabel{help4}
\end{equation} 
\begin{equation}
\begin{split}
&f_2(x,\r)=1+\r^2\biggl[a^{(2,0)}(x)
+\ln\r\ a^{(2,1)}(x)\biggr]\\
&\qquad+\r^4\biggl[a^{(4,0)}(x)
+\ln\r\ a^{(4,1)}(x)+\ln^2\r\ a^{(4,2)}(x)+\ln^3\r\ a^{(4,3)}(x)\biggr]\\
&\qquad+\calo(\r^6\ln^5\r)
\end{split}
\eqlabel{help5}
\end{equation} 
\begin{equation}
\begin{split}
&f_3(x,\r)=1+\r^2\biggl[b^{(2,0)}(x)
+\ln\r\ b^{(2,1)}(x)\biggr]\\
&\qquad+\r^4\biggl[b^{(4,0)}(x)
+\ln\r\ b^{(4,1)}(x)+\ln^2\r\ b^{(4,2)}(x)+\ln^3\r\ b^{(4,3)}(x)\biggr]\\
&\qquad+\calo(\r^6\ln^5\r)
\end{split}
\eqlabel{help6}
\end{equation} 
where the leading order terms are specified by the parameters
$G_{ij}^{(0)}(x)$, $\pz$ and $\kz$ as above.  Note that there is
a residual reparametrization ambiguity associated with the choice of $h$
in \eqref{5metric}.
We (partially) fix $h$ order by order in the perturbative expansion in
such a way that at each order in $\r$ all fields are given by a finite order
polynomial in $\ln\r$, as indicated in
\eqref{help1}-\eqref{help6}. This still leaves some diffeomorphisms,
of the form
\begin{equation}
\r \to \hr = \r\biggl[1+\r^2\biggl(\dd_{20}+\dd_{21}\ln\r\biggr)+\r^4\biggl(
\dd_{40}+\dd_{41}\ln\r+\dd_{42}\ln^2\r+\dd_{43}\ln^3\r\biggr) + \cdots \biggr],
\eqlabel{reparam}
\end{equation}
unfixed, and this will result in some freedom in the solutions which we will
find.

We can now find the solution by solving the equations of motion of appendix
\ref{neweom} order by order in $\rho$. We have found the general solution
with the boundary conditions described above up to fourth order in
$\rho$, and it is explicitly given in appendix \ref{asssol}. At the
second order in $\rho$ we find that the solution depends on two
arbitrary functions of $x$. This is related to the reparametrization
freedom \eqref{reparam}, which involves two arbitrary constant
parameters at second order. The fact that we find two arbitrary
functions rather than two arbitrary parameters is related to the fact
that $x$-derivatives of fields show up in the equations at higher
orders in $\rho$ than the fields themselves, so we expect the
non-constant part of these functions to be determined at the next
order, and indeed it is (as described in appendix
\ref{asssol}). At the fourth order we similarly find four arbitrary
functions associated to the reparametrization freedom, and we also find
additional arbitrary functions associated (as usual in holographic
renormalization) to the one-point functions of the dimension four operators
(the two scalar operators and the traceless part of the stress-energy
tensor) which are not determined by the UV expansion near the boundary
(but which must be determined by the behavior of the solution at large
values of $\rho$, which we do not discuss here).

The solution we found is rather complicated, and its precise form is
not very illuminating. It is useful to check this solution using some
of the symmetries of the problem. First, the reduced type IIB action
\eqref{5action} is invariant under shifting the dilaton together
with rescaling $P$ : $P \to \alpha P$, $e^{\Phi} \to \alpha^{-2} e^{\Phi}$.
This means that our solution must also be invariant under the same
transformation, and all fields except the dilaton cannot depend on $P$
and $\pz$ separately but only on the combination $P^2 \pz$; it is easy to
verify that this is indeed the case. Two additional symmetries involve
reparametrizations. The reparametrization \eqref{reparam} is a symmetry of
our ansatz and boundary condition, so it must take one solution to another,
and we verify in appendix \ref{checksymms} that this is indeed the case.
We can also consider a scaling symmetry $\rho \to \lambda \rho$; this
does not leave our asymptotic solution invariant, but we can make it into
a symmetry if we also give an appropriate transformation to the metric
$G_{ij}^{(0)}$, $\pz$ and $\kz$. In appendix \ref{checksymms} we verify that
this symmetry is also satisfied by our solution of appendix \ref{asssol}.
The truncated action \eqref{5action} also has
another interesting scaling property : 
it scales by a factor of $\beta^2$ when we take
\begin{equation}
K \to \beta K,\quad
e^{\Phi} \to \beta e^{\Phi},\quad
\Omega_1^4 \to \beta \Omega_1^4,\quad
\Omega_2^4 \to \beta \Omega_2^4,\quad
g_{\mu \nu} \to \beta^{1/2} g_{\mu \nu}.
\eqlabel{symmetry3}
\end{equation}
Since this rescales the action by a constant (without acting on the
coordinates, just on the fields), it is a symmetry of the equations of
motion so it should also be a symmetry of our solution.  In appendix
\ref{checksymms} we verify that the symmetry
\eqref{symmetry3} is satisfied by our solution of appendix
\ref{asssol} as well.

 It is easy to verify that the solution of appendix \ref{asssol} has a
good $P\to 0$ limit, where it describes the asymptotic behavior of the
conformal field theory of \cite{kw} compactified on an arbitrary
four-manifold. In this limit the solution simplifies considerably;
no logarithms appear at second order,
while at fourth order some logarithmic terms can appear 
in the expansions of fields which are dual to operators of dimension four.
The reparametrization freedom
\eqref{reparam} is reduced in this limit just to the terms with no logs,
and correspondingly we have less arbitrariness in the solutions (we should
set the functions $a^{(2,1)}$, $a^{(4,1)}$, $a^{(4,2)}$ and $a^{(4,3)}$
which appear in the solution to zero, and we should also
set $(a^{(4,0)}-b^{(4,0)})=\calo(P)$).

\section{Holographic renormalization}

In this section we describe the holographic renormalization of the
cascading gauge theories.  We begin by regularizing the action and
computing the divergences of
the regularized action.  We find that in addition
to the familiar power-law divergences of asymptotically anti-de Sitter 
(AdS) geometries
one encounters various logarithmic divergences. The logarithmic
divergences are represented by {\it finite} order polynomials in
$\ln\r$, at least in the specific reparametrization
choice\footnote{Recall that $h$ is fixed order-by-order in the
perturbative solution in such a way that only a finite number of
powers of $\ln\r$ appears at each order in $\r$, see
\eqref{help1}-\eqref{help6}.} of $h$ that we made in \eqref{5metric}.

In the previous section we obtained 
the asymptotic solution of the holographic dual to the cascading gauge
theory for arbitrary finite (not infinitesimal)
boundary metric $G_{ij}^{(0)}(x)$ and for constant parameters
$\{\pz, \kz \}$, up to order $\r^4$. 
As discussed above, since the sources we introduced are not arbitrary,
even if we find counter-terms which give rise to a finite regularized
action we are not guaranteed that all correlation functions (given by
derivatives of the action with respect to arbitrary sources) will be
finite. Instead, we have to directly check that we can find counter-terms
that will make all the correlation functions finite. 
Of course, using just the asymptotic solutions that we computed above
we cannot compute arbitrary correlation functions, since generic $n$-point
functions depend on knowing the full solution and not just its asymptotic
form. However, in order to compute one-point functions the asymptotic
solutions are actually enough\footnote{The one-point functions will
often depend on arbitrary functions appearing in the asymptotic solutions
related to expectation values as mentioned above, but no additional
information about the large $\rho$ behavior of the solution
is needed to compute the one-point functions.}, since they are
just given by the derivatives of the action with respect to infinitesimal
sources which can be translated into derivatives of the action with respect
to fields near the boundary. In particular, the one-point functions
corresponding to the first three sources in \eqref{sources} are simply
given by the variation of the action with respect to the parameters
$G_{ij}^{(0)}(x)$, $\kz$ and $\pz$, respectively.

So, our procedure to determine the counter-terms
will be to compute the one-point functions of the operators coupling to the
sources \eqref{sources} and to require that they are all finite. This will not
determine the counter-terms uniquely, and we will see that some ambiguities
will remain even in the one-point functions which we compute, but some general
properties will be independent of these ambiguities and we expect them to
be true for any consistent counter-terms (including the correct ones which
renormalize the theory for arbitrary sources). Note that, as discussed above,
some of our sources correspond to dimension four operators ($T^{ij}$
coupling to ${\tilde G}_{ij}$, $\calo_{\pz}$ coupling to ${\tilde p}$ 
and $\calo_{\kz}$
coupling to $\tilde K$) and we will be
able to compute their one-point functions using our solution directly. On
the other hand, the standard holographic
operator-state correspondence (generalized to our background)
implies that in order to compute the one-point
correlation functions of the operators $\calo_6$ and $\calo_8$ 
coupling to the sources $\alpha_6(x)$ and $\alpha_8(x)$, respectively,
one needs to know the
asymptotic holographic background to order $\rho^6$ and
$\r^8$, respectively. Since we know the supergravity
geometry only to order $\r^4$, we will not be able to
compute these one-point functions, but we can still require that the
contributions to them from the terms we computed should not lead to
divergences. So, we can require
that the renormalized one-point correlation functions of the subtracted
operators\footnote{To be defined in subsection \ref{1point}.} $\calo_6^s$
and $\calo_8^s$ satisfy (up to possible logarithmic corrections)
\begin{equation}
\vev{\calo_6^s}=\calo(1),
\qquad \vev{\calo_8^s}=\calo(\r^{-2}),\qquad {\rm as}\ \r\to 0,
\eqlabel{oconst}
\end{equation} 
which is equivalent to saying that these operators do not have the leading
and the first two subleading power-law divergences.  
We find that requiring that all dimension four one-point functions
are finite and that there are no
power-law divergences in $\calo_6^s$ and
$\calo_8^s$ significantly reduces the ambiguities in the
renormalized one-point functions of the stress-energy tensor.  

We begin by regularizing the action in subsection \ref{regaction}, 
and the
computation of the regularized one-point correlation
functions is explained in subsection \ref{1point}.
In subsection \ref{loccounter} we discuss the local counter-terms that
are needed for the renormalization of $\vev{T_{ij}}$, $\vev{\calo_{\pz}}$,
 $\vev{\calo_{\kz }}$ and for the cancellation of 
power-law divergences in $\vev{\calo_6}$ and
 $\vev{\calo_8}$. In this subsection we use a particular ansatz for the
counter-terms which we call the `minimal subtraction scheme'.

In subsection \ref{1pointren} we present results for the one-point
 correlation functions of the operators $T_{ij}$, $\calo_{\pz}$,
 $\calo_{\kz }$ in the minimal subtraction renormalization
 scheme. 
We also discuss the $P\to 0$ limit of the minimal subtraction 
regularization scheme.
Already in the minimal subtraction scheme there are some ambiguities in
the results, and in more general renormalization schemes additional
ambiguities appear.
In subsection \ref{renamb} we comment on the ambiguities which appear in
general schemes for the renormalization.

Finally, in subsection \ref{rendiv} we discuss other possible renormalization
prescriptions. 
Our main result in this section is that the 
cascading gauge theory can be renormalized with finite one-point functions,
and in particular with a finite stress-energy
tensor. In section \ref{application} we show that this does not contradict
the expectation\footnote{This was originally proposed in \cite{bhkt1},
and further evidence was presented in \cite{bhkt3}.} that at high
temperature the number of effective degrees of freedom
of the cascading gauge theory grows as $K_{eff}^2\propto 
\ln^2(T/\Lambda)$, where $\Lambda$ is the strong coupling scale of
the cascading gauge theory. One might naively think (given the known
thermodynamic properties of the cascading gauge theories
\cite{bhkt1,bhkt2,bhkt3}) that a different renormalization scheme
would be more natural, in which the renormalized one-point functions
are not finite but rather depend on the combination\footnote{Notice
that in this prescription certain $\ln\r$ divergences are allowed, as
long as they come from $K(\r)$.}  $K(\r)=\kz -2 P^2\ln\r$ (evaluated
at the cutoff) rather than on $\kz $, but this does not appear to be
possible.  It would be interesting to explore this second
renormalization scheme in more detail.

\subsection{The regularized action}
\label{regaction}

We write the effective action \eqref{5action} as 
\begin{equation}
S_5=\frac{1}{16\pi G_5}\int_{\calm_5}vol_{\calm_5}\call_{(5)}
\eqlabel{ladef}
\end{equation}
where, when evaluated on a solution to the equations of motion,
\begin{equation}
\begin{split}
\call_{(5)}=&\ 2\om_1\om_2^4\left(\square_5\ [\ln\om_2]+4(\del[\ln\om_2])^2+
\del[\ln\om_1]\del[\ln\om_2]
\right)\\
&+4\om_1(\om_1^2-3\om_2^2).
\end{split}
\eqlabel{laeval}
\end{equation} 
We wish to regularize the theory by imposing a cutoff on the space $\calm_5$,
putting in a boundary $\del \calm_5$ at some $\rho=\rho_0$.
In some cases the effective action \eqref{5action}
evaluated on the equations of motion 
is a total derivative -- for instance, this is true 
for the cascading gauge theory 
on $\reals \times S^3$ or
$\reals \times S^1\times S^2$ -- and in such cases we can rewrite 
\eqref{5action} as an integral just over the boundary, but in general (for
instance on $dS_2\times S^2$ or
$dS_4$) this is not the case. With the ansatz \eqref{5metric} we find
that we can write the action 
evaluated on a solution to the equations of motion as
\begin{equation}
\begin{split}
\sqrt{-g}\call_{(5)}=&\ \frac 12\biggl[\r^{-3}\sqrt{-G}f_2^{1/2}f_3^2[\ln
h]'\biggr]'\\
&+\r^{-5}\sqrt{-G}f_2^{1/2}f_3^2\biggl\{\delta_0+\delta_2
+\delta_4+\delta_6\biggr\},
\end{split}
\eqlabel{laevalexp}
\end{equation} 
where the prime denotes a derivative with respect to $\rho$ and the
subscript in $\delta_i$ indicates the power-law 
scaling of the terms as $\r\to 0$, \ie, $\delta_i\propto \r^{i}$. 
From here on derivative operators and Laplacians will be with respect
to the four dimensional metric $G_{ij}$ rather than the five dimensional
metric. We find 
\begin{equation}
\delta_0=4f_3^{-2}\left(f_2-3f_3\right),
\eqlabel{d0}
\end{equation}
\begin{equation}
\delta_2=\r^2\left(f_3^{-1}f_3''-3\r^{-1}f_3^{-1}f_3'+\frac 12 \square
h\right),
\eqlabel{d2}
\end{equation}
\begin{equation}
\begin{split}
\delta_4=&\r^2\biggl(f_3^{-1}f_3'[\ln\sqrt{-G}]'+f_3^{-2}(f_3')^2+2f_3^{-1}
\del f_3\del h+hf_3^{-1}\square f_3\\
&\qquad +\frac  12
f_2^{-1}f_3^{-1}f_2'f_3'+\frac 14 f_2^{-1}\del f_2\del h
\biggr),
\end{split}
\eqlabel{d4}
\end{equation}
\begin{equation}
\begin{split}
\delta_6=&\r^2h\biggl(f_3^{-2 }(\del f_3)^2+\frac 12
f_2^{-1}f_3^{-1}\del f_2\del f_3
\biggr).
\end{split}
\eqlabel{d6}
\end{equation}

Once we introduce the cutoff as a 
boundary, in order to get consistent equations
of motion we must introduce also a
generalized Gibbons-Hawking (GH) term
\begin{equation}
S_{GH}=\frac{1}{8\pi G_{5}}
\int_{\del\calm_5}d^4x\ \sqrt{-\det(\gamma_{\mu\nu})}
\ \Omega_1\Omega_2^4\ \biggl( 
\nabla_\mu n^{\mu}+n^{\mu}\nabla_\mu\ln\left(\Omega_1\Omega_2^4\right)\bigg),
\eqlabel{bterm}
\end{equation}
where $n^\mu$ is a unit space-like vector orthogonal to the 
four-dimensional boundary $\del\calm_5$, and $\gamma_{\mu\nu}$
is the induced metric on $\del\calm_5$
\begin{equation}
\gamma_{\mu\nu}\equiv g_{\mu\nu}-n_\mu n_\nu.
\eqlabel{indm}
\end{equation}
Note that $\nabla_\mu n^{\mu}$  is nothing but the 
extrinsic curvature of the 
boundary $\del\calm_5$, calculated with the metric  \eqref{5met},
and that the whole boundary term \eqref{bterm} 
coincides with the Kaluza-Klein reduction of the standard  
Gibbons-Hawking term for the action \eqref{10action} with a
nine dimensional boundary $\del\calm_{10}$.
In the ansatz \eqref{5metric} we have
\begin{equation}
\begin{split}
n^{\mu}=&-\delta_\r^\mu\ \r h^{-1/4},\\
\gamma_{ij}=&\r^{-2}h^{-1/2} G_{ij},\\
\sqrt{-\det(\gamma_{\mu\nu})}=&\r^{-4}h^{-1}\sqrt{-G},
\end{split}
\end{equation}
(evaluated at $\rho=\rho_0$) and we find 
\begin{equation}
S_{GH}=\frac{1}{16\pi G_5}\int_{\del\calm_5}d^4x \sqrt{-G}\call_{GH},
\eqlabel{ghlag}
\end{equation}
where 
\begin{equation}
\call_{GH}=-2\r^{-3}f_2^{1/2}f_3^2\biggl\{
\frac 14 [\ln h]'-4\r^{-1}+[\ln\sqrt{-G}]'+\frac 12 f_2^{-1}f_2'+2f_3^{-1}f_3'
\biggr\}.
\eqlabel{callgh}
\end{equation}

The total regularized effective action is
\begin{equation}
\begin{split}
S^\r_4=&\frac{1}{16\pi
G_5}\int_{\del\calm_5}d^4x\sqrt{-G}\call_{(4)}^\r,\\
\sqrt{-G}\call_{(4)}^\r=&\sqrt{-G}\call_{GH}+\int_{\r_0} 
d\r\ \sqrt{-g}\call_{(5)}.
\end{split}
\eqlabel{call4}
\end{equation} 
Generally this will diverge, and we will need to add to it
some counter-term Lagrangian. We define the 
subtracted action to be
\begin{equation}
\begin{split}
S_{tot}=&\frac{1}{16\pi
G_5}\int_{\del\calm_5}d^4x\sqrt{-G}\biggl(\call_{(4)}^\r+
h^{-1}\r^{-4}\call^{counter}
\biggr),
\end{split}
\eqlabel{4eff}
\end{equation} 
where the (local) counter-term Lagrangian must be chosen in such a
way that correlation functions computed from $S_{tot}$ remain finite 
in the limit $\r_0 \to 0$.
The renormalized  action is then simply 
\begin{equation}
S_{eff}=\lim_{\r_0 \to 0} S_{tot}.
\eqlabel{ren4eff}
\end{equation}
As explained in \cite{bfs2}, one should distinguish between $S_{eff}$
and $S_{tot}$, as the variations required to obtain correlation
functions should be performed {\it before} the limit $\r_0 \to 0$ is
taken. This is necessary in order to implement the subtraction
covariantly.

\subsection{Regularized one-point correlation functions}\label{1point}

As explained in the previous subsection, we can write
the subtracted effective action as
\begin{equation}
S_{tot}=S_5+S_{GH}+ S_{ct}
\eqlabel{stot}
\end{equation}
where $S_5$ is the bulk term \eqref{ladef}, $S_{GH}$ is the generalized
Gibbons-Hawking term \eqref{bterm}, and we still need to determine the
counter-term action
\begin{equation}
S_{ct}=\frac{1}{16\pi G_5}\int_{\del \calm_5}d^4x\sqrt{-\ga}\call^{counter}.
\eqlabel{sct}
\end{equation} 
The holographic renormalization is implemented by assuming that
$\call^{counter}$ is a local functional of the fields
$\{\ga_{ij},K,\Phi,\om_1,\om_2\}$ on the regularization boundary
$\del\calm_5$.  Under a generic
variation of the fields in the action we have
\begin{equation}
\begin{split}
\dd S_{tot}=&\int_{\calm_{5}}\sqrt{-g}\biggl\{\ [\cdots]_{\mu\nu}\ \dd g^{\mu\nu}
+[\cdots]\dd\Phi+[\cdots]\dd K+[\cdots]\dd\om_1+[\cdots]\dd\om_2
\biggr\}\\
&+\int_{\del\calm_5}\sqrt{-\ga}\biggl\{[\cdots]_{ij}\ \dd \ga^{ij}+[\cdots]\dd\Phi+[\cdots]\dd K
+[\cdots]\dd\om_1+[\cdots]\dd\om_2
\biggr\},
\end{split}
\eqlabel{h1}
\end{equation}
where $[\cdots]$ in the bulk $\calm_5$ integral stand for the
corresponding five dimensional equations of motion
\eqref{keq}-\eqref{r5}, while the $[\cdots]$ in the boundary $\del\calm_5$ 
integral in \eqref{h1} 
involve only the boundary metric $\ga_{ij}$
and the boundary values of the fields $K,\Phi,\om_1,\om_2$.  Clearly, 
evaluated on a solution to the bulk
equations of motion, $\dd S_{tot}$ does not depend on $\dd
g_{55}$, and thus we have the general expression
\begin{equation}
\dd S_{tot}=\dd S_{tot}\biggl[\dd \ga_{ij},\dd K, \dd \Phi, 
\dd \om_1,\dd \om_2\biggr]
\eqlabel{genvarstot}
\end{equation}
depending only on the values of the fields on $\del \calm_5$.  

In order to compute one-point functions we need to take derivatives of
this action with respect to our sources. As mentioned above, the
one-point function of the stress-energy tensor is the variation with
respect to $G_{ij}^{(0)}$, and we can write the one-point functions of
$\calo_{\pz}$ and $\calo_{\kz}$ as variations with respect to $\pz$
and $\kz$, respectively. Notice that $\{\dd \gamma_{ij},
\dd K, \dd \Phi, \dd \om_1,\dd
\om_2\}$ depend implicitly on the variation of the source boundary
metric $\dd G_{ij}^{(0)}$ and on $\dd \pz$ , $\dd \kz $.  Given
\eqref{genvarstot}, the subtracted one-point correlation functions of
the operators dual to $\{ G_{ij}^{(0)}, \kz , \pz\}$ can be evaluated as 
\begin{equation}
\begin{split}
\vev{\calo_{\pz}^s}\equiv 
\frac{16\pi G_5}{\sqrt{-G}}\ \frac{\dd S_{tot}}{\dd \pz}=&
\frac{16\pi G_5}{\sqrt{-\gamma}}\frac{1}{\r^4 h}\ 
\biggl[\frac{\dd S_{tot} }{\dd \Phi}\frac{\dd \Phi}{\dd \pz}
+\frac{\dd S_{tot} }{\dd K}\frac{\dd K}{\dd \pz}\\ &+\frac{\dd S_{tot}
}{\dd \ga_{ij}}\frac{\dd \ga_{ij}}{\dd \pz} +\frac{\dd S_{tot} }{\dd
\om_1}\frac{\dd \om_1}{\dd \pz}+\frac{\dd S_{tot} }{\dd
\om_2}\frac{\dd \om_2}{\dd \pz}\biggr],
\end{split}
\eqlabel{op0}
\end{equation} 
\begin{equation}
\begin{split}
\vev{\calo_{\kz }^s} \equiv 
\frac{16\pi G_5}{\sqrt{-G}}\ \frac{\dd S_{tot}}{\dd \kz }=&
\frac{16\pi G_5}{\sqrt{-\gamma}}\frac{1}{\r^4 h}\ 
\biggl[\frac{\dd S_{tot} }{\dd \Phi}\frac{\dd \Phi}{\dd \kz }
+\frac{\dd S_{tot} }{\dd K}\frac{\dd K}{\dd \kz }\\
&+\frac{\dd S_{tot} }{\dd \ga_{ij}}\frac{\dd \ga_{ij}}{\dd \kz }
+\frac{\dd S_{tot} }{\dd \om_1}\frac{\dd \om_1}{\dd \kz }+\frac{\dd S_{tot} }
{\dd \om_2}\frac{\dd \om_2}{\dd \kz }\biggr],
\end{split}
\eqlabel{oK0}
\end{equation} 
\begin{equation}
\begin{split}
\vev{\calo_{G^{(0)}}^{ij\ s}}\equiv
\frac{16\pi G_5}{\sqrt{-G}}\ \frac{\dd S_{tot}}{\dd G^{(0)}_{ij}}=&
\frac{16\pi G_5}{\sqrt{-\gamma}}\frac{1}{\r^4 h}\ 
\biggl[\frac{\dd S_{tot} }{\dd \Phi}\frac{\dd \Phi}{\dd G^{(0)}_{ij}}
+\frac{\dd S_{tot} }{\dd K}\frac{\dd K}{\dd G^{(0)}_{ij}}\\
&+\frac{\dd S_{tot} }{\dd \ga_{kl}}\frac{\dd \ga_{kl}}{\dd G^{(0)}_{ij}}
+\frac{\dd S_{tot} }{\dd \om_1}
\frac{\dd \om_1}{\dd G^{(0)}_{ij}}+\frac{\dd S_{tot} }
{\dd \om_2}\frac{\dd \om_2}{\dd G^{(0)}_{ij}}\biggr].
\end{split}
\eqlabel{ot0}
\end{equation} 
The renormalized one-point correlation functions of the corresponding
operators are then simply evaluated as
\begin{equation}
\begin{split}
\vev{\calo_{\pz}}=&\lim_{\r_0 \to 0} \vev{\calo_{\pz}^s},\\
\vev{\calo_{\kz }}=&\lim_{\r_0 \to 0} \vev{\calo_{\kz }^s},\\
8\pi G_5\ \vev{T^{ij}}=& \lim_{\r_0 \to 0} \vev{\calo_{G^{(0)}}^{ij\ s}},
\end{split}
\eqlabel{ren1point}
\end{equation}
where we have defined a standard normalization for $T^{ij}$.  For each
subtracted correlator we find it convenient to separate the
regularized contribution from $S_5+S_{GH}$, and the counter-term
contribution from $S_{ct}$. We can do this separation separately
for every term in the equations -- for example, we can write
\begin{equation}
\begin{split}
\calo_\Phi^s&\equiv\frac{16\pi G_5}{\sqrt{-G}}\ 
\frac{\dd S_{tot}}{\dd \Phi}\equiv 
\frac{1}{\r^4 h}\biggl(\calo_\Phi^\r+\calo_\Phi^c\biggr)\\
&= \frac{1}{\r^4 h}\biggl(\frac{16\pi G_5}{\sqrt{-\ga}}\ \frac{\dd(S_5+S_{GH})}{\dd \Phi}
+\frac{16\pi G_5}{\sqrt{-\ga}}\ \frac{\dd S_{ct}}{\dd \Phi}
\biggr).
\end{split}
\eqlabel{exphi}
\end{equation}
Using \eqref{5action}, \eqref{bterm} we then find that the contributions from
the original action are given by
\begin{equation}
\calo_{\Phi}^\r=\frac{16\pi G_5}{\sqrt{-\ga}}
\frac{\dd (S_5+S_{GH})}{\dd \Phi}=\r h f_2^{1/2}f_3^2 \Phi',
\eqlabel{dsdphi}
\end{equation}
\begin{equation}
\calo_{K}^\r=\frac{16\pi G_5}{\sqrt{-\ga}}\frac{\dd (S_5+S_{GH})}{\dd K}=\frac{f_2^{1/2}\r K'}{2P^2 e^\Phi},
\eqlabel{dsdK}
\end{equation}
\begin{equation}
\calo_{\om_1}^\r=\frac{16\pi G_5}{\sqrt{-\ga}}\frac{\dd (S_5+S_{GH})}{\dd \om_1}=-2\r h^{3/4} f_3^2\biggl([\ln\sqrt{-G}]'+2f_3^{-1}f_3'-4
\r^{-1}\biggr),
\eqlabel{dsdom1}
\end{equation}
\begin{equation}
\begin{split}
\calo_{\om_2}^\r=\frac{16\pi G_5}{\sqrt{-\ga}}
\frac{\dd (S_5+S_{GH})}{\dd \om_2}=-8\r h^{3/4} f_2^{1/2}f_3^{3/2}
\biggl(&\ [\ln\sqrt{-G}]'+\frac 12f_2^{-1}f_2'\\&+\frac 32f_3^{-1}f_3'
-4 \r^{-1}\biggr),
\end{split}
\eqlabel{dsdom2}
\end{equation}
and
\begin{equation}
\begin{split}
\calo_{\ga\ ij}^\r=& \left(-\Theta_{ij}+\Theta\ga_{ij}\right)\Omega_1\Omega_2^4+
n^\lambda\nabla_\lambda\left(\Omega_1\Omega_2^4\right)\ \ga_{ij}\\
=&\r^{-2}h^{1/2}\bigg\{ f_2^{1/2}f_3^2\bigg\{ \frac 12 \r G_{ij}'
+G_{ij}\bigg(3-\r\left[\ln\sqrt{-G}\right]'+\frac 34 \r [\ln h]'\\
&\qquad\qquad\qquad\qquad\qquad\qquad\qquad -\r\left[\ln
\left(h^{5/4}f_2^{1/2}f_3^2\right)\right]'\bigg)
\bigg\}\bigg\}
\end{split}
\eqlabel{tijgh}
\end{equation}
where all expressions should be evaluated at $\rho=\rho_0$ and
\begin{equation}
\Theta^{ij}=\frac 12 \left(\nabla^{i}n^j+\nabla^{j}n^i\right),\qquad 
\Theta=\Theta^{ij}\ga_{ij}.
\eqlabel{thetadef}
\end{equation}
Note that all the regularized correlation functions contain power-law
and logarithmic divergences as the cutoff is removed, $\r_0 \to 0$. Thus,
counter-terms must be determined to remove these divergences.  Once the
counter-term Lagrangian $\call^{counter}$ is specified, one can compute
counter-term contributions to the subtracted operators such as
$\calo_\Phi^s$. Then, using the asymptotic solution explicitly given in
appendix \ref{asssol}, the remaining variational derivatives 
can be evaluated to obtain the subtracted one-point functions. For example,
\begin{equation}
\begin{split}
\vev{\calo_{\pz}^s}=\frac{1}{\r^4 h}\biggl[&\ 
\biggl(\calo_\Phi^\r+\calo_\Phi^c\biggr)\frac{\dd \Phi}{\dd \pz}
+\biggl(\calo_K^\r+\calo_K^c\biggr)\frac{\dd K}{\dd \pz}
+\biggl(\calo_{\om_1}^\r+\calo_{\om_1}^c\biggr)\frac{\dd \om_1}{\dd
\pz}\\ &+\biggl(\calo_{\om_2}^\r+\calo_{\om_2}^c\biggr)\frac{\dd
\om_2}{\dd \pz} +\biggl(\calo_{\ga}^{ij\ \r}+\calo_{\ga}^{ij\
c}\biggr)\ \frac{\dd \ga_{ij}}{\dd \pz}\biggr],
\end{split}
\eqlabel{calp}
\end{equation}
with similar expressions for $\calo_{\kz }^s$, $\calo_{G^{(0)}}^{ij\ s}$.

Similarly we can analyze
the subtracted one-point correlation functions of the operators $\calo_6$ and
$\calo_8$.  Given the 
constant infinitesimal sources $\alpha_6$ and $\alpha_8$ of \eqref{sources}
for the corresponding
dual supergravity fields, we have
\begin{equation}
\begin{split}
\vev{\calo_{8}^s}=\frac{1}{\r^4 h}\biggl[&\ 
\biggl(\calo_\Phi^\r+\calo_\Phi^c\biggr)\frac{\dd \Phi}{\dd \a_8}
+\biggl(\calo_K^\r+\calo_K^c\biggr)\frac{\dd K}{\dd \a_8}
+\biggl(\calo_{\om_1}^\r+\calo_{\om_1}^c\biggr)\frac{\dd \om_1}{\dd \a_8}\\
&+\biggl(\calo_{\om_2}^\r+\calo_{\om_2}^c\biggr)\frac{\dd \om_2}{\dd \a_8}
+\biggl(\calo_{\ga}^{ij\ \r}+\calo_{\ga}^{ij\ c}\biggr)\ 
\frac{\dd \ga_{ij}}{\dd \a_8}\biggr],
\end{split}
\eqlabel{o8}
\end{equation} 
\begin{equation}
\begin{split}
\vev{\calo_{6}^s}=\frac{1}{\r^4 h}\biggl[&\ \biggl(\calo_\Phi^\r+
\calo_\Phi^c\biggr)\frac{\dd \Phi}{\dd \a_6}
+\biggl(\calo_K^\r+\calo_K^c\biggr)\frac{\dd K}{\dd \a_6}
+\biggl(\calo_{\om_1}^\r+\calo_{\om_1}^c\biggr)\frac{\dd \om_1}{\dd \a_6}\\
&+\biggl(\calo_{\om_2}^\r+\calo_{\om_2}^c\biggr)\frac{\dd \om_2}{\dd \a_6}
+\biggl(\calo_{\ga}^{ij\ \r}+
\calo_{\ga}^{ij\ c}\biggr)\ \frac{\dd \ga_{ij}}{\dd \a_6}\biggr],
\end{split}
\eqlabel{o6}
\end{equation} 
where the boundary field variations represent the response of the fields
to turning on infinitesimal sources $\dd \a_8$ and $\dd \a_6$.
Note that in equation \eqref{sources} we only wrote down the source
at leading order in $\r^2$, while naively higher orders in the source
will also contribute to the one-point functions using the equations
above. However, since the divergences have to cancel order by order
in $\r^2$, it is easy to see that after we have canceled the
divergences (in {\it all} the operators) at some order in $\r$, the higher
order terms in the sources (which naively could contribute to a divergence
at the next order) multiply a vanishing expression, so they do not contribute.
The first contribution of the higher order terms in the sources comes with
one higher power of $\r^2$ than the contribution of the leading terms, but
since the latter is required to be finite as $\r_0 \to 0$, the higher order
terms never contribute in this limit.

In the next subsection we describe the construction of the counter-terms
that lead to finite one-point correlation functions
\eqref{ren1point}. As we explained earlier in the section, since we
know the asymptotic solution of the dual supergravity background only
to order $\r^4$, we cannot compute precisely the
subtracted operators $\calo_6^s$ and $\calo_8^s$: at best, we expect
to be able to remove only the leading ($\calo
(\r^{-6}\ln^\#\r)$ and $\calo (\r^{-8}\ln^\#\r)$),
next-to-leading ($\calo (\r^{-4}\ln^\#\r)$ and $\calo
(\r^{-6}\ln^\#\r)$), and next-to-next-to-leading ($\calo
(\r^{-2}\ln^\#\r)$ and $\calo (\r^{-4}\ln^\#\r)$) power-law
divergences in their one-point correlation functions, see
\eqref{oconst}.

\subsection{Local counter-terms}\label{loccounter}

The counter-term Lagrangian in \eqref{sct} must be a local functional of
the fields on $\del\calm_5$. It is useful to separate the dependence
of the counter-terms on the metric $\gamma_{ij}$ 
from the dependence on the other
fields. Given the structure
of the divergences of the regularized correlation functions
\eqref{dsdphi}-\eqref{tijgh},  it is clear that 
the most general form of $\call^{counter}$ must be
\begin{equation}
\begin{split}
\call^{counter}=\call_0+\call_{\calr}\ \calr_\ga+\call_{\calr^2}\ \calr^2_\ga+\call_{\calr ic^2}\ 
\calr_{ab\ \ga}\calr^{ab}_\ga+\call_{\square\calr}\ \square_\ga \calr_\ga+\call_{kinetic},
\end{split}
\eqlabel{lcounterg}
\end{equation} 
where $\{\call_0,\call_{\calr},\call_{\calr^2},\call_{\calr
ic^2},\call_{\square\calr}\}$ are functions of any local fields at
the boundary except for the metric $\ga_{ij}$. Notice that
\eqref{lcounterg} contains counter-terms proportional to $\square_\ga \calr$.
Even though for constant $\kz ,\ \pz$ these are total
derivatives (up to order $\calo(\r^4\ln^\#\r)$), so they do
not contribute to the stress-energy one-point correlation function,
they do contribute to the renormalization of the $\calo_{\pz}$ and
$\calo_{\kz }$ operators. Finally, $\call_{kinetic}$, which scales as
$\calo(\r^4\ln^\#\r)$, contains `kinetic' invariants of the boundary
scalars $\calf_i=\{\Phi,K,\om_1,\om_2\}$ of the type\footnote{Counter-terms
of the form $\calf_1 \del \calf_2 \del \calf_3$ may also be added.}
\begin{equation}
\calf_1\square_\ga \calf_2 = \rho^2  \calf_3\square_\ga R 
+\cdots = \square_\ga\  \bigg[\rho^2 \calf_3 R\bigg] +\cdots
\eqlabel{au0}
\end{equation}
(where we used the form of the solution in which the leading
non-constant terms in the scalar fields are proportional to the
curvature $R$) and, thus, it is also a total derivative to order
$\calo(\r^4\ln^\#\r)$.  Again, even though counter-terms of the type
\eqref{au0} do not contribute to the stress-energy one-point function,
they are necessary for removing $\ln\r$ divergences in the one-point
functions of $\calo_{\pz}$ and $\calo_{\kz }$.

Let us discuss in detail the evaluation of $\calo_{\ga}^{ij\ c}$ 
($\calo_{\Phi}^c$, $\calo_{K}^c$, $\calo_{\om_1}^c$
and $\calo_{\om_2}^c$ can be evaluated analogously).  
We will need the following asymptotic expansions:
\begin{equation}
\begin{split}
\calr_\ga=&h^{1/2}\r^2\biggl(R_\r+\frac 32 \square \ln h\biggr)+
\calo(\r^6\ln^\#\r),\\
\calr_{ij\ \ga}=&R_{ij\ \r}+\frac 12 h^{-1}\nabla_i\nabla_j h+
\frac 14 h^{-1}G_{ij}\square h+\calo(\r^4\ln^\#\r),\\
\calr^2=&h\r^4 R^2+\calo(\r^6\ln^\#\r),\\
\calr_{ab\ \ga}\calr^{ab}_\ga=&h\r^4 R_{ab}R^{ab}+\calo(\r^6\ln^\#\r),
\end{split}
\eqlabel{au1}
\end{equation} 
where 
\begin{equation}
\begin{split}
R_{ij\ \r}=&R_{ij} + {\rho }^2\,\Bigg( \frac{h}{4}\,
         \left( \square R_{ij} -
           2\,\left( R_{iabj}R^{ab} + R_{ik}R^{k}_{j} \right)
          \right) \\
     &\qquad\qquad 
     + \left(\frac{\nabla_i\nabla_j R}{96}+\frac{G_{ij}^{(0)} \square
     R}{192}\right)
         \left( -2\,\kz - P^2\,\pz +
           4\,P^2\,\pz\,\ln \rho \right) \\
   &\qquad\qquad +
      \frac{1}{4}\left({2\,\nabla_i\nabla_j a^{(2,0)} +
         \nabla_i \nabla_j a^{(2,1)} \,\left( 1 + 2\,\ln \rho  \right) }
       \right)\\
      &\qquad\qquad +\frac{1}{8}G_{ij}^{(0)}\left(
         2 \square a^{(2,0)} +
            \square a^{(2,1)}
             \left( 1 + 2 \ln \rho \right) \right)
             \Bigg)
+\calo(\r^4)
\end{split}
\eqlabel{au2}
\end{equation}
and 
\begin{equation}
R_\r=G^{ij} R_{ij \ \r}.
\eqlabel{au3}
\end{equation}
In \eqref{au1} and \eqref{au2} the differential operators on the right
hand side are evaluated with the metric $G_{ij}$.

Given  \eqref{lcounterg} we can compute the contribution of  $S_{ct}$ to the 
subtracted operator $\calo_{\ga}^{ij\ s}$,
\begin{equation}
\begin{split}
\calo_{\ga}^{ij\ c}=&\frac{16\pi G_5}{\sqrt{-\ga}}\ 
\frac{\dd S_{ct}}{\dd \ga_{ij}}\\
=&
\frac 12 \call_0\ \ga^{ij}+
\left(-\calr^{ij}_\ga+\frac 12\calr_\ga 
\ga^{ij}
-\gamma^{ij} \Box_\gamma +
    {\nabla}_{\gamma}^i
    \nabla_{\gamma}^j \right)
\call_\calr
\\
&+\call_{\calr ic^2}\left(\frac 12 \calr_{ab\ \ga}\calr^{ab}_\ga\ga^{ij}+
\nabla^i_\ga\nabla_\ga^j\calr_\ga -\square_\ga 
\calr^{ij}_\ga-\frac 12  \ga^{ij}\square_\ga \calr_\ga+
2\calr^{aijb}_\ga\calr_{ab\ \ga}\right)\\
&+\call_{\calr^2}\left(\frac 12
\calr_\ga^2\ga^{ij}-2\calr_\ga\calr^{ij}_\ga+2\nabla_\ga^i\nabla_\ga^j\calr_\ga
-2\ga^{ij}\square_\ga \calr_\ga
\right)
+\calo(\r^8\ln^\#\r).
\end{split}
\eqlabel{tijcounter}
\end{equation}
Notice that the terms which are relevant for removing divergences
in the one-point functions of $\calo_{\pz}$, $\calo_{\kz}$ and $T^{ij}$
are terms up to order $\r^4$ in $\call_0$, terms up to order $\r^2$ in
$\call_{\calr}$, and terms up to order $\r^0$ in $\call_{\calr^2}$,
$\call_{\calr ic^2}$, $\call_{\square \calr}$ and $\call_{kinetic}$.

Ideally we would like all the counter-terms to be functions just of
the local fields at the boundary. However, already in the
asymptotically AdS case it turns out that it is not possible to do
this, and terms which explicitly involve $\ln \r$ are necessary; these
terms are related to the conformal anomaly. Since our theory reduces
to an asymptotically AdS theory as $P\to 0$, we expect that such
explicit $\r$-dependent terms will be required in our case as well, so
we will allow them in our ansatz. We will find that terms involving up
to three powers of $\ln \r$ are required for renormalizing the
cascading gauge theories.

In this subsection we will discuss a specific ansatz for the
counter-terms which turns out to suffice for obtaining finite
one-point functions; more general possibilities will be discussed
in subsection \ref{renamb}.
We begin by noting that the simple counter-term Lagrangian 
\begin{equation}
\begin{split}
\call^{counter}_{power}=&K-2\om_1^4-8\om_2^4+\calr_\gamma\ \om_1^2
\left(-\frac{1}{12}K+
\frac{1}{12}P^2e^\Phi-\frac16\om_1^4
\right)
\end{split}
\eqlabel{e1power}
\end{equation}
removes all power law divergences in $\calo_{\pz}^s$, $\calo_{\kz}^s$ 
and $\calo_{G_{ij}^{(0)}}^s$, leaving only logarithmic divergences which
still need to be canceled.  Thus, it is convenient to parameterize the
complete counter-term Lagrangian as follows (making a specific choice
for the form of $\call_{kinetic}$ that will turn out to be sufficient) :
\begin{equation}
\begin{split}
\call^{counter}=&\ K-2\om_1^4-8\om_2^4+\cala_4+\cala_6+
\calr_\gamma\ \om_1^2\left(-\frac{1}{12}K+
\frac{1}{12}P^2e^\Phi-\frac16\om_1^4+\calb_2+\calb_4
\right)\\
&+\calr_\gamma^2\ (\call_{\calr^2}^0+\call_{\calr^2}^2)
+\calr_{ab\ \ga}\calr^{ab}_\ga\
(\call_{\calr ic^2}^0+\call_{\calr ic^2}^2)+\square_\ga \calr_\ga\
(\call_{\square\calr}^0 + \call_{\square\calr}^2)\\
&+\delta_1\ \ln\r\ \om_1^6  \square_\ga \Phi+\delta_2\ \om_1^6  \square_\ga \Phi,
\end{split}
\eqlabel{e1}
\end{equation}
where the subscript in $\cala,\calb$ and the superscript in $\call$
indicates the scaling near the boundary of the 
local field configuration represented by
that coefficient, for instance $\cala_n\propto \calo(\r^n\ln^\#\r)$.
Note that the counter-terms
containing $\{\cala_6,\calb_4, \call_{\calr^2}^2,\call_{{\calr
ic}^2}^2,
\call_{\square\calr}^2\}$ do not affect the one-point functions of
$\calo_{\pz}$,  $\calo_{\kz }$ and 
 $\calo_{G_{ij}^{(0)}}$, but they
can contribute to the renormalization of power-law divergences
in the one-point functions of the irrelevant
operators $\calo_6$ and $\calo_8$. Counter-terms scaling as higher
powers of $\r$ do not contribute to any of the one-point functions
we compute so we ignore them.

Since, as discussed above, we do not have a systematic way to determine
the counter-terms, we will write down an ansatz for the counter-terms
and check if it can lead to finite one-point functions (including 
\eqref{oconst}).
There are two scaling symmetries which we can use to constrain the form
of the counter-terms. As mentioned in the previous section, the
type IIB action has a scaling symmetry of $P\to \alpha P$, $e^{\Phi}
\to \alpha^{-2} e^{\Phi}$, and this will also be a property of the
divergent terms in the action, so we can choose our counter-terms to
depend on the dilaton and on $P$ only through the combination $P^2 e^{\Phi}$.
Furthermore, both the truncated action \eqref{5action}
and the generalized Gibbons-Hawking term \eqref{bterm}
have weight two 
under the scaling symmetry \eqref{symmetry3},
$S_5\to\, \beta^2\, S_5$ and $S_{GH}\to\, \beta^2\, S_{GH}$.
 Thus, the divergences will have the
same scaling, so in order to cancel them we need to have the same
scaling also for the counter-terms that we add -- this means that
$\call^{counter}$ in \eqref{sct} should scale with a factor of $\beta$
under this transformation, and we will use this to constrain our
counter-terms. 

In addition we will assume that the counter-terms
contain only non-negative integer powers of $K$ and of $P^2$; again this
is consistent with the structure of the divergences of the action, so
it is reasonable to assume that the ``correct'' counter-terms (which
render all correlation functions finite) should have the same property.

It is convenient to introduce the following short-hand notations
\begin{equation}
\begin{split}
X_a &\equiv \left(1-\frac{\om_2^2}{\om_1^2}\right),\\
X_b &\equiv K-4\om_1^4+\frac 12 P^2 e^\Phi,
\end{split}
\eqlabel{o2}
\end{equation}
for field configurations that scale as $\r^2$.
Using the asymptotic solution of
section \ref{asssol} we can verify that as $\r\to 0$
\begin{equation}
X_a=\calo(\r^2),\qquad X_b=\calo(\r^2\ln\r).
\eqlabel{xaxby}
\end{equation}
We can replace any dependence of the counter-terms on $\Omega_1$ and
$\Omega_2$ by a dependence on $X_a$ and $X_b$, and it will be convenient
to do this in many places because of the scaling \eqref{xaxby}.

Our ``minimal subtraction ansatz'' for the counter-terms involves choosing
the kinetic terms to take the specific form they take in \eqref{e1}. In
addition, when we use the counter-terms \eqref{e1power}
we find that the one-point functions
of $\calo_6^s$ and $\calo_8^s$
do not contain divergent terms with negative powers of $h$ that are
proportional to $R^2$ or $R_{ab}R^{ab}$, and
we require that all counter-terms that we add should preserve this property.
This turns out to restrict $\cala_4$ to be proportional to $X_a^2
\Omega_1^4$ and $\calb_2$ to be proportional to $X_a$, and also to
restrict $\cala_6=\calb_4=0$.
Finally, we restrict all the counter-terms to grow no faster than $\ln^3\r$
(multiplying the appropriate power of $\r$) near the boundary,
consistent with the fact that this is the scaling of the divergences in the
action.
Together with the scaling symmetries described above, this leads to an ansatz
containing $82$
independent coefficients (including $\dd_1,\dd_2$ in \eqref{e1}). By
computing the one-point functions using
the counter-terms in this ansatz we find that 
it is possible to get finite
one-point functions for the operators $O_{\kz },O_{\pz},T_{ij}$ and to satisfy
\eqref{oconst}; these requirements give $75$ constraints on the $82$
coefficients of the ansatz, leading to a $7$-parameter ambiguity in the
counter-terms. In the parameterization \eqref{e1} 
the resulting counter-terms take the form :
\begin{equation}
\begin{split}
\delta_1=&-\frac{50}{21},\\
\cala_4=&\frac{18}{5}\ X_a^2 \om_1^4,\\
\calb_2=& X_a\ \biggl(\frac 16 K-\frac{1}{30}P^2e^\Phi\biggr),
\end{split}
\eqlabel{resf1}
\end{equation}
\begin{equation}
\begin{split}
\call_{\calr^2}^0=&-\frac{1}{144} P^4 e^{2\Phi} \ln^3\r\ -
\frac{1}{96} P^2 e^\Phi \ln^2\r\  K-\frac{1}{192} \ln\r\  K^2\\
&+\left(\frac{1}{96}+4\kappa_1\right) P^4 e^{2\Phi} \ln^2\r\ 
+\left(\frac{1}{96}+4\kappa_1\right) P^2 e^\Phi \ln\r\  K\\
&+\left(\kappa_1+\frac{1}{1152}\right) K^2
+\left(2\kappa_2-\frac{43}{2304}\right) P^4 e^{2\Phi} \ln\r\ 
\\ &+\left(\kappa_2
-\frac{13}{1152}\right) P^2 e^\Phi K+
\kappa_3 P^4 e^{2\Phi},\\
\end{split}
\eqlabel{resf11}
\end{equation}
\begin{equation}
\begin{split}
\call_{{\calr ic}^2}^0=&\frac{1}{48} P^4 e^{2\Phi} \ln^3\r\ +
\frac{1}{32} P^2 e^\Phi \ln^2\r\  K
+\frac{1}{64} \ln\r\  K^2+\left(-\frac{1}{32}-12\kappa_1\right) 
P^4 e^{2\Phi} \ln^2\r\ \\
&+\left(-\frac{1}{32}-12\kappa_1\right) P^2 e^\Phi \ln\r\  K+
\left(-\frac{1}{256}-3\kappa_1\right) K^2
\\
&+\left(\frac{43}{768}-6\kappa_2\right) P^4 e^{2\Phi} \ln\r\  
+\left(\frac{5}{192}-3\kappa_2\right) P^2 e^\Phi K\\
&+\left(\frac{541}{138240}-3 \kappa_3\right) P^4 e^{2\Phi},
\end{split}
\eqlabel{resf12}
\end{equation}
\begin{equation}
\begin{split}
\call_{\square\calr}^0=&\frac{1}{144} P^4 e^{2\Phi} \ln^3\r\ +
\frac{1}{96} P^2 e^\Phi \ln^2\r\  K+\frac{1}{192} \ln\r\  
K^2\\
&+\left(\frac{383}{5760}+4\kappa_4\right) P^4 e^{2\Phi} \ln^2\r\ +
\left(\frac{2231}{40320}+4\kappa_4\right) 
P^2 e^\Phi \ln\r\  K\\ &+\left(\kappa_4+\frac{1}{64}\right) K^2
+\left(-\frac{17}{26880}+2\kappa_5+\frac{7}{320}\dd_2\right) 
P^4 e^{2\Phi} \ln\r\ \\
&+\left(\kappa_5+\frac{1}{64}\dd_2+\frac{29}{2160}\right) 
P^2 e^\Phi K+\kappa_6 
P^4 e^{2\Phi},
\end{split}
\eqlabel{resf2}
\end{equation}
\begin{equation}
\begin{split}
\call_{\calr^2}^2=&X_a \biggl(-\frac{1}{240} P^4 e^{2\Phi} \ln^2\r\ -
\frac{1}{240} P^2 e^\Phi \ln\r\  K-
\frac{1}{720} K^2+P^4 e^{2\Phi} 
\left(\frac{1}{240}+\frac 85 \kappa_1\right) \ln\r\ \\
&\qquad +P^2 e^\Phi \left(\frac 45\kappa_1+\frac{1}{1152}\right) K
+P^4 e^{2\Phi} \left(\frac 25\kappa_2+\frac{43}{34560}\right)\biggr)\\
&+X_b \biggl(-\frac{1}{1152} K+\frac{1}{2304} P^2 e^\Phi\biggr),
\end{split}
\eqlabel{resf21}
\end{equation}
\begin{equation}
\begin{split}
\call_{\calr ic^2}^2=&X_a \biggl(\frac{1}{80} P^4 e^{2\Phi} \ln^2\r\ +
\frac{1}{80} P^2 e^\Phi \ln\r\  K
+\frac{1}{240} K^2+P^4 e^{2\Phi} 
\left(-\frac{1}{80}-\frac{24}{5}\kappa_1\right) \ln\r\ \\
&\qquad -\frac{12}{5}P^2 e^\Phi K \kappa_1-\frac 65 P^4 e^{2\Phi} \kappa_2\biggr)+
X_b \biggl(\frac{1}{384} K-\frac{1}{384} P^2 e^\Phi
\biggr),
\end{split}
\eqlabel{resf22}
\end{equation}
\begin{equation}
\begin{split}
\call_{\square\calr}^2=&X_a \biggl(\frac{1}{240} P^4 e^{2\Phi} \ln^2\r\ +
\frac{1}{240} P^2 e^\Phi \ln\r\   K
-\frac{1}{480} K^2\\
&+P^4 e^{2\Phi} 
\left(\frac 85\kappa_4+\frac{533}{14400}\right) \ln\r\ 
 +\frac{4}{5}P^2 e^\Phi \kappa_4 K+\frac{2}{5}P^4 e^{2\Phi} \kappa_5\biggr)\\
&+
X_b \biggl(\frac{25}{672}P^2e^\Phi\ln\r-\frac{1}{768} K-\left(\frac{167}{23040}+\frac{1}{64}\dd_2\right) 
P^2 e^\Phi\biggr).
\end{split}
\eqlabel{a4b2}
\end{equation}
This result depends on seven parameters : $\kappa_i$ ($i=1,\cdots,6$)
and $\delta_2$.
In appendix \ref{shift} we give a simple argument explaining why the
coefficients $\kappa_i$ turned out to be ambiguous.
In addition, we expect to find an ambiguity in the counter-terms  
corresponding to reparametrizing $\r \to \lambda \r$, because of the
explicit $\r$-dependence in the counter-terms. This is present already
in the asymptotic AdS case, and it explains why the parameter $\delta_2$
turned out to be ambiguous (since this reparametrization shifts
$\delta_2$, in addition to modifying the $\kappa_i$ parameters).
 
\subsection{Renormalized one-point correlation functions}\label{1pointren}

In this subsection we describe in detail our results
in the minimal subtraction renormalization scheme; more general schemes
are described in the next subsection.
\nxt
The one-point function of the stress energy tensor is given by
\begin{equation}
\begin{split}
8\pi G_5 \vev{T_{ij}}=&G_{ij}^{(0)} 
\biggl(R_{ab}R^{ab} \left(\frac{1921}{276480} \pz^2 P^4-\frac{1}{512} \kz^2+\frac{1}{96} \kz  P^2 \pz
\right)\\
&\qquad -R^2 \left(\frac{1}{4608} \kz^2+\frac{337}{51840} \pz^2 P^4+\frac{175}{27648} \kz  P^2 \pz\right)\\
&\qquad +R\left(\frac{1}{16}\kz a^{(2,0)}+\frac{1}{128}P^2\pz a^{(2,0)}+\frac{5}{256}P^2\pz a^{(2,1)}\right)\\
&\qquad +\square R \left(\frac{391}{82944} \pz^2 P^4-\frac{53}{23040} \kz^2+\frac{323}{46080} \kz  P^2 \pz
\right)\biggr)\\
&+R_{aijb} R^{ab} \left(\frac{17}{8640} \pz^2 P^4-\frac{1}{32} \kz^2+\frac{7}{192} \kz  P^2 \pz\right)\\
&-R_i^{\ a} R_{aj} \left( \frac{1}{64} \kz^2+\frac{1}{256} \pz^2 P^4+\frac{1}{64} \kz  P^2 \pz\right)\\
&+R R_{ij} \left(\frac{1691}{103680} \pz^2 P^4-\frac{1}{576} \kz^2+\frac{13}{432} \kz  P^2 \pz\right)\\
&-R_{ij}\left(\frac{1}{16}P^2\pz a^{(2,1)}+\frac 14 \kz a^{(2,0)}\right) \\
&-\nabla_i\nabla_j R \left( \frac{2773}{207360} \pz^2 P^4
+\frac{5}{3456} \kz  P^2 \pz+\frac{7}{1152} \kz^2\right)\\
&+\square R_{ij} \left(-\frac{17}{17280} \pz^2 P^4-\frac{7}{384} \kz  P^2 \pz+\frac{1}{64} \kz^2
\right)\\
&-\nabla_i\nabla_j a^{(2,0)}\left(\frac{1}{16}P^2\pz+\frac{1}{16}\kz\right)
+\nabla_i\nabla_j a^{(2,1)}\left(\frac{7}{128}P^2\pz+\frac{3}{64}\kz\right)\\
&+2 G_{ij}^{(4,0)}-\frac 12 G_{ij}^{(0)} G_a^{(4,0)a}+\frac 32 G_{ij}^{(0)} \left(b^{(4,0)}-a^{(4,0)}\right)
\\
&+T_{ij}^{ambiguity},
\end{split}
\eqlabel{tijfinal}
\end{equation}
where 
\begin{equation}
\begin{split}
T_{ij}^{ambiguity}=&\left(\frac 12\pz^2 P^4 \kappa_3+\frac 12  
\pz P^2 \kappa_2 \kz +\frac 12 \kappa_1 \kz^2\right)\times \\
&\biggl(-2 \nabla_i\nabla_j R+6 \square R_{ij}-12 R_{aijb} R^{ab}-3 G_{ij}^{(0)} R_{ab}R^{ab}+R^2 G_{ij}^{(0)}
-4 R R_{ij}\\
&-\square R G_{ij}^{(0)}\bigg).
\end{split}
\eqlabel{ambiguity}
\end{equation}
\nxt 
The one-point function of the trace of the stress-energy  tensor in
the ``minimal subtraction'' ansatz is 
unambiguously given by
\begin{equation}
\begin{split}
8\pi G_5 \vev{T_{i}^i}=&R_{ab}R^{ab} \biggl(
-\frac{1}{96} \kz  P^2 \pz+\frac{101}{4608}\pz^2P^4+\frac{1}{128}\kz^2
\biggr)\\
&+R^2\biggl(
\frac{11}{2304} \kz  P^2 \pz-\frac{67}{6912}\pz^2P^4-\frac{1}{384}\kz^2
\biggr)\\
&+RP^2\pz\biggl(
\frac{1}{32} a^{(2,0)}+\frac{1}{64}a^{(2,1)}
\biggr)\\
&+\square R \biggl(
\frac{43}{2304} \kz  P^2 \pz+\frac{151}{11520}\pz^2P^4+\frac{1}{384}\kz^2
\biggr)\\
&+6(b^{(4,0)}-a^{(4,0)}).
\end{split}
\eqlabel{traceres}
\end{equation} 
\nxt The one-point function of $\calo_{\pz}$ is given by
\begin{equation}
\begin{split}
\vev{\calo_{\pz}}=&R_{ab}R^{ab} \left(\frac{263}{34560} \pz P^4+
\frac{3}{64} \kz  P^2\right)
+R^2 \left(-\frac{79}{4608} \kz  P^2-\frac{83}{138240} \pz P^4\right)\\
&+RP^2\left(-\frac{19}{192}a^{(2,0)}+\frac{13}{384}a^{(2,1)}
\right)+\square R \left(\frac{77}{3456} \kz  P^2+\frac{33}{1280} \pz P^4\right)\\
&+\frac{1}{\pz} \left(4 p^{(4,0)}-3 \left(b^{(4,0)}-a^{(4,0)}\right)\right)
+\calo_{\pz}^{ambiguity},
\end{split}
\eqlabel{opfinal}
\end{equation}
where 
\begin{equation}
\begin{split}
\calo_{\pz}^{ambiguity}=&(3 R_{ab}R^{ab}-R^2) \left(-2 P^4 \kappa_3 \pz- P^2 \kappa_2
\kz \right)\\
&+\square R \left(2 P^4 \pz \kappa_6+ P^2 \kappa_5 \kz 
+\dd_2 P^2\left(\frac{7}{320}\kz-\frac{3}{640}P^2\pz\right)
\right).
\end{split}
\eqlabel{opambiguity}
\end{equation}
\nxt The one-point function of $\calo_{\kz}$ is given by
\begin{equation}
\begin{split}
\vev{\calo_{\kz }}=&R_{ab}R^{ab} \left(-\frac{1}{192} \kz +\frac{7}{1152} P^2 \pz\right)
+R^2 \left(\frac{1}{2304} \kz +\frac{5}{13824} P^2 \pz\right)\\
&-R\left(\frac{1}{32}a^{(2,0)}+\frac{1}{64}a^{(2,1)}\right)+\square R \left(\frac{53}{1920} \kz +\frac{11}{34560} P^2
\pz\right)\\
&+ \frac{6}{P^2 \pz} \left(a^{(4,0)}
-b^{(4,0)}\right)
+\calo_{\kz }^{ambiguity},
\end{split}
\eqlabel{okfinal}
\end{equation}
where 
\begin{equation}
\begin{split}
\calo_{\kz }^{ambiguity}=&(3 R_{ab}R^{ab}-R^2) \left(-2 \kz  \kappa_1
-\kappa_2 \pz P^2 \right)\\ &+\square R \left(2
\kz  \kappa_4+ \pz P^2 \kappa_5\right).
\end{split}
\eqlabel{okambiguity}
\end{equation}
\nxt
The conformal anomaly is the transformation of the action under
scaling transformations (see \cite{Deser:1996na} for a review). It is
generally defined as (for non-conformal theories that have some beta
functions $\beta_j$ for the couplings to operators $\calo_j$)
\begin{equation}
{\rm conformal\ anomaly} = \vev{T_i^i}-\frac{1}{2} 
\sum_j \beta_j \vev{\calo_j}.
\end{equation}
In our case the form of the asymptotic solution \eqref{ktsol} (and the
analysis in the appendix of the scaling transformation \eqref{sym1}) indicates
that the coupling 
$\kz$ depends on the scale, and that its derivative with respect
to the logarithm of the scale is given by $(-2P^2 \pz)$. Thus, we have
\begin{equation}
{\rm conformal\ anomaly} = \vev{T_i^i}+P^2 \pz \vev{\calo_{\kz }}.
\end{equation}
Using the previous results we find that this is given by 
\begin{equation}
\begin{split}
{\rm conformal\ anomaly} =&\left(3 R_{ab}R^{ab}-R^2\right)
\left(\frac{1}{384}\kz^2-\frac{1}{192}\kz P^2\pz+
\frac{43}{4608}\pz^2 P^4\right)\\
&+\square R\left(\frac{1}{384}\kz^2+\frac{533}{11520}\kz P^2\pz+
\frac{29}{2160}\pz^2 P^4\right)\\
&+P^2\pz\ \calo_{\kz }^{ambiguity}.
\end{split}
\label{ourconfanom}
\end{equation}
Note that, as expected of the conformal anomaly, this is independent
of any parameters (like ($a^{(4,0)}-b^{(4,0)}$)) associated to the IR
behavior of the theory, and that in our ansatz 
all the ambiguities in the conformal
anomaly are related to the ambiguities in the one-point function of
$\calo_{\kz}$. Somewhat surprisingly, we find a finite result for
the conformal anomaly, though its precise value is ambiguous because
of the ambiguities in our counter-terms.

In any local field theory
(see, for instance, \cite{Deser:1996na}) the conformal anomaly is a
linear combination of the Euler density, the Weyl tensor squared
and $\square R$. 
As in other theories dual to gravitational backgrounds the
conformal anomaly that we computed 
does not contain terms proportional to $R_{abcd} R^{abcd}$,
so this requires the conformal anomaly to be
a linear combination of a term proportional to
$(3 R_{ab} R^{ab} - R^2)$ and a term proportional to $\square R$.
We find that in our minimal subtraction ansatz this is indeed the case,
even though it did not necessarily have to be the case because of the
explicit $\r$-dependence of our counter-terms.
\nxt It is straightforward to analyze the $P\to 0$ limit of the 
minimal subtraction renormalization scheme; this provides a holographic
renormalization for a truncation of the conformal field theory of \cite{kw}. 
The $P\to 0$ limit of the
asymptotic solution was discussed in section \ref{eom}. 
It is easy to verify that the one-point functions of the stress tensor and
of $\calo_{\pz}$ have a good $P\to 0$ limit, but we should be more careful
with $\calo_{\kz}$.
Note that in this limit equations \eqref{kdef} and \eqref{fdil} imply
that we should change variables from $K(y)$ to the variable $\tilde{k}(y)$
related to
the two-form, given by $K(y) = \tilde{K}_0 + P \tilde{k}(y)$, which remains
finite in the $P\to 0$ limit.
Let us denote by $\vev{\calo_{\tilde{k}}}$ the operator dual to
$\tilde{k}(y)$. Then, it is easy to see from
\eqref{okfinal} that  $\vev{\calo_{\tilde{k}}}$ is finite, provided that
as $P\to 0$
\begin{equation}
a^{(4,0)}=b^{(4,0)}+\frac 16 P\pz {\tilde k}^{(4,0)}+\calo(P^2),
\eqlabel{b40p0}
\end{equation}
where the parameter ${\tilde k}^{(4,0)}$ is precisely the expectation value 
\begin{equation}
\vev{\calo_{\tilde{k}}}={\tilde k}^{(4,0)}.
\eqlabel{defkv}
\end{equation}

Notice that in the $P\to 0$ limit of the minimal subtraction scheme,
with the scaling \eqref{b40p0}, the scalar one-point functions
$\vev{\calo_{\tilde{k}}}$, $\vev{\calo_{\pz}}$ and the conformal
anomaly $\vev{T_i^i}$ are unambiguous. The conformal anomaly we find
reproduces the known results in the AdS/CFT correspondence 
\cite{Henningson:1998gx}, up
to the well-known \cite{bk} ambiguity in the term in
the conformal anomaly proportional to $\square R$.  This ambiguity
arises from finite (in the $P\to 0$ limit) counter-terms $\call_{\calr
ic^2}^0,
\call_{\calr^2}^0$, and it is proportional to 
\begin{equation}
\vev{T_i^i}\propto \left(\call_{\calr ic^2}^0+
3 \call_{\calr^2}^0\right)\, \square R.
\eqlabel{traceminp0}
\end{equation}
In the minimal subtraction scheme, even though both $\call_{\calr
ic^2}^0$ and $\call_{\calr^2}^0$ are ambiguous,
there is no ambiguity in the combination \eqref{traceminp0}.
If we work directly in the conformal $P=0$ theory, the ambiguity can be
reintroduced by simply shifting
\begin{equation}
\begin{split}
&\call_{\calr^2}^0\to\, \call_{\calr^2}^0+\dd_{\calr^2}\\
&\call_{\calr ic^2}^0\to\, \call_{\calr ic^2}^0+\dd_{\calr ic^2}
\end{split}
\eqlabel{shiftll}
\end{equation}  
where $\dd_{\calr ic^2}$ and $\dd_{\calr^2}$ are arbitrary constants
with $\dd_{\calr ic^2}+3\dd_{\calr^2}\ne 0$.  However, such a simple
modification is not possible in the $P\to 0$ limit of the holographic
renormalization of the cascading gauge theories.  The problem is that
the one-point functions of the irrelevant operators (see \eqref{o8},
\eqref{o6})  are sensitive 
to the renormalized $\vev{T_{i}^i}$, and thus to its ambiguity, which is
$\propto \square R$.  So, if a given set of counter-terms renormalizes
the irrelevant operators, one would expect that a generic shift as in
\eqref{shiftll} would reintroduce divergences $\propto \square R$ in
these one-point functions.
This is indeed what we find. We would like to emphasize that the fact
that we find an 
unambiguous $\vev{T_i^i}$ is only a feature of the minimal subtraction
scheme -- an ambiguity proportional to $\square R$
does appear in more general renormalization
schemes discussed in the next subsection.  

Finally, we would like to
mention that the $P\to 0$ limit of the counter-terms we found in 
our holographic renormalization
correctly reproduces the unambiguous counter-terms of the conformal
holographic renormalization
\cite{Henningson:1998gx,bk}, including also the 
counter-terms with explicit $\ln\r$ dependence \cite{bfs1}.
To see this it is useful to note that in this limit $\Omega_1^4=\Omega_2^4
=h=K/4$, 
so the first line of \eqref{e1} is simply $-\frac{3}{2} K -
\frac{1}{8} K R$, while the second line includes the terms
$\ln(\r^2) K \frac{1}{32} (R_{ab} R^{ab} - \frac{1}{3} R^2)$,
in agreement with equation (5.42) of 
\cite{bfs1} up to overall normalization factors in the metric and in $G_5$
(recall that $K$ is a constant in the $P\to 0$ limit).

\subsection{Ambiguities in the choice of counter-terms}
\label{renamb}

In the previous subsection we presented the results of the ``minimal
subtraction ansatz'' which leads to specific finite one-point functions;
this ansatz and the resulting one-point
functions have a $7$-parameter ambiguity. The ansatz we used is the
simplest one we could find, but we do not have a good argument that it
is correct (in the sense that it leads to all correlation functions being
finite); in particular it is not hard to find more general choices for
$\call_{kinetic}$, and to add terms with more negative powers of the
$\Omega$'s, in a way which preserves the finiteness of all one-point
functions\footnote{One rather ugly feature of our choice of 
$\call_{kinetic}$ is that when we include non-constant sources for the
scalars, there is a term with an explicit $\ln \r$ dependence appearing
already at order $\r^2$. It is possible to choose other forms of
$\call_{kinetic}$ which do not have an explicit dependence on $\r$, but
they do not have a good $P\to 0$ limit. The correct choice should
presumably be determined by introducing more general sources for the
scalar operators, which we hope to do in future work.}.

Since with our limited choice of the sources we do not have a way
to uniquely determine the counter-terms, we have studied various
possibilities for the counter-terms in order to see which of our
results for the one-point functions are modified by more general
counter-terms and which remain true\footnote{We have studied the
most general possible counter-terms which do not contain very large
explicit powers of $\ln \r$ and do not contain large negative powers
of the $\Omega$ fields; we believe that these should be properties of
the ``correct'' counter-terms but we do not have a rigorous argument
for this.}. We find that in a flat
background (with $R_{ij}=0$) the one-point functions are completely
independent of the choice of counter-terms; they are finite and given
by the unambiguous results that we found in the previous subsection
when $R_{ij}=0$. In curved space there are ambiguities in some
one-point functions. Recall that already in asymptotically anti-de
Sitter spaces the stress-energy tensor has a 2-parameter ambiguity
\cite{bk} related to a freedom in the definition of the stress-energy
tensor in curved space. In the minimal subtraction ansatz we found
one of these two ambiguities in the stress tensor \eqref{ambiguity}
(multiplied by an arbitrary linear combination of $P^4 p_0^2$,
$P^2 p_0 K_0$ and $K_0^2$), and in more general renormalization
schemes we find also the other ambiguity (again multiplied by
an arbitrary linear combination of $P^4 p_0^2$,
$P^2 p_0 K_0$ and $K_0^2$), which contributes also a term proportional
to $\square R$ to $\vev{T^i_i}$. It turns out that once these ambiguities
are determined (by choosing a specific definition for the stress-energy
tensor in curved space) the other one-point functions are also
determined, up to a freedom in shifting the one-point functions
of $\calo_{p_0}$ and $\calo_{K_0}$ by terms proportional to
$\square R$; presumably this freedom can also be interpreted as some sort
of ambiguity in the definition of these operators in curved space.

Overall, with the most general ansatz for the counter-terms that we
checked we find an $11$-parameter ambiguity in the results for the
one-point functions. We expect that some of the ambiguities would
remain also once all the correlation functions are renormalized, but
some may disappear. One ambiguity which will always remain in any
theory of this type (involving an explicit $\ln \r$-dependence in the
counter-terms) corresponds to the freedom of redefining the radial
coordinate by $\r \to \lambda \r$. Using the most general
counter-terms we find that the conformal anomaly is not necessarily
given by a linear combination of $3R_{ab}R^{ab}-R^2$ and $\square R$,
as it must be in any local field theory with a gravitational dual; as
mentioned above this is consistent because of the explicit
$\r$-dependence of our counter-terms. If we impose this as an
additional constraint on the counter-terms we find only a
$9$-parameter ambiguity in the one-point functions. Generic
counter-terms do not necessarily have a good $P\to 0$ limit, and this
could also be imposed as an additional constraint.

In summary, in flat space we find unambiguous results for all one-point
functions, while in curved space there is a finite number of ambiguity
parameters (as in asymptotically AdS space). The conformal anomaly
turns out to be finite but ambiguous. This is not surprising given that
it is related to the number of degrees of freedom in the theory at
high energies (it is independent of the IR behavior), which seems to
be ill-defined in the cascading theories -- the surprise is that in
any specific renormalization scheme we actually find a finite answer for the
conformal anomaly, as well as for all other one-point functions.

\subsection{Other possible renormalization schemes}\label{rendiv}
 
The most intriguing feature of 
our results
is that we find
finite results (without any $\ln^\#\r_0$
divergences) for all one-point correlation functions. This is somewhat
surprising since some one-point functions, such as the conformal
anomaly, are supposed to count the number of degrees of freedom in the
theory, which is believed to diverge at high energies.
In particular, in the conformal theory with $P=0$ one has (up to
$\square R$ ambiguities)
\begin{equation}
\begin{split}
8\pi G_5 \vev{T_i^{\ i\ conformal}}= - 
\frac{\kz^2}{16}\biggl[ -\frac 18 R_{ab}R^{ab}+\frac{1}{24}R^2\biggr],\\
\end{split}
\eqlabel{paradox}
\end{equation}
and the coefficient is proportional to
the central charge (in general there are two independent coefficients in
the conformal anomaly, but in theories with gravitational duals they are
always equal to each other). On the other hand, the behavior of the 5-form
flux in the solution of \cite{kt} suggests that the number of degrees of
freedom in the cascading theories diverges at high energies, and also
thermodynamical studies of cascading gauge
theories \cite{bhkt1,bhkt3} suggest that the number of effective
degrees of freedom of the theory accessed at
temperature $T$ is proportional to $K_{eff}^2 \propto 
(P^2 \pz\ln\frac {T}{\Lambda})^2$.
Thus, it may be natural to guess that the conformal anomaly of
the cascading gauge theory would be as in \eqref{paradox}, but with a
replacement
\begin{equation}
\kz \longrightarrow K_{eff}(\r_0)\equiv \kz -2P^2 \pz\ln\r_0 
\eqlabel{replace} 
\end{equation} 
depending explicitly on the cutoff scale. This is quite different
from the finite result that we found above \eqref{ourconfanom}. This leads to a
natural question :
in the context of the
holographic renormalization, is it possible to `renormalize' the cascading
gauge theory in such a way that all one-point correlation functions
contain $K_{eff}$ instead of $\kz$ ? Here, by `renormalize' we
mean that there are no explicit $\ln^\#\r_0$ divergences in one-point
correlation functions, apart from the ones appearing implicitly in
$K_{eff}$. With the tool-set of counter-terms as in section
\ref{loccounter}, requiring that all the counter-terms have a good limit
as $P\to 0$, and that the leading and the first two subleading
power-law divergences of $\calo_6$ and $\calo_8$ are removed
\eqref{oconst}, it is possible to show that such a renormalization
prescription is not possible. Specifically, one finds that the stress
energy one-point function contains certain $\ln^\#\r_0$ terms (even on
manifolds with $\square R=\square R_{ij}=\nabla_i\nabla_j R=0$) which
cannot be subtracted by any counter-terms.  Needless to say, it would
be very interesting to explore these issues further.

\section{Application : cascading gauge theories at finite temperature}
\label{application}

In this section we study the high-temperature thermodynamics of
cascading gauge theories, and we verify that the finite results which
we found in the previous section (which are unambiguous for the
thermodynamics of the theory in flat space) are consistent with
the expectation that the cascading theories will have an effective
``running rank''
$K_{eff}\propto P^2 \pz\ln\frac {T}{\Lambda}$.

The thermodynamics of cascading gauge theories was studied in
\cite{bhkt1,bhkt2,bhkt3}. It was noted there that the $\zet_{2P}$ chiral 
symmetry
is restored in the black brane solutions which are dual to the cascading
theories at high temperatures (compared to the strong coupling 
scale $\Lambda$), and thus the high temperature solutions can be described
using the ansatz we use in this paper. In the previous studies the 
stress-energy tensor was not renormalized, so
the only thermodynamic property which could be extracted
was the entropy density (which depends only on the horizon area). The
high-temperature solutions involve a parameter $K_{\star}$ which is the
value of the five-form flux at the horizon (the minimal value of the
radial coordinate); the solution is constructed in a perturbation
expansion in $P^2/K_{\star}$ (valid at high temperatures), and the leading
term in this expansion is explicitly known \cite{bhkt3}.
The parameter $K_{\star}$ should in principle be determined
in terms of the temperature, and it appears in the computation of
\cite{bhkt3} of the entropy density. 
The authors of \cite{bhkt3} used physically
reasonable (albeit somewhat ad-hoc) arguments to argue that
\begin{equation}
K_\star=2 P^2 p_0 \ln\frac{T}{\Lambda}+\cdots
\eqlabel{kst}
\end{equation} 
where $\cdots$ indicate sub-dominant terms in the high temperature
limit $T\gg \Lambda$. In this section we will use the renormalized 
one-point functions
of the stress-energy tensor to determine rigorously the relation between
$K_\star$ and $T$, and thus the high-temperature thermodynamics of the
cascading gauge theories.
 
Given the results of the previous section for the renormalized
one-point functions of the stress energy tensor, one can compute the
ADM mass-density (the energy density) and the pressure of the black
brane solution which is holographically dual to the cascading gauge
theory at finite temperature, in addition to the entropy density, to
leading order in $P^2/K_*$ (as in \cite{bhkt3}). This allows us to
explicitly verify (to leading order in $P^2/K_{\star}$) the relation
\footnote{Recall that in the absence of a chemical
potential, the free energy density equals minus the pressure.}
\begin{equation}
f=-\calp=\epsilon-T s,
\eqlabel{fedef}
\end{equation}
where $f$ and $\epsilon$ are the free energy and the energy densities, $s$
is the entropy density, and $\calp$ is the pressure. Additionally,
requiring the first law of thermodynamics
\begin{equation}
d\epsilon=T ds
\eqlabel{1stlaw}
\end{equation}
gives an equation which leads to \eqref{kst}. 

The rest of this section is organized as follows. In subsection
\ref{thermo} we discuss the thermodynamics of cascading gauge theories
to leading order in $P^2/K_{\star}$. In subsection \ref{hydro} we
briefly comment on the hydrodynamical properties of the cascading gauge
theory plasma.

\subsection{Thermodynamics of cascading gauge theories}
\label{thermo}

Throughout this section we will use the notations and results of
\cite{bhkt3} \footnote{We have independently verified all the results 
which we actually use.}. In particular, we use $K_\star$ to denote the
5-form flux $K$ evaluated at the horizon, and $a$ for the non-extremality
parameter (which will be related to the temperature below). 
The ten dimensional Einstein frame metric of the
non-extremal cascading solution is
\begin{equation}
\begin{split}
ds_{10}^2=&\left(\frac{8a}{K_\star v}\right)^{1/2} 
e^{2P^2\eta}\left[-(1-v)(dx_0)^2+(dx_{\alpha})^2\right]
+\frac{\sqrt{K_\star}}{32}e^{-2P^2(\eta-5\xi)}\ \frac{dv^2}{v^2(1-v)}\\
&+\frac{\sqrt{K_\star}}{2}e^{-2P^2(\eta-\xi)}\left[e^{-8P^2\omega}e^2_\psi+
e^{2P^2\omega}\left(e_{\theta_1}^2+e_{\phi_1}^2+
e_{\theta_2}^2+e_{\phi_2}^2\right)\right],
\end{split}
\eqlabel{bhmet11}
\end{equation} 
where $\alpha=1,2,3$, 
$v$ is a radial coordinate such that $v\to 1_-$ at the horizon
and $v\to 0_+$ at the boundary, and $\eta, \xi, \omega$ are functions
of $v$. To leading order in $P^2/K_\star$ the 5-form and dilaton are
given by
\begin{equation}
\begin{split}
K=&K_{\star}-\frac{P^2}{2}\ \ln v,\\
\Phi=&\Phi_\star+\frac{1}{4K_{\star}} {\rm Li}_2(1-v),
\end{split}
\eqlabel{bhsol111}
\end{equation} 
where $\Phi_\star$ is the dilaton value at the horizon, and ${\rm
Li}_n(z)$ is the polylogarithm function. As in \cite{bhkt3}, we choose
the dilaton to vanish at the boundary, corresponding to choosing $p_0=1$ in
the notations of the previous sections; we will write our results in
this section using this choice of $p_0$, and the $p_0$-dependence can
always be reinstated by recalling that factors of $p_0$ come together
with factors of $P^2$. The functions $\{\eta, \xi,
\omega\}$ satisfy the following ordinary differential equations :
\begin{equation}
\begin{split}
&v(1-v)\omega''-v \omega'-\frac{3}{4v}\omega=\frac{1}{40K_{\star}},\\
&v(1-v)\xi''-v \xi'-\frac{2}{v}\xi=-\frac{1}{40K_{\star}},\\
&v(1-v)\eta''-v \eta'-\frac{2}{v}\eta=\frac{1}{16K_{\star}v}
\left(2-v-4\ln v\right).
\end{split}
\eqlabel{equations11}
\end{equation} 
The singularity-free solution with the correct asymptotics in the UV is
uniquely determined in terms of $\{a,K_{\star}\}$. For our purposes it
will be necessary to know the asymptotics of the solution near the
boundary.  As computed in \cite{bhkt3}, near the boundary $v\to 0$
\begin{equation}
\begin{split}
&\xi\sim \frac{v}{80K_{\star}}+\cdots,\\
&\eta\sim \frac{\ln v-1}{8K_{\star}}+\cdots,\\
&\omega\sim -\frac{1}{30K_{\star}}v+\cdots,\\
&\Phi\sim \frac{v}{4K_\star}(\ln v-1)+\cdots, 
\end{split}
\eqlabel{bha1}
\end{equation}
where $\cdots$ denote sub-dominant terms.  It is straightforward to
compute the Hawking temperature and the entropy density of the black
brane solution (to leading order in $P^2/K_\star$). We find
\begin{equation}
\begin{split}
&T=\frac{(2a)^{1/4}}{2\pi K_\star^{3/2}}\left(4K_\star-P^2\right),\\
&\frac{s}{T^3}=\frac{\pi^3 K_\star}{64G_5}(K_\star+P^2).
\end{split}
\eqlabel{themoaa1}
\end{equation}
In order to compute the energy and the pressure of the black brane we
need to relate the $v$ coordinate to the $\rho$ coordinate that we
used in our solution 
to evaluate the stress tensor. This is done by comparing the
product of warp factors in front of\footnote{ Notice that the product
of warp factors in \eqref{matcha} does not contain $\ln\r$ factors,
and thus can be consistently evaluated using the (uniformly small in
$\r$) $\calo(P^2)$ solution \eqref{bha1} for $\xi$ and $\omega$.  }
$dx_0^2$ and $e_\psi^2$
\begin{equation}
\begin{split}
h^{-1/2}\r^{-2}\times h^{1/2} f_2\ &=\
\left(\frac{2a}{v}\right)^{1/2}(1-v)\ e^{2P^2(\xi-4\omega)},\\
\r^{-2}\left(1+\cdots\right)\ &=\ \left(\frac{2a}{v}\right)^{1/2} 
\left(1+\cdots\right).
\end{split}
\eqlabel{matcha}
\end{equation}
Thus, to compare we need to define the radial coordinate $\r$ by
\begin{equation}
\r^4\equiv \frac{v}{2a}.
\eqlabel{defra}
\end{equation} 
Given \eqref{defra}, by translating the asymptotic form of the solution 
\eqref{bha1} to our ansatz \eqref{help1}-\eqref{help6} we find 
\begin{equation}
\begin{split}
b^{(4,0)}-a^{(4,0)}=&-\frac{2a P^2}{3K_\star},\\
p^{(4,0)}=&\frac{a(\ln (2a) -1)}{2K_{\star}},\\
G_{00}^{(4,0)}=&-\frac{P^2 a}{4K_\star},\\
G_{\alpha \alpha}^{(4,0)}=&\frac{a}{4K_\star} (8K_\star+P^2).
\end{split}
\eqlabel{asspara}
\end{equation}
For the black brane geometry the renormalized one-point function of the
stress tensor \eqref{tijfinal} is given by
\begin{equation}
8\pi G_5 \vev{T_{ij}}=-\frac{1}{2}G_{ij}^{(0)}\ G_{a}^{(4,0)a}+2G_{ij}^{(4,0)}+
\frac 32 G_{ij}^{(0)}\ (b^{(4,0)}-a^{(4,0)}),
\eqlabel{stressbha}
\end{equation}
where in this background $G_{ij}^{(0)}=\eta_{ij}={\rm diag}(-1,1,1,1)$ is
the Minkowski metric.  We are now in a position to compute the energy
density $\epsilon$ and the pressure $\calp$. We find
\begin{equation}
\begin{split}
\epsilon=&\frac{a}{8\pi G_5 K_\star}(P^2+3K_\star),\\
\calp=&\frac{a}{8\pi G_5 K_\star}(K_\star-P^2).
\end{split}
\eqlabel{enprea}
\end{equation}
With \eqref{themoaa1}, \eqref{enprea} it is straightforward to verify
\eqref{fedef} (to leading order in $P^2/K_{\star}$).  
Given \eqref{themoaa1} we can evaluate $a$ in terms of
$T$ and $K_{\star}$.  We expect that $K_{\star}=K_{\star}(T)$.  Enforcing
the first law of thermodynamics \eqref{1stlaw} leads to a differential
equation on $K_\star$
\begin{equation}
0=2P^2- T \frac{dK_\star}{dT},
\end{equation} 
which leads to (at leading order in $P^2/K_{star}$) 
\begin{equation}
K_\star(T)=2P^2 \ln\frac{T}{\Lambda}
\eqlabel{ksta}
\end{equation}
for some constant $\Lambda$,
as found from different considerations in \cite{bhkt3}.  This allows
us to write the energy density and the pressure \eqref{enprea} purely
in terms of the temperature (to leading order in $P^2/K_\star$), and
they exhibit the expected behavior of an almost-conformal theory with
a number of degrees of freedom proportional to $K_\star(T)^2$.

The fact
that we obtain a finite result for the free energy density should be
useful in analyzing the deconfinement phase transition in this theory
(of course, this requires going beyond the limit of $P^2 \ll K_\star$,
but our renormalization works independently of this limit).

\subsection{Hydrodynamics of cascading gauge theories}
\label{hydro}

Small low-energy deviations from the thermodynamic equilibrium  in
a strongly coupled gauge theory plasma are expected to be well described by
hydrodynamics. In this paper we advocated the definition of cascading
gauge theories in terms of the dual string theory. Thus, the appropriate
description of relaxation processes in cascading gauge theory plasma
is in terms of ``holographic hydrodynamics'', introduced for conformal
gauge theory plasma in \cite{ne2,ne4}. The effective hydrodynamic
description of relaxation of density perturbations in plasma is
completely specified by two viscosity coefficients\footnote{Not to be
confused with $\{\eta,\xi\}$ of the previous section.}, the shear
viscosity $\eta$ and the bulk viscosity $\xi$, and the speed of sound
waves $c_s$. In a conformal gauge theory plasma $\epsilon=3\calp$, and thus
using the well-known relation
\begin{equation}
c_s^2=\frac{\del \calp}{\del\cal\epsilon}
\eqlabel{csdef}
\end{equation}
we find
\begin{equation}
c_s^{conformal}=\frac{1}{\sqrt{3}}.
\eqlabel{csconf}
\end{equation} 
Furthermore, conformal invariance guarantees that the bulk viscosity vanishes 
\begin{equation}
\xi^{conformal}=0,
\eqlabel{n4xi}
\end{equation}
while holographic hydrodynamics predicts \cite{ne1} the ratio of the
shear viscosity to entropy density  in the planar limit and at strong 't Hooft coupling
(namely, in the gravity approximation)\footnote{Finite 't Hooft
coupling corrections to \eqref{esconf} were discussed in \cite{bls}.} to be
\begin{equation}
\frac{\eta}{s}\bigg|^{conformal}=\frac{1}{4\pi}.
\eqlabel{esconf}
\end{equation} 
It is interesting to generalize these results to non-conformal
theories such as the cascading gauge theories. In particular, recall
that QCD has a mass scale, and in some regimes the quark-gluon plasma of
QCD could be described
by strongly coupled hydrodynamics. It is thus of practical
importance\footnote{A possible application is in  hydrodynamics
models describing elliptic flows in heavy ion collision experiments at
RHIC \cite{r1,r2,r3}.} to obtain hydrodynamic predictions for strongly
coupled non-conformal gauge theory plasma (even though the theory we
discuss here is of course very different from QCD).

Somewhat surprisingly, the ratio of shear viscosity to entropy density
in all gauge theory plasmas which are dual to gravitational 
theories (including the cascading
gauge theories), was found \cite{bl1,kss1,bh2} to be universal
in the supergravity approximation,\footnote{Of course,
neither the entropy nor the shear viscosity by itself is universal.}
and given by \eqref{esconf}. On the contrary, the speed of sound
and the sound wave attenuation (which is determined in part by the
bulk viscosity) are not expected to be universal. Indeed, in
\cite{bbs} it was found that explicit breaking of conformal
invariance in strongly coupled gauge theory plasma by fermionic (bosonic) mass
terms $m_f$ 
($m_b$) leads to a modified dispersion relation, with
the speed of sound given in the high temperature ($T\gg m_f$, $T\gg m_b$) 
regime by
\begin{equation}
c_s=\frac{1}{\sqrt{3}}\left(1-\dd_f\frac{m_f^2}{T^2}-\dd_b 
\frac{m_b^4}{T^4}\right).
\eqlabel{csn2star}
\end{equation}
  In
\eqref{csn2star}, $\dd_f$ and $\dd_b$ are positive coefficients.
The cascading gauge theory plasma differs from the non-conformal plasma
discussed in \cite{bbs} in that its scale invariance is broken by
a dynamically generated scale $\Lambda$.  At high temperatures the
cascading gauge theory plasma is expected to resemble a conformal
plasma. Indeed, using the results of the previous subsection
we find that the speed of sound in this plasma is given by
\begin{equation}
\begin{split}
c_s^2=\frac{\del \calp}{\del\cal\epsilon}=\frac{\frac{\del\calp}{\del
T}}{\frac{\del\cal\epsilon}{\del T}} =&\frac 13
-\frac{4P^2}{9K_\star}+\calo(P^4) =\frac 13
-\frac{2}{9\ln\frac{T}{\Lambda}}+\calo(P^4).
\end{split}
\eqlabel{cskt}
\end{equation}
It is amusing to note that the appearance of $\ln T$ in this
correction, suggesting that the cascading gauge theory plasma has (at
least some) hydrodynamic properties similar to those of weakly
interacting relativistic systems (since such a correction is expected
to arise in asymptotically free gauge theories at high temperature,
\ie, at weak coupling).

In would be very interesting to further study the hydrodynamics of
cascading gauge theory plasma, and in particular to evaluate its bulk
viscosity.

\section{Conclusions}

In this paper we have performed a holographic renormalization of the
cascading gauge theory of Klebanov and Tseytlin \cite{kt} compactified
on an arbitrary four-manifold, assuming that the $\zet_{2P}$ global
symmetry is unbroken; we have found counter-terms that can be
added to the action of this theory so that the one-point functions of
all operators (in the truncated action) are finite (on an arbitrary
four-manifold). As discussed in detail above, the holographic
renormalization in these theories is complicated by the fact that we
cannot introduce arbitrary sources for the fields in the truncated
action, since some of them correspond to irrelevant operators.  This
is an interesting problem already for asymptotically anti-de Sitter
spaces, where it is also not clear how to perform a holographic
renormalization for correlation functions of irrelevant operators; the
difference in our case is that in the cascading theory these operators
mix with the metric so we cannot consistently ignore them. Because of
this complication we have performed the analysis with sources for only
some of the operators, and we were not able to uniquely determine the
counter-terms. However, we have proved that there exist counter-terms
that make all the one-point functions finite. The choice of these
counter-terms is ambiguous, and we believe that this ambiguity can be
resolved (up to the usual freedom related to redefinitions of operators)
by requiring that arbitrary correlation functions are finite --
it would be interesting to renormalize more general correlation functions
and verify that this is correct. Within our limited ansatz for the
counter-terms we found that some one-point functions were uniquely
determined, but others were ambiguous -- we conjecture that the
unambiguous one-point functions would remain the same for any consistent
choice of counter-terms (including the correct one which renders all
correlation functions finite), again it would be interesting to verify this
\footnote{Note that even though we discussed the ambiguity in the language
of one-point functions, the ambiguity in the one-point functions that we
found is independent of the state, so it may be thought of as
an ambiguity in the definition of the operators themselves.}.

As we discussed in the introduction, our main result is that the
renormalization of these theories leads to finite one-point functions
despite the infinite number of high-energy degrees of freedom in these
theories; we discussed in section 4 how this is consistent with the
finite temperature behavior. Another possible renormalization of the
cascading gauge theories involves flowing to them (at some finite
scale $\mu$) from finite rank $N$ gauge theories as discussed in 
\cite{Hollowood:2004ek},
assuming that the construction of \cite{Hollowood:2004ek} is valid also
at strong coupling where the gravity approximation that we have been working
in is valid. It may be possible to define the confining gauge theories as
a limit of the construction of \cite{Hollowood:2004ek} in which $N$ and
$\mu$ are both taken to infinity in a correlated way. Then, if one would
compute correlation functions keeping the cutoff scale always above the
scale $\mu$ one would get infinite results (say, for the conformal anomaly)
because of the diverging rank of the high-energy group. However, keeping
the cutoff scale above $\mu$ is problematic in the limit in which $\mu$
goes to infinity; 
we believe that our prescription in which the
cutoff scale goes to infinity only at the end is more natural (and more
analogous to the usual holographic renormalization performed in
asymptotically AdS backgrounds). It would be interesting to study further
various alternative renormalization schemes and to understand where they
agree and where they disagree, in order to understand better how to define
the cascading gauge theories. In the absence of an alternative definition
for the cascading gauge theories we suggest that they can be defined in
terms of their correlation functions, which one can compute using the
procedure we described in this paper (at least in the gravity approximation;
it should be possible to generalize this to the full string theory, but
this involves understanding string theory in Ramond-Ramond backgrounds).

There are many interesting generalizations of our results. It would be
interesting to analyze more general correlation functions; of course,
the precise computation of higher $n$-point functions involves additional
information beyond just the asymptotic solution that we found, but the
counter-terms needed for the finiteness of these correlation functions
can be found purely by using our asymptotic solutions. More precisely,
$n$-point functions of the stress-energy tensor can be analyzed using our
solution, while $n$-point functions of other operators require the
generalization of our solution to more general sources. We expect the
resulting $2$-point functions to agree with the results of 
\cite{Krasnitz:2000ir,Krasnitz:2002ct}
which were computed without carefully regulating the theory. Another direction
is to add additional fields to our truncated action; in particular, in
order to study backgrounds like that of \cite{ks} where the $\zet_{2P}$
symmetry is spontaneously broken, we would need to add to our action
fields that are charged under this symmetry. We believe that it should
be possible in such backgrounds to add a finite number of fields to our
effective action \eqref{5action} and to perform the holographic
renormalization as we did in this paper; it would be interesting to verify
this. Finally, the asymptotic form of many other cascading backgrounds
(generalizing the work of \cite{kt}) has recently been found
\cite{Herzog:2004tr}, and it should be
possible to generalize our results to these backgrounds.

Even without these generalizations there are several interesting
applications of our results, which allow us to compute the (finite)
stress-energy tensor in any solution corresponding to the compactified
cascading theory which preserves the $\zet_{2P}$ symmetry. One example
of such a solution is the cascading theory at finite temperature, in the
high temperature phase in which the $\zet_{2P}$ chiral symmetry is
unbroken. We have computed the thermodynamical properties of this
theory at very high temperatures (compared to the strong coupling scale)
in section 4 above. The solution for arbitrary temperatures is not known,
but given such a solution (which can in principle be found numerically)
our results allow us to precisely compute its free energy. In particular,
our results would allow us to determine the temperature at which the
free energy vanishes, which should be interpreted as the deconfinement
temperature of the cascading gauge theory (as in \cite{Witten:1998zw};
note that in the case analyzed in \cite{Witten:1998zw} a simple subtraction
of the action in the two competing backgrounds was sufficient to renormalize
the action, but we do not expect this to be true in more complicated
backgrounds such as those of the cascading gauge theories\footnote{It is known 
that background subtraction as a method for computing the free energy 
does not work for charged black holes in 
$AdS_5$, and for the supergravity dual to mass 
deformed $\caln=4$ SYM theory \cite{bpaz,bln2,n2hydro}.}). Other
interesting backgrounds, analyzed in \cite{bt}, describe the confining
gauge theory on $S^3\times \reals$, $S^4$ and $dS_4$. Again, given any
solution of this type our results allow us to precisely compute the
stress-energy tensor of that solution (for example, the Casimir energy
of the cascading gauge theory on $S^3$, which our results guarantee will
be finite despite the infinite number of high-energy degrees of freedom).

\section*{Acknowledgments}

It is a pleasure to thank Vijay Balasubramanian, Marcus Berg, Nick Dorey,
Eduard Gorbar, Michael Haack, Gary Horowitz,
Igor Klebanov, Volodya Miransky, Rob Myers,
Kostas Skenderis, Dam Son, Andrei Starinets, Matt Strassler, Marika
Taylor and Arkady Tseytlin for valuable discussions. This work was
supported in part by the Albert Einstein Minerva Center for Theoretical
Physics. OA would
like to thank the Aspen Center for Physics, the Kavli Institute for
Theoretical Physics, and the Perimeter Institute for hospitality
during various stages of this project.  The work of OA was supported
in part by the Israel-U.S. Binational Science Foundation, by the
Israel Science Foundation (grant number 1399/04), by the
Braun-Roger-Siegl foundation, by the European network
HPRN-CT-2000-00122, by a grant from the G.I.F., the German-Israeli
Foundation for Scientific Research and Development, and by Minerva.
AB would like to thank the Weizmann Institute of Science, the University of
Pennsylvania, the Aspen Center for Physics and the Kavli Institute for
Theoretical Physics for hospitality during various stages of this
project. Research at Perimeter Institute is supported in part by
funds from NSERC of Canada. AB acknowledges support by an NSERC
Discovery grant. AY would like to thank the Kavli Institute for
Theoretical Physics for hospitality during this
project. The work of AY is supported in part by a Kreitman
foundation fellowship.

\appendix

\section{Details of the solution}

\subsection{Equations of motion}
\label{neweom}

In what follows we denote by a prime the derivative with respect to
$\r$ and by $\del_i$, $i=0,\cdots,3$, the partial derivative with
respect to $x^i$. Also, we denote for arbitrary functions $g_1, g_2$
on $\calm_5$
\begin{equation}
\begin{split}
&\del g_1\del g_2\equiv G^{ij}\ \del_i g_1\ \del_j g_2,\\
&\square g_1\equiv\frac{1}{\sqrt{-G}}
\del_i\left[\sqrt{-G}G^{ij}\del_j g_1\right].
\end{split}
\eqlabel{notations}
\end{equation}
Note that this is a different notation than the one we used in the
beginning of section \ref{eom}.

The equations of motion of the five dimensional supergravity \eqref{5action}
dual to the cascading gauge theory on an arbitrary curved manifold
$\del\calm_5$, in the metric parameterization \eqref{5metric}, are :
\begin{equation}
\begin{split}
0=&\biggl[e^{-2\Phi}(-G)f_2h^{-2}\r^{-6}\left(K'\right)^2\biggr]'
+e^{-\Phi}\sqrt{-G}K'h^{-1}\r^{-6}\biggl\{
2f_2\del_i\left(e^{-\Phi}\sqrt{-G}G^{ij}\del_jK\right)\\
&+e^{-\Phi}\sqrt{-G}\del f_2\del K\biggr\} -4P^2e^{-\Phi}(-G)KK'
h^{-3}f_3^{-2}\r^{-8}
\end{split}
\eqlabel{keq1}
\end{equation}
\begin{equation}
\begin{split}
0=&\biggl[(-G)f_2f_3^4\r^{-6}\left(\Phi'\right)^2\biggr]'
+\sqrt{-G}\Phi'f_3^{2}\r^{-6}\biggl\{
2f_2\del_i\left(hf_3^2\sqrt{-G}G^{ij}\del_j\Phi\right)\\
&+hf_3^2\sqrt{-G}\del f_2\del \Phi\biggr\} +\frac 12 (-G)\Phi'
P^{-2} h^{-1}f_3^2\r^{-8}
\biggl\{e^{-\Phi}f_2\r^2\left((K')^2+h(\del K)^2\right)\\
&-4e^{\Phi}P^4 \biggr\}
\end{split}
\eqlabel{peq1}
\end{equation}
\begin{equation}
\begin{split}
&\frac{1}{\sqrt{-G}}\biggl[\sqrt{-G}h^{-2}\r^{-3}(hf_2^2)'\biggr]'
-\frac 32 f_2^{-1}h^{-2}\r^{-3}f_2'(hf_2^2)'-\r^{-3}(\del f_2)^2
\\
&+2\r^{-3}f_2\square f_2+\frac 12 \r^{-3}f_2h^{-1}\del f_2\del h
+f_2^2\r^{-3}\square \ln h\\
&=-3P^2e^\Phi\r^{-5}h^{-2}f_2f_3^{-2}-\r^{-5}K^2h^{-3}f_2f_3^{-4}\\
&+\frac 14 h^{-2}\r^{-3}f_2^2f_3^{-2}P^{-2}e^{-\Phi}\biggl\{
(K')^2+h(\del K)^2
\biggr\}\\
&-\r^{-3}h^{-3}f_3^{-2}\biggl\{ (hf_2^2)'(hf_3^2)'+h\del (h
f_2^2)\del(h f_3^2)\biggr\} +16\r^{-5}h^{-1} f_3^{-2}f_2^3
\end{split}
\eqlabel{om11}
\end{equation}
\begin{equation}
\begin{split}
&\frac{1}{\sqrt{-G}}\biggl[\sqrt{-G}h^{-2}\r^{-3}(hf_3^2)'\biggr]'
-\frac 32 f_3^{-1}h^{-2}\r^{-3}f_3'(hf_3^2)'-\r^{-3}(\del f_3)^2
\\
&+2\r^{-3}f_3\square f_3+\frac 12 \r^{-3}f_3h^{-1}\del f_3\del h
+f_3^2\r^{-3}\square \ln h\\
&= -P^2e^\Phi\r^{-5}h^{-2}f_2^{-1}-\r^{-5}K^2h^{-3}f_2^{-1}f_3^{-2}
-\frac 14 h^{-2}\r^{-3}P^{-2}e^{-\Phi}\biggl\{ (K')^2+h(\del K)^2
\biggr\}\\
&-\frac 14\r^{-3}h^{-3}f_2^{-2}\biggl\{
(hf_2^2)'(hf_3^2)'+h\del (h f_2^2)\del(h f_3^2)\biggr\}\\
&-\frac 34\r^{-3}h^{-3}f_3^{-2}\biggl\{ (hf_3^2)'(hf_3^2)'+h\del (h
f_3^2)\del(h f_3^2)\biggr\} +4\r^{-5}h^{-1}(6f_3-2f_2)
\end{split}
\eqlabel{om21}
\end{equation}
\begin{equation}
\begin{split}
&R_{5ij}=\ G_{ij}\biggl\{ \frac 13
\r^{-2}h^{-2}f_2^{-1}f_3^{-2}P^2e^{\phi}+\frac 16\r^{-2}
h^{-3}f_2^{-1}f_3^{-4}K^2+\frac{1}{12}\r^3h^{-1}f_2^{-2}f_3^{-2}\biggl(
\\
&\frac{1}{\sqrt{-G}}\left[\sqrt{-G}\r^{-3}f_3^{-6}h^{-5}[h^5f_2^2
f_3^8]'\right]'+\frac{1}{\sqrt{-G}}\del_i\left[
\sqrt{-G}\r^{-3}f_3^{-6}h^{-4}G^{ij}\del_j[h^5f_2^2
f_3^8]\right]\\
&-\frac 32 \r^{-3}f_3^{-6}h^{-5}f_2^{-1}f_2'[h^5f_2^2 f_3^8]'-\frac
32 \r^{-3}f_3^{-6}h^{-4}f_2^{-1}\del f_2\del[h^5f_2^2
f_3^8]\biggr)\\
&-8\r^{-2}h^{-1}f_3^{-1}+\frac 43 \r^{-2}
h^{-1}f_2f_3^{-2}\biggr\}\\
&+\biggl\{\nabla_i\nabla_j\left[\frac 54 \ln h+\frac 12 \ln
f_2+2\ln f_3\right]+\frac 58h^{-2}\del_ih\del_jh\\
&+\frac 18
h^{-1}f_2^{-1}\left(\del_ih\del_jf_2+\del_if_2\del_jh\right) +\frac
12
h^{-1}f_3^{-1}\left(\del_ih\del_jf_3+\del_if_3\del_jh\right)\\
&-G_{ij}\left(\frac{5}{16}h^{-2}(\del h)^2+\frac 18h^{-1}f_2^{-1}
\del h\del f_2+\frac 12 h^{-1}f_3^{-1}\del h \del f_3\right)\\
&+\left(\frac 12 h^{-1}G_{ij}' -\frac 14
h^{-2}G_{ij}h'-\r^{-1}h^{-1}G_{ij} \right)\left(\frac 54
h^{-1}h'+\frac 12 f_2^{-1}f_2'+2f_3^{-1}f_3'\right)
\biggr\}\\
&+\biggl\{\frac {5}{16} h^{-2}\del_ih\del_jh+\frac 18
h^{-1}f_2^{-1}\left(\del_ih\del_jf_2+\del_if_2\del_jh\right)
\\
&+\frac 12
h^{-1}f_3^{-1}\left(\del_ih\del_jf_3+\del_if_3\del_jh\right) +\frac
14 f_2^{-2}\del_if_2\del_jf_2 +f_3^{-2}\del_if_3\del_jf_3
\biggr\}\\
&+\biggl\{ \frac 14
P^{-2}h^{-1}f_3^{-2}e^{-\Phi}\del_iK\del_jK+\frac 12 \del_i\Phi
\del_j\Phi \biggr\}
\end{split}
\eqlabel{ricciij}
\end{equation}
\begin{equation}
\begin{split}
&R_{5i\r}=\biggl\{
\del_i\left[\frac 54 h^{-1}h'+\frac 12 f_2^{-1}f_2'+2f_3^{-1}f_3'\right]\\
&-\frac 12 G^{kn}G_{ni}'\left(\frac 54 h^{-1}
\del_kh+\frac 12 f_2^{-1}\del_kf_2+2f_3^{-1}\del_kf_3\right)\\
&+\left(\frac 14 h^{-1}h'+\r^{-1}\right)\left(\frac 54 h^{-1}
\del_ih+\frac 12 f_2^{-1}\del_if_2+2f_3^{-1}\del_if_3\right)\\
&-\frac 14 h^{-1}\del_ih \left(\frac 54 h^{-1}h'+\frac 12
f_2^{-1}f_2'+2f_3^{-1}f_3'\right)
\biggr\}\\
&+\biggl\{ \frac{5}{16}h^{-2}\del_ihh'+\frac 18 h^{-1}f_2^{-1}\left(
\del_ihf_2'+\del_if_2h'\right)+\frac 12 h^{-1}f_3^{-1}\left(
\del_ihf_3'+\del_if_3h'\right)\\
&+\frac 14 f_2^{-2}\del_if_2f_2'+ f_3^{-2}\del_if_3f_3'
\biggr\}\\
&+\biggl\{ \frac 14 h^{-1}f_3^{-2}P^{-2}e^{-\Phi}\del_iKK'+\frac 12
\del_i\Phi\Phi' \biggr\}
\end{split}
\eqlabel{ricciip}
\end{equation}
\begin{equation}
\begin{split}
&R_{5\r\r}=\biggl\{ \frac 13
\r^{-2}h^{-1}f_2^{-1}f_3^{-2}P^2e^{\phi}+\frac 16\r^{-2}
h^{-2}f_2^{-1}f_3^{-4}K^2+\frac{1}{12}\r^3f_2^{-2}f_3^{-2}\biggl(
\\
&\frac{1}{\sqrt{-G}}\left[\sqrt{-G}\r^{-3}f_3^{-6}h^{-5}[h^5f_2^2
f_3^8]'\right]'+\frac{1}{\sqrt{-G}}\del_i\left[
\sqrt{-G}\r^{-3}f_3^{-6}h^{-4}G^{ij}\del_j[h^5f_2^2
f_3^8]\right]\\
&-\frac 32 \r^{-3}f_3^{-6}h^{-5}f_2^{-1}f_2'[h^5f_2^2 f_3^8]'-\frac
32 \r^{-3}f_3^{-6}h^{-4}f_2^{-1}\del f_2\del[h^5f_2^2
f_3^8]\biggr)\\
&-8\r^{-2}f_3^{-1}+\frac 43 \r^{-2}
f_2f_3^{-2}\biggr\}\\
&+\biggl\{
\left[\frac 54 h^{-1}h'+\frac 12 f_2^{-1}f_2'+2f_3^{-1}f_3'\right]'\\
&+\left(\frac{5}{16}h^{-1}(\del h)^2+\frac 18 f_2^{-1}\del h\del
f_2+
\frac 12 f_3^{-1}\del h \del f_3\right)\\
&-\left(\frac{5}{16}h^{-2}( h')^2+\frac 18 f_2^{-1}h^{-1}h'f_2'+
\frac 12 f_3^{-1}h^{-1} h' f_3'\right)\\
&+\left(\frac{5}{4}\r^{-1}h^{-1}h'+\frac 12 f_2^{-1}\r^{-1}f_2'+ 2
f_3^{-1}\r^{-1}  f_3'\right)
\biggr\}\\
&+\biggl\{ \left(\frac 14 h^{-1}h'+\frac 12 f_2^{-1} f_2'\right)^2+
4\left(\frac 14 h^{-1}h'+\frac 12 f_3^{-1} f_3'\right)^2
\biggr\}\\
&+\biggl\{ \frac 14 h^{-1}f_3^{-2}P^{-2}e^{-\Phi}(K')^2+\frac 12
(\Phi')^2 \biggr\}
\end{split}
\eqlabel{riccirr}
\end{equation}

\subsection{Coefficients of the asymptotic solution}\label{asssol}

\subsubsection{Next-to-leading order solution, $\calo(\r^2)$}

To next-to-leading order, the solution with the ansatz
\eqref{help1}-\eqref{help6} depends on two undetermined functions,
which we choose to be $a^{(2,0)}(x)$ and $a^{(2,1)}(x)$. We find
that the other coefficients in the solution are then given by
\begin{equation}
\begin{split}
    G^{(2,1)}_{ij}=&-\frac{1}{24}\left(12\,a^{(2,1)}\,G_{ij}^{(0)} -
    P^2\,\pz\,\left( 6\,R_{ij} -
       G_{ij}^{(0)}\,R \right)\right)
\end{split}
\eqlabel{rij21} 
\end{equation}
\begin{equation}
\begin{split}
    G^{(2,0)}_{ij}=&-\frac{1}{96}\left(48\,a^{(2,0)}\,G_{ij}^{(0)}
    +24\,a^{(2,1)}\,G_{ij}^{(0)}+
    \left( 2\,\kz + P^2\,\pz \right) \,
     \left( 6\,R_{ij} -
       G_{ij}^{(0)}\,R \right) \right)
\end{split}
\eqlabel{rij20} 
\end{equation}
\begin{equation}
\begin{split}
    p^{(2,1)}=&\ 0 
\end{split}
\eqlabel{rp21} 
\end{equation}
\begin{equation}
\begin{split}
   p^{(2,0)}=&-\frac{1}{24}P^2 \pz R
\end{split}
\eqlabel{pr20} 
\end{equation}
\begin{equation}
\begin{split}
    h^{(2,2)}=&\ P^2 \pz a^{(2,1)}
\end{split}
\eqlabel{rh22} 
\end{equation}
\begin{equation}
\begin{split}
    h^{(2,1)}=&\ P^2\,\pz\,a^{(2,0)} - \frac{1}{288}
  \left(36\,\left( 4\,\kz + P^2\,\pz \right) \,
      a^{(2,1)} +
     5\,P^4\,\pz^2\,R \right)
\end{split}
\eqlabel{rh21}
\end{equation}
\begin{equation}
\begin{split}
    h^{(2,0)}=\frac{1}{3456} \biggl(&-432\,\left( 4\,\kz + P^2\,\pz \right) \,
     a^{(2,0)}\\&+
    P^2\,\pz\,\left( -216\,a^{(2,1)} +
       5\,\left( 6\,\kz + 11\,P^2\,\pz \right) \,
        R \right) \biggr)
\end{split}
\eqlabel{rh20}
\end{equation}
\begin{equation}
\begin{split}
    b^{(2,1)}=&\ a^{(2,1)}
\end{split}
\eqlabel{rb21}
\end{equation}
\begin{equation}
\begin{split}
    b^{(2,0)}=&\ a^{(2,0)} - \frac{1}{96} P^2\,\pz\,
     R
\end{split}
\eqlabel{rb20}
\end{equation}
\begin{equation}
\begin{split}
    K^{(2,1)}=&-\frac{1}{24} P^2\,\pz\,
      \left( -12\,a^{(2,1)} +
        P^2\,\pz\,R \right)
\end{split}
\eqlabel{rk21}
\end{equation}
\begin{equation}
\begin{split} 
K^{(2,0)}=&\frac{1}{96} P^2\,\pz\,\left( 48\,a^{(2,0)} -
      24\,a^{(2,1)} +
      \left( 2\,\kz + 7\,P^2\,\pz \right) \,
       R \right)
\end{split}
\eqlabel{rk20}
\end{equation}
where the Ricci tensor $R_{ij}$ and scalar $R$ are computed using
the boundary metric $G_{ij}^{(0)}$.

\subsubsection{Next-to-next-to-leading order solution, $\calo(\r^4)$}

At this order, we have several arbitrary functions, which we choose
to be $p^{(4,0)}$, $a^{(4,3)}$, $a^{(4,2)}$, $a^{(4,1)}$,
$a^{(4,0)}$,  $b^{(4,0)}$, and $G_{ij}^{(4,0)}-\frac 14 G_{ij}^{(0)}
G_k^{(4,0)k}$ (with $G_k^{(4,0)k}\equiv G_{ij}^{(4,0)}G^{(0)ij}$;
note that $G_k^{(4,0)k}$ is not arbitrary but will be determined
below). On the other hand, in order to find a solution we find that
the Laplacians of the second order coefficients
$\{a^{(2,0)},a^{(2,1)}\}$ are constrained to satisfy
\begin{equation}
\begin{split}
\square  a^{(2,0)}=&-\frac{1}{2160}(161P^2\pz+78\kz ){\square R,}\\
\square  a^{(2,1)}=&\frac{13}{180}P^2\pz {\square R},
\end{split}
\eqlabel{constraints4}
\end{equation}
where $\square$ is the Laplacian operator with the metric\footnote{
We make a convention that $\square $ is evaluated with respect to
$G_{ij}^{(0)}(x)$ whenever it acts on the coefficients of the
perturbative solution \eqref{help1}-\eqref{help6}, rather than
on the supergravity fields.  In the latter case the
operator $\square $ is evaluated with respect to the {\it full} four
dimensional metric $G_{ij}(x,\r)$, as in the equations of motion
\eqref{keq1}-\eqref{riccirr}.  } $G^{(0)}_{ij}$. The constant parts
of $a^{(2,0)}$ and $a^{(2,1)}$ remain unfixed, as expected.  The
remaining functions in the solution are then given by
\begin{equation}
\begin{split}
  &G^{(4,3)}_{ij}=\frac{1}{96}\, P^4\,\pz^2\, \square R_{ij} -
  \frac{1}{288}\, P^4\,\pz^2\, \nabla_i \nabla_j R -
  \frac{1}{48}\, P^4\,\pz^2\,R_{aijb}R^{ab}  -
  \frac{1}
   {144}\, P^4\,\pz^2\,R_{ij}\,R \\
   &\qquad+ G_{ij}^{(0)}\,
   \left( -\frac{1 }{576}\, P^4\,
          \pz^2\, \square R -
     \frac{1}{192}\, P^4\,\pz^2\,R_{ab}R^{ab} -
     \frac{3}{4}\, a^{(4,3)} +
     \frac{1}{576}\, P^4\,\pz^2\,{R}^2
     \right)
\end{split}
\eqlabel{rij43}
\end{equation}
\begin{equation}
\begin{split}
    &G^{(4,2)}_{ij}=-\frac{1}{64}
\,P^2\,\pz\,
       \left( \kz + P^2\,\pz \right)\, \square R_{ij}
 +
  \frac{1}{576}\,P^2\,\pz\,
     \left( 3\,\kz + 4\,P^2\,\pz \right)\, \nabla_i\nabla_j R  \\
     &+
  \frac{1}{32}\, P^2\,\pz\,\left( \kz + P^2\,\pz
       \right) \,R_{aijb}R^{ab} +
  \frac{1}{32}\, P^4\,\pz^2\,R_{ia}R^{a}_{j} -
  \frac{1}{16}\, P^2\,\pz\,\nabla_i\nabla_j a^{(2,1)} \\
  &+
  R_{ij}\,\left( -\frac{1}{4}\, P^2\,\pz\,
          a^{(2,1)} +
     \frac{1}{288}\, P^2\,\pz\,
        \left( 3\,\kz - 2\,P^2\,\pz \right) \,
        R \right)  \\
  &+
  G_{ij}^{(0)}\,\biggl( \frac{1 }{1152}\, P^2\,
        \pz\,\left( 3\,\kz +
          4\,P^2\,\pz \right)\, \square R +
     \frac{1}{128}\, P^2\,\pz\,
        \left( \kz + P^2\,\pz \right) \,
        R_{ab}R^{ab} \\
     &\qquad+
     \frac{1}{16}\left(-12\,a^{(4,2)} +
        7\,(a^{(2,1)})^2 - 3\,a^{(4,3)}\right)
      + \frac{1}{24}\, P^2\,\pz\,a^{(2,1)}\,
        R\\
   &\qquad-
     \frac{1}{2304}\, P^2\,\pz\,
        \left( 6\,\kz + P^2\,\pz \right) \,
        {R}^2 \biggr)
\end{split}
\eqlabel{rij42}
\end{equation}
\begin{equation}
\begin{split}
    &G^{(4,1)}_{ij}=\frac{1 }{13824}\, \left( -36\,{\kz}^2 -
       96\,\kz\,P^2\,\pz - 149\,P^4\,\pz^2
       \right)\, \nabla_i\nabla_j R \\
    &+
       \frac{1}{512}\, 
     \left( 4\,{\kz}^2 + 8\,\kz\,P^2\,\pz +
       5\,P^4\,\pz^2 \right)\, \square R_{ij}  \\
    &+
  \frac{1}{256}\, \left( -4\,{\kz}^2 -
       8\,\kz\,P^2\,\pz - 5\,P^4\,\pz^2
       \right) \,R_{aijb}R^{ab}\\
  &-
  \frac{1}{64}\, P^2\,\pz\,\left( 2\,\kz +
       P^2\,\pz \right) \,R_{ia}R^{a}_{j} \\
   &+
  \frac{1}{64}\left(-4\, P^2\,\pz \nabla_i\nabla_j a^{(2,0)}+
     \left( 2\,\kz + 5\,P^2\,\pz \right) \,
      \nabla_i\nabla_j a^{(2,1)}\right) \\
  &+
  R_{ij}\,\left( \frac{1}{8}\left(
-2\,P^2\,\pz\,
         a^{(2,0)} + \kz\,a^{(2,1)}
\right) + \frac{1}
        {6912}\, \left( -36\,{\kz}^2 +
          48\,\kz\,P^2\,\pz +
          79\,P^4\,\pz^2 \right) \,R \right) \\
  &+ G_{ij}^{(0)}\,
   \biggl( \frac{
     1}{27648}\,  \left(-36\,{\kz}^2 -
          96\,\kz\,P^2\,\pz -
          119\,P^4\,\pz^2 \right)\, \square R  \\
   &\qquad-
     \frac{1}{3072}\left( 12\,{\kz}^2 +
          24\,\kz\,P^2\,\pz + P^4\,\pz^2
          \right) \,R_{ab}R^{ab} \\
   &\qquad+
     \frac{1}{32}\left(-24\,a^{(4,1)} - 4\,a^{(4,2)} +
        28\,a^{(2,0)}\,a^{(2,1)} +
        7\,(a^{(2,1)})^2 +  3\,a^{(4,3)}\right) \\
   &\qquad+
      \frac{1}{48}\left( 2\,P^2\,\pz\,a^{(2,0)} -
          \kz\,a^{(2,1)} \right) \,
        R
   +
     \frac{1}{27648}\left( 36\,{\kz}^2 +
          12\,\kz\,P^2\,\pz -
          85\,P^4\,\pz^2 \right) \,
        {R}^2 \biggr)
\end{split}
\eqlabel{rij41}
\end{equation}
\begin{equation}
\begin{split}
&G_{k}^{(4,0)k}=
    \frac{1}{27648}\,P^2\,\pz\,
     \left( -6\,\kz + 31\,P^2\,\pz \right)\, \square R\\
&+
  \frac{1}{3072}\left( 12\,{\kz}^2 +
       12\,\kz\,P^2\,\pz - 5\,P^4\,\pz^2
       \right) \,R_{ab}R^{ab}+\biggl(
  \frac{1}{256} \Big( 448\,(a^{(2,0)})^2 - 64\,a^{(4,1)}\\
& +
     32\,a^{(4,2)} 
+
     224\,a^{(2,0)}\,a^{(2,1)} +
     8\,(a^{(2,1)})^2 - 24\,a^{(4,3)}
    -768\,b^{(4,0)} +
    8\,\kz\,\square a^{(2,0)} \\
&+
     8\,P^2\,\pz\,\square a^{(2,0)}-
     6\,\kz\,\square a^{(2,1)} -
     7\,P^2\,\pz\,\square a^{(2,1)} \biggr) \\
  &+
  \frac{1}{384}\, \biggl( 2\,\left( 8\,\kz - 3\,P^2\,\pz \right)
          \,a^{(2,0)} +
       P^2\,\pz\,a^{(2,1)} \biggr) \,
     R \\
&+ \frac{1}{27648}\left( -24\,{\kz}^2 -
       54\,\kz\,P^2\,\pz - 37\,P^4\,\pz^2
       \right) \,{R}^2
\end{split}
\eqlabel{rij40}
\end{equation}
\begin{equation}
\begin{split}
    &p^{(4,3)}=\frac{1}{576}\, P^4\,\pz^2\,
    \left( \square R + 3\,R_{ab}R^{ab} -
      {R}^2 \right)
\end{split} 
\eqlabel{p43}
\end{equation}
\begin{equation}
\begin{split}
    &p^{(4,2)}=-\frac{ 1}
     {768}\, P^2\,\pz\,
       \left( 2\,\kz + 11\,P^2\,\pz \right)\, \square R - \frac{1}{256}
P^2\,\pz\,
     \left( 2\,\kz + P^2\,\pz \right) \,
     R_{ab}R^{ab}\\
 &\qquad+
  \frac{1}{768}
P^2\,\pz\,\left( 2\,\kz +
       P^2\,\pz \right) \,{R}^2
\end{split}
\eqlabel{p42}
\end{equation}
\begin{equation}
\begin{split}
    &p^{(4,1)}=\frac{1}{27648}\,P^2\,\pz\,
     \left( 138\,\kz + 373\,P^2\,\pz \right)\, \square R  \\ 
&+
   \frac{1}{4608}P^2\,\pz\,\left( 42\,\kz -
       19\,P^2\,\pz \right) \,R_{ab}R^{ab}
   +
  3\,\left( a^{(4,0)} - b^{(4,0)} \right)\\ &+
  \frac{1}{
     384}P^2\,\pz\,\left( -6\,a^{(2,0)} +
       5\,a^{(2,1)} \right) \,R
+ \frac{1}{27648}\, P^2\,\pz\,
     \left( -102\,\kz + 59\,P^2\,\pz \right) \,
     {R}^2
\end{split}
\eqlabel{p41}
\end{equation}
\begin{equation}
\begin{split}
    &h^{(4,4)}=P^2\,\pz\,a^{(4,3)}
\end{split}
\eqlabel{h44}
\end{equation}
\begin{equation}
\begin{split}
    &h^{(4,3)}=-\frac{1}{512}\,P^6\,{\pz}^3\, \square R -
  \frac{3}{512}\,P^6\,{\pz}^3\,R_{ab}R^{ab}+
  P^2\,\pz\,a^{(4,2)}+
  \frac{1}{512}P^6\,{\pz}^3\,{R}^2 \\
&\qquad+
  \frac{1}{16}\left(-24\,P^2\,\pz\,(a^{(2,1)})^2 -
     \left( 8\,\kz + 3\,P^2\,\pz \right) \,
      a^{(4,3)}\right) 
\end{split}
\eqlabel{h43}
\end{equation}
\begin{equation}
\begin{split}
&h^{(4,2)}=\frac{1}{110592}\,P^4\,\pz^2\,
     \left( 324\,\kz + 965\,P^2\,\pz \right)\, \square R   +
  \frac{25}{576}\,P^4\,\pz^2\,a^{(2,1)}\,
     R\\
   &+ \frac{1}{36864} P^4\,\pz^2\,
     \left( 324\,\kz + 23\,P^2\,\pz \right) \,
     R_{ab}R^{ab}+
  \frac{1}{12288}P^4\,\pz^2\,
     \left( -36\,\kz + 5\,P^2\,\pz \right) \,
     {R}^2\\
   &+
  \frac{1}{64} \biggl( 64\,P^2\,\pz\,a^{(4,1)} -
     4\,\left( 8\,\kz + 3\,P^2\,\pz \right) \,
      a^{(4,2)} -
     192\,P^2\,\pz\,a^{(2,0)}\,
      a^{(2,1)} \\
  &\qquad+
     48\,\kz\,(a^{(2,1)})^2 +
     5\,P^2\,\pz\,(a^{(2,1)})^2 -
     3\,P^2\,\pz\,a^{(4,3)} \biggr)
\end{split}
\eqlabel{h42}
\end{equation}
\begin{equation}
\begin{split}
&h^{(4,1)}=-\frac{1}{442368}\,P^2\,\pz\,
       \left( 408\,{\kz}^2 +
         2140\,\kz\,P^2\,\pz +
         2259\,P^4\,\pz^2 \right)\square R\ \\
       &+
  \frac{1}{147456}P^2\,\pz\,\left( -408\,{\kz}^2 -
       412\,\kz\,P^2\,\pz + 11\,P^4\,\pz^2
       \right) \,R_{ab}R^{ab} \\
  &+
  \frac{1}{128} \biggl( -192\,P^2\,\pz\,(a^{(2,0)})^2
  -
     32\,P^2\,\pz\,a^{(4,0)}
  -
     64\,\kz\,a^{(4,1)}
  -
     24\,P^2\,\pz\,a^{(4,1)} \\
  &-
     4\,P^2\,\pz\,a^{(4,2)}
  +
     4\,\left( 48\,\kz + 5\,P^2\,\pz \right) \,
      a^{(2,0)}\,a^{(2,1)} +
     19\,P^2\,\pz\,(a^{(2,1)})^2
  +
     3\,P^2\,\pz\,a^{(4,3)}\\
& +
     160\,P^2\,\pz\,b^{(4,0)} \biggr)-
  \frac{1}{
     13824}5\,P^2\,\pz\,
     \left( -138\,P^2\,\pz\,a^{(2,0)} +
       \left( 60\,\kz + 101\,P^2\,\pz \right) \,
        a^{(2,1)} \right) \,R \\
& +
     \frac{1
     }{1327104}P^2\,\pz\,
     \left( 1224\,{\kz}^2 +
       780\,\kz\,P^2\,\pz -
       1057\,P^4\,\pz^2 \right) \,{R}^2
\end{split}
\eqlabel{h41}
\end{equation}
\begin{equation}
\begin{split}
&h^{(4,0)}=\frac{1}{3538944}\,P^2\,\pz\,
     \left( 480\,{\kz}^2 +
       3244\,\kz\,P^2\,\pz +
       4657\,P^4\,\pz^2 \right)\square R  \\
     &+
  \frac{1}{1179648}P^2\,\pz\,\left( 600\,{\kz}^2 +
       908\,\kz\,P^2\,\pz - 59\,P^4\,\pz^2
       \right) \,R_{ab}R^{ab} \\
   &+
  \frac{1}{512} \biggl(8\,\left( 48\,\kz + 5\,P^2\,\pz \right) \,
      (a^{(2,0)})^2 +
     8\,\left( 2\,\kz + 13\,P^2\,\pz \right) \,
      a^{(4,0)} -
     8\,P^2\,\pz\,a^{(4,1)} \\
  &+
     4\,P^2\,\pz\,a^{(4,2)}
  +
     76\,P^2\,\pz\,a^{(2,0)}\,
      a^{(2,1)} -
     3\,P^2\,\pz\,(a^{(2,1)})^2 -
     3\,P^2\,\pz\,a^{(4,3)} \\
  &-
     272\,\kz\,b^{(4,0)}
  -
     200\,P^2\,\pz\,b^{(4,0)}
  -
     32\,P^2\,\pz\,p^{(4,0)} \biggr) \\
  &+
  \frac{1}{
     221184}P^2\,\pz\,\biggl( -2\,
        \left( 2706\,\kz + 4001\,P^2\,\pz \right) \,
        a^{(2,0)} +
       \left( 174\,\kz + 31\,P^2\,\pz \right) \,
        a^{(2,1)} \biggr) \,R \\
  &+ \frac{1
     }{10616832}P^2\,\pz\,
     \left( -1392\,{\kz}^2 +
       220\,\kz\,P^2\,\pz +
       5231\,P^4\,\pz^2 \right) \,{R}^2
\end{split}
\eqlabel{h40}
\end{equation}
\begin{equation}
b^{(4,3)}=a^{(4,3)}
\eqlabel{b43} 
\end{equation}
\begin{equation}
b^{(4,2)}=-\frac{1}{576}\,P^4\,\pz^2\, \square R  -
  \frac{1}{192}P^4\,\pz^2\,R_{ab}R^{ab} +
  a^{(4,2)} + \frac{1}{576}P^4\,\pz^2\,
     {R}^2
\eqlabel{b42}
\end{equation}
\begin{equation}
\begin{split}
&b^{(4,1)}=\frac{1}{6912}\,P^2\,\pz\,
     \left( 12\,\kz + 43\,P^2\,\pz \right)\,\square R  +
  \frac{1}{576}P^2\,\pz\,\left( 3\,\kz -
       2\,P^2\,\pz \right) \,R_{ab}R^{ab} \\
&+
  a^{(4,1)}
- \frac{1}{192}P^2\,\pz\,
     a^{(2,1)}\,R +
  \frac{1}{
     6912}P^2\,\pz\,\left( -12\,\kz +
       11\,P^2\,\pz \right) \,{R}^2
\end{split}
\eqlabel{b41}
\end{equation}
\begin{equation}
K^{(4,3)}=-\frac{1}{288}\,P^6\,{\pz}^3 \square R-
  \frac{1}{96}P^6\,{\pz}^3\,R_{ab}R^{ab} +
  \frac{1}{4}P^2\,\pz\,a^{(4,3)} +
  \frac{1}{288}P^6\,{\pz}^3\,{R}^2
\eqlabel{k43}
\end{equation}
\begin{equation}
\begin{split}
&K^{(4,2)}=\frac{1}{192}\,P^4\,\pz^2\,
     \left( \kz + 4\,P^2\,\pz \right)\square R  +
  \frac{1}{256}P^4\,\pz^2\,
     \left( 4\,\kz + 3\,P^2\,\pz \right) \,
     R_{ab}R^{ab}
\\
  &+
  \frac{1}{16}P^2\,\pz\,\biggl( 4\,a^{(4,2)} -
       3\,\left( (a^{(2,1)})^2 +
          a^{(4,3)} \right)  \biggr)  +
  \frac{1}{48}P^4\,\pz^2\,a^{(2,1)}\,
     R\\
  &-
  \frac{1}{384}P^4\,\pz^2\,
     \left( 2\,\kz + P^2\,\pz \right) \,
     {R}^2
\end{split}
\eqlabel{k42}
\end{equation}
\begin{equation}
\begin{split}
&K^{(4,1)}=-\frac{1}{13824}\,P^2\,\pz\,
       \left( 18\,{\kz}^2 +
         159\,\kz\,P^2\,\pz +
         308\,P^4\,\pz^2 \right)\square R \\
       &-
  \frac{1}{9216}P^2\,\pz\,\left( 36\,{\kz}^2 +
       156\,\kz\,P^2\,\pz + 7\,P^4\,\pz^2
       \right) \,R_{ab}R^{ab} \\
  &+
  \frac{1}{32}P^2\,\pz\, \biggl( -96\,a^{(4,0)} +
       8\,a^{(4,1)} - 4\,a^{(4,2)} -
       12\,a^{(2,0)}\,a^{(2,1)} +
       3\,(a^{(2,1)})^2 + 3\,a^{(4,3)} \\
  &+
       96\,b^{(4,0)} \biggr)
  +
  \frac{1}{384}P^2\,\pz\,\biggl( 14\,P^2\,\pz\,
        a^{(2,0)} -
       \left( 4\,\kz + 11\,P^2\,\pz \right) \,
        a^{(2,1)} \biggr) \,R
   \\
   &+ \frac{1}{13824}P^2\,\pz\,
     \left( 18\,{\kz}^2 +
       69\,\kz\,P^2\,\pz - 29\,P^4\,\pz^2
       \right) \,{R}^2
\end{split}
\eqlabel{k41}
\end{equation}
\begin{equation}
\begin{split}
&K^{(4,0)}=\frac{1}{221184}\,P^2\,\pz\,
     \left( 132\,{\kz}^2 +
       944\,\kz\,P^2\,\pz +
       1371\,P^4\,\pz^2 \right)\, \square R  \\
     &+
  \frac{1}{36864} P^2\,\pz\,\left( 84\,{\kz}^2 +
       124\,\kz\,P^2\,\pz - 21\,P^4\,\pz^2
       \right) \,R_{ab}R^{ab}\\
  &+
  \frac{1}{128} \biggl(-24\,P^2\,\pz\,(a^{(2,0)})^2 +
     16\,\left( 6\,\kz + 11\,P^2\,\pz \right) \,
      a^{(4,0)} -
     8\,P^2\,\pz\,a^{(4,1)} +
     4\,P^2\,\pz\,a^{(4,2)}\\
  &+
     12\,P^2\,\pz\,a^{(2,0)}\,
      a^{(2,1)}
     -
     3\,P^2\,\pz\,(a^{(2,1)})^2 -
     3\,P^2\,\pz\,a^{(4,3)} 
  -
     96\,\kz\,b^{(4,0)} -
     144\,P^2\,\pz\,b^{(4,0)}\\ &
  -64\,P^2\,\pz\,p^{(4,0)} \biggr) 
  -\
  \frac{1}{3072} P^2\,\pz\,\left( 2\,\kz +
       3\,P^2\,\pz \right) \,
     \left( 22\,a^{(2,0)} - 5\,a^{(2,1)}
       \right) \,R\\
  &+
  \frac{1}
     {663552}P^2\,\pz\,\left( -396\,{\kz}^2 -
       252\,\kz\,P^2\,\pz+
       713\,P^4\,\pz^2 \right) \,{R}^2
\end{split}
\eqlabel{k40}
\end{equation}
where all the differential operators are evaluated with respect to
$G_{ij}^{(0)}$.

\subsection{Symmetries of the asymptotic solution}\label{checksymms}

There are three symmetries we can use to check our general solution to
\eqref{keq1}-\eqref{riccirr}. \nxt First, there is a scaling
symmetry
\begin{equation}
\begin{split}
&\r\to \la \r,\qquad G_{ij}\to \la^2 G_{ij},\qquad h\to h\\
&K\to K,\qquad \Phi\to \Phi,\qquad f_2\to f_2,\qquad f_3\to f_3
\end{split}
\eqlabel{sym1}
\end{equation}
which translates into the following scaling symmetry of the
parameters of the asymptotic solution \eqref{help1}-\eqref{help6}
\begin{equation}
\pz\to \pz
\end{equation}
\begin{equation}
\kz \to \kz +2P^2\pz\ln\la
\eqlabel{sym10}
\end{equation}
\begin{equation}
G_{ij}^{(0)} \to \la^2 G_{ij}^{(0)}
\end{equation}
\begin{equation}
a^{(2,1)}\to {\frac{1}{{\lambda }^2}}\, a^{(2,1)}
\end{equation}
\begin{equation}
a^{(2,0)}\to {\frac{1}{{\lambda}^2}}\, \left(a^{(2,0)} - a^{(2,1)}\,\ln\la\right)
\eqlabel{sym12}
\end{equation}
\begin{equation}
    a^{(4,3)}\to\frac{1}{\lambda^4}\, a^{(4,3)}
\eqlabel{syma43}
\end{equation}
\begin{equation}
    a^{(4,2)}\to {\frac{1}{{\lambda }^4}\left(a^{(4,2)} -
        3\,a^{(4,3)}\,\ln\la\right)}
\eqlabel{syma42}
\end{equation}
\begin{equation}
a^{(4,1)}\to {\frac{1}{{\lambda }^4}\left(a^{(4,1)} -
        2\,a^{(4,2)}\,\ln\la +
        3\,a^{(4,3)}\,{\ln^2\la}\right)}
\eqlabel{syma41}
\end{equation}
\begin{equation}
a^{(4,0)}\to {\frac{1}
        {{\lambda }^4}\left(a^{(4,0)} -
        \ln\la\,\left( a^{(4,1)} +
           \ln\la\,\left( -a^{(4,2)} +
              a^{(4,3)}\,\ln\la \right)  \right) \right)}
\eqlabel{syma40}
\end{equation}
\begin{equation}
\begin{split}
b^{(4,0)}\to& \frac{1}{{\lambda }^4}\, b^{(4,0)} -
  \frac{1}{{\lambda }^4}\, a^{(4,1)}\,\ln\la -
  \frac{\ln\la}{576\,{\lambda }^4}\, P^2\,\pz
     \left( 3\,\kz - 2\,P^2\,\pz +
       3\,P^2\,\pz\,\ln\la \right)\,R_{ab}R^{ab}  \\
  &-
   \frac{\ln\la }{6912\,{\lambda }^4}\,P^2\,\pz\,
     \left( 12\,\kz + 43\,P^2\,\pz +
       12\,P^2\,\pz\,\ln\la \right)\square R
\\
  &+ \frac{{\ln\la}^2\,
      }{{\lambda }^4}\, \left( a^{(4,2)} -
       a^{(4,3)}\,\ln\la \right) +
  \frac{\ln\la}{192\,{\lambda }^4}\, P^2\,\pz\,a^{(2,1)}\,
     R
 \\
  &+
  \frac{\ln\la}{6912\,{\lambda }^4}\, P^2\,\pz\,
     \left( 12\,\kz - 11\,P^2\,\pz +
       12\,P^2\,\pz\,\ln\la \right) \,
     {R}^2
\end{split}
\eqlabel{symb40}
\end{equation}
\begin{equation}
\begin{split}
G_{ij}^{(4,0)}&\to
        \frac{1}{\lambda^2}G_{ij}^{(4,0)}+
    \frac{\ln{\lambda}}{\lambda^2}\Bigg(\frac{1}{16}\,P^2\,\pz\, \nabla_i\nabla_j a^{(2,0)} +
  \frac{1 }{64} \,\left( -2\,\kz -
       5\,P^2\,\pz \right)\, \nabla_i\nabla_j a^{(2,1)}\\
  & +
  \frac{1 }{512}\,\left( -4\,{\kz}^2 -
       8\,\kz\,P^2\,\pz - 5\,P^4\,\pz^2
       \right)\, \square R_{ij} +
  \frac{1}{8} \,\left( 2\,P^2\,\pz\,
        a^{(2,0)} - \kz\,a^{(2,1)}
       \right) R_{ij}\\
  & + \frac{1 }{27648} \,
     \left( 36\,{\kz}^2 +
       96\,\kz\,P^2\,\pz + 119\,P^4\,\pz^2
       \right)\,G_{ij}^{(0)}\, \square R\\
  & + \frac{1 }{13824}
     \left( 36\,{\kz}^2 +
       96\,\kz\,P^2\,\pz + 149\,P^4\,\pz^2
       \right)\, \nabla_i\nabla_j R\\
  & + \frac{1}{3072} \,
     \left( 12\,{\kz}^2 +
       24\,\kz\,P^2\,\pz + P^4\,\pz^2
       \right) \,G_{ij}^{(0)}\, R_{ab}R^{ab}\\
  & +
  \frac{1}{256} \left( 4\,{\kz}^2 +
       8\,\kz\,P^2\,\pz + 5\,P^4\,\pz^2
       \right) \,R_{iabj}R^{ab}+
  \frac{1}{64} P^2\,\pz\,\left( 2\,\kz +
       P^2\,\pz \right) \,R_{ia}R^{a}_j\\
  & +
  \frac{1}{27648}\,\left( -36\,{\kz}^2 -
       12\,\kz\,P^2\,\pz + 85\,P^4\,\pz^2
       \right) \,G_{ij}^{(0)}{R}^2
    \\
  & + R\,
   \left( \frac{1}{6912}\biggl( 36 {\kz}^2 -
          48 \kz P^2 \pz -
          79\,P^4 \pz^2 \right) R_{ij} \\
&+
     \frac{1 }{48} G_{ij}^{(0)} \left( -2 P^2 \pz
           a^{(2,0)} +
          \kz a^{(2,1)} \right)\biggr) 
    +
  \frac{1 }{32}
( 24\,a^{(4,1)} +
       4\,a^{(4,2)} -
       28\,a^{(2,0)}\,a^{(2,1)} \\
&-
       7\,{a^{(2,1)}}^2 - 3\,a^{(4,3)}
       )G_{ij}^{(0)}
\Bigg)+
  \frac{\ln^2\lambda}{\lambda^2}
  \Bigg(-\frac{1}{16}\,P^2\,\pz\,  \nabla_i\nabla_j a^{(2,1)}\\
&-
  \frac{1 }{64}\,P^2\,\pz\,
     \left( \kz + P^2\,\pz \right)\, \square R_{ij} 
 +
  \frac{1}{576}\,P^2\,\pz\,
     \left( 3\,\kz + 4\,P^2\,\pz \right)\\
&   +
  \frac{1 }{1152}\,P^2\,\pz\,
     \left( 3\,\kz + 4\,P^2\,\pz \right) \,G_{ij}^{(0)}\square R
  +
  \frac{1}{128} \,P^2\,\pz\,
     \left( \kz + P^2\,\pz \right) \,
     G_{ij}^{(0)} R_{ab}R^{ab}\\
&+
  \frac{1}{32}\, P^2\,\pz\,\left( \kz + P^2\,\pz
       \right) \,R_{iabj}R^{ab} +
  \frac{1}{32}\, P^4\,\pz^2\,R_{ia}R^{a}_j \\
 &-
  \frac{1}{2304}\,P^2\,\pz\,
     \left( 6\,\kz + P^2\,\pz \right) \,
     G_{ij}^{(0)}{R}^2 -
  \frac{1}
   {4}P^2\,\pz\,a^{(2,1)}R_{ij} \\
   &+ R\,\left( \frac{1}{288}\, P^2\,\pz\,
        \left( 3\,\kz - 2\,P^2\,\pz \right) \,
        R_{ij} +
     \frac{1}
      {24} \,P^2\,\pz\,a^{(2,1)}\, G_{ij}^{(0)}\right) \\
   & + \frac{1}{16}\,
     \left( -12\,a^{(4,2)} +
       7\,{a^{(2,1)}}^2 - 3\,a^{(4,3)}
       \right)\, G_{ij}^{(0)} \Bigg)
   +
   \frac{\ln^3\lambda}{\lambda^2}
   \Bigg(-\frac{1}{96} \,P^4\,\pz^2\, \square R_{ij} \\
  &+
  \frac{1}{288}\,P^4\,\pz^2\, \nabla_i\nabla_j R +
  \frac{1}
   {576}\,G_{ij}^{(0)}\,P^4\,\pz^2\, \square R + \frac{1}{192}\,P^4\,\pz^2\,
     G_{ij}^{(0)}\, R_{ab}R^{ab}\\
& +
  \frac{1}{48} P^4\,\pz^2\,R_{iabj}R^{ab}
  +
  \frac{1}{144}P^4\,\pz^2\,R\,R_{ij} -
  \frac{1}
   {576} \,P^4\,\pz^2\,G_{ij}^{(0)}\, {R}^2+ \frac{3}{4}\,
a^{(4,3)}\,G_{ij}^{(0)}
\Bigg)
\end{split}
\eqlabel{symgij402}
\end{equation}
\begin{equation}
\begin{split}
p^{(4,0)}\to&
    -
  \frac{\ln\lambda}{4608\,{\lambda }^4}\,
     \biggl( 42\,\kz - 19\,P^2\,\pz +
       18\,\left( 2\,\kz + P^2\,\pz \right) \,
        \ln\lambda \\
  &+
  24\,P^2\,\pz\,{\ln^2\lambda}
       \biggr)P^2\,\pz\,R_{ab}R^{ab}\,
  -
  \frac{\ln\lambda}{27648\,{\lambda }^4}\,
     \biggl( 138\,\kz +
  373\,P^2\,\pz \\
  &+
       36\,\left( 2\,\kz + 11\,P^2\,\pz \right) \,
        \ln\lambda + 48\,P^2\,\pz\,{\ln\lambda}^2
       \biggr)\,P^2\,\pz\,\square R
  +
  \frac{p^{(4,0)}}{{\lambda }^4} \\
  &+
    \frac{3\,\ln\lambda}{{\lambda
    }^4}\left(b^{(4,0)}-a^{(4,0)}\right) + \frac{  \ln\lambda}{\lambda^4}
  \left( \frac{1}{64}P^2\,\pz\,a^{(2,0)}
         -
     \frac{5
        }{384}\,P^2\,\pz\,a^{(2,1)} \right) 
  R\\
  &+
  \frac{\ln\lambda}{27648\,{\lambda }^4}
     \biggl( 102\,\kz - 59\,P^2\,\pz +
       36\,\left( 2\,\kz + P^2\,\pz \right) \,
        \ln\lambda \\
  &+ 48\,P^2\,\pz\,{\ln^2\lambda}
       \biggr)P^2\,\pz\,{R}^2
\end{split}
\eqlabel{symp40}
\end{equation}
The solution transforms covariantly under this transformation, and
in addition, the constraints
\eqref{constraints4} and \eqref{rij40} are also invariant under the
symmetry transformations \eqref{sym1}. \nxt Another symmetry is the
residual gauge freedom associated with unfixed diffeomorphisms
\begin{equation}
\begin{split}
\r&\to \hr\equiv
\r\biggl[1+\r^2\biggl(\dd_{20}+\dd_{21}\ln\r\biggr)+\r^4\biggl(
\dd_{40}+\dd_{41}\ln\r+\dd_{42}\ln^2\r+\dd_{43}\ln^3\r\biggr)\biggr]\\
K&\to K,\qquad \Phi\to \Phi,\qquad h\to h
\left(\frac{\hr}{\r}\right)^4\left(\frac{\del\hr}{\del\r}\right)^{-4}
,\qquad f_2\to f_2 \left(\frac{\hr}{\r}\right)^{-2}\left(\frac{\del\hr}{\del\r}\right)^{2}\\
f_3&\to f_3
\left(\frac{\hr}{\r}\right)^{-2}\left(\frac{\del\hr}{\del\r}\right)^{2},\qquad
G_{ij}\to G_{ij}
\left(\frac{\hr}{\r}\right)^{4}\left(\frac{\del\hr}{\del\r}\right)^{-2}
\end{split}
\eqlabel{sym2}
\end{equation}
where $\{\dd_{20},\cdots,\dd_{43}\}$ are arbitrary constants. This
symmetry is realized on the parameters of the asymptotic
solution as follows
\begin{align}
\pz\to\, & \pz\eqlabel{sym2p0}\\
\kz\to\, & \kz\eqlabel{sym2k0}\\
G_{ij}^{(0)}\to\, & G_{ij}^{(0)}\eqlabel{sym2gij0}\\
    a^{(2,0)}\to\, & 4\,\dd_{20} + 2\,\dd_{21} + a^{(2,0)} \eqlabel{sym20}\\
    a^{(2,1)}\to\, & {4\,\dd_{21} + a^{(2,1)}} \eqlabel{sym22}\\
    a^{(4,3)}\to\, & {8\,\dd_{43} + a^{(4,3)}} \eqlabel{sym2a43}\\
    a^{(4,2)}\to\, & -8\,{\dd_{21}}^2 + 8\,\dd_{42} +
      6\,\dd_{43} + a^{(4,2)} +
      2\,\dd_{21}\,a^{(2,1)}
        \eqlabel{sym2a42}\\
    a^{(4,1)}\to\, & {-16\,\dd_{20}\,\dd_{21} - 6\,{\dd_{21}}^2 +
      8\,\dd_{41} + 4\,\dd_{42} +
      2\,\dd_{21}\,a^{(2,0)} +
      a^{(4,1)} +
      2\,\dd_{20}\,a^{(2,1)} +
      \dd_{21}\,a^{(2,1)}}
        \eqlabel{sym2a41} \\
    a^{(4,0)}\to\, &  {-8\,{\dd_{20}}^2 - 6\,\dd_{20}\,\dd_{21} +
      {\dd_{21}}^2 + 8\,\dd_{40}+ 2\,\dd_{41} +
      2\,\left( \dd_{20} + \dd_{21} \right) \,
       a^{(2,0)} + a^{(4,0)} -
      \dd_{20}\,a^{(2,1)}}
        \eqlabel{sym2a40} \\
    b^{(4,0)}\to\, &
        -8\,{\dd_{20}}^2 - 6\,\dd_{20}\,\dd_{21} +
      {\dd_{21}}^2 + 8\,\dd_{40}+
      2\,\dd_{41} +
      2\,\left( \dd_{20} + \dd_{21} \right) \,
       a^{(2,0)} +
      b^{(4,0)}
    - \dd_{20}\,a^{(2,1)}
\notag \\
    &-
      \frac{1}{48}P^2\,\pz\,
         \left( \dd_{20} + \dd_{21} \right) \,
         R
         \eqlabel{sym2b40} \\
    p^{(4,0)}\to\, &  p^{(4,0)}+\frac{1}{12}P^2\pz\, \dd_{20}\, R
    \eqlabel{sym2p40}
\end{align}
\begin{equation}
\begin{split}
G_{ij}^{(4,0)}\to\, & G_{ij}^{(4,0)} +
    \frac{1 }{8}\,\left( P^2\,\pz\,\dd_{21} +
       2\,\kz\,\left( 2\,\dd_{20} +
          \dd_{21} \right)  \right)\, R_{ij}\\
    &+
  G_{ij}^{(0)}\,\Bigg( -\frac{1}{48}
        \left( P^2\,\pz\,\dd_{21}  +
          2\,\kz\,\left( 2\,\dd_{20} +
             \dd_{21} \right) \right)\, R +
     \left( 2\,\dd_{20} + \dd_{21} \right) \,
      a^{(2,0)} \\
  &\qquad +
    13\,{\dd_{20}}^2 +
        16\,\dd_{20}\,\dd_{21} +
        3\,{\dd_{21}}^2 - 6\,\dd_{40} -
        2\,\dd_{41} +
        \frac 12 \left( 3\,\dd_{20} + \dd_{21} \right) \,
         a^{(2,1)} \Bigg)
\end{split}
\eqlabel{sym2gij40}
\end{equation}
Again, in addition to the solution transforming covariantly, the constraints
\eqref{constraints4} and \eqref{rij40} are also invariant under the
symmetry transformations \eqref{sym2}.
\nxt Finally, there is a symmetry which rescales the action 
(and thus is a symmetry of any solution to the equations 
of motion) given by \eqref{symmetry3}, which in terms of our variables
is given by
\begin{equation}
\begin{split}
&G_{ij}\to\ \beta\, G_{ij},\qquad h\to\, \beta\, h,\qquad 
e^\Phi\to\, \beta\, e^{\Phi},\\
&K\to\, \beta\, K,\qquad f_2\to\, f_2,\qquad f_3\to\, f_3.
\end{split}
\eqlabel{symmetry31}
\end{equation}
This symmetry is realized on the parameters of the asymptotic
solution as follows
\begin{equation}
\pz\to\, \beta\, \pz,\qquad 
\kz\to\, \beta\, \kz,\qquad
G_{ij}^{(0)} \to \, \beta\, G_{ij}^{(0)},\qquad
G_{ij}^{(4,0)}\to\, \beta\, G_{ij}^{(4,0)},
\end{equation}
with all other parameters remaining unchanged.
Again, in addition to the solution transforming covariantly, the constraints
\eqref{constraints4} and \eqref{rij40} are also invariant under the
symmetry transformations \eqref{symmetry31}. The symmetry transformation
\eqref{symmetry3} is very useful in constraining the possible counter-terms 
in the holographic renormalization of the cascading gauge theories, as
discussed in section \ref{loccounter}.

\subsection{Ambiguities of the minimal subtraction}\label{shift}

In this appendix we discuss a specific simple ambiguity of the 
counter-term action, which is present in our
minimal subtraction ansatz.

Consider a counter-term ansatz of the form
\begin{equation}
\dd\call^{counter}=\dd\call_1(K,P^2e^{\Phi})\ 
\left(\calr_\gamma^2-3 \calr_{ab\ \ga}\calr_\ga^{ab}\right)+
\dd\call_2(K,P^2e^{\Phi})\ \square_\ga \calr\,.
\eqlabel{sh5}
\end{equation}
This ansatz was chosen so that it does not contribute to $\vev{T^i_i}$,
and it is easy to verify that (since $\alpha_8$ is a source only for $h$) it
also does not contribute to $\vev{\calo_8}$.

It is straightforward to verify that \eqref{sh5} leads to
\begin{equation}
\vev{\dd\calo_{\pz}}=\frac{\del \dd\call_1}{\del \pz}\ 
\left(R^2-3 R_{ab}R^{ab}\right)+\frac{\del \dd\call_2}{\del \pz}\ \square R\,,
\eqlabel{sh6}
\end{equation}
\begin{equation}
\vev{\dd\calo_{\kz}}=\frac{\del \dd\call_1}{\del \kz}\ 
\left(R^2-3 R_{ab}R^{ab}\right)+\frac{\del \dd\call_2}{\del \kz}\ \square R\,.
\eqlabel{sh7}
\end{equation}
From \eqref{sh6}, \eqref{sh7} it follows that $\vev{\dd\calo_{\pz}}$ and 
$\vev{\dd\calo_{\kz}}$ will be finite if the $\dd\call_i$ are finite in the 
limit $\r\to 0$.
Under the symmetry \eqref{symmetry3} we must have the scaling 
\begin{equation}
\dd\call_i\to\, \beta^2\ \dd\call_i\,.
\eqlabel{sh8}
\end{equation}
The requirement of finiteness in the $\r\to 0$ limit, along with \eqref{sh8},
lead to the following choice of counter-terms :
\begin{equation}
\begin{split}
\dd\call_1=&\kappa_1 \left(K+2 P^2e^\Phi\ln\r\right)^2+\kappa_2 \left(K+2 P^2e^\Phi\ln\r\right)P^2 e^\Phi+
\kappa_3 P^4 e^{2\Phi}\,,\\
\dd\call_2=&\kappa_4 \left(K+2 P^2e^\Phi\ln\r\right)^2+\kappa_5 \left(K+2 P^2e^\Phi\ln\r\right)P^2 e^\Phi+
\kappa_6 P^4 e^{2\Phi}\,.
\end{split}
\eqlabel{sh9}
\end{equation}

The arguments above show that the only possible divergences arising
from the counter-terms \eqref{sh9} are in 
$\vev{\calo_6^s}$.
We find that these counter-terms indeed lead to extra divergences
in   $\vev{\calo_6^s}$, but they can
be removed (preserving what has been achieved thus far) by
\begin{equation}
\dd\call^{counter}\to\, \dd\call^{counter}+\dd\call_{1extra}
\left(\calr_\gamma^2-3 \calr_{ab\ \ga}\calr_\ga^{ab}\right)+
\dd\call_{2extra}\ \square_\ga \calr\,,
\eqlabel{sh10}
\end{equation}
with 
\begin{equation}
\begin{split}
\dd\call_{1extra}=&X_a\left(\frac 45 \kappa_1 \left(K+2 P^2e^\Phi\ln\r\right)P^2 e^\Phi
+\frac 25 \kappa_2 P^4 e^{2\Phi}\right)\,,\\ 
\dd\call_{2extra}=&X_a\left(\frac 45 \kappa_4 \left(K+2 P^2e^\Phi\ln\r\right)P^2 e^\Phi
+\frac 25 \kappa_5 P^4 e^{2\Phi}\right)\,. 
\end{split}
\eqlabel{sh11}
\end{equation}

Thus, we can always add to our action the counter-terms \eqref{sh9}
and \eqref{sh10} without generating any divergences.


\begin{thebibliography}{99}



\bibitem{mal}
J.~M.~Maldacena,
``The large N limit of superconformal field theories and supergravity,''
Adv.\ Theor.\ Math.\ Phys.\  {\bf 2}, 231 (1998)
[Int.\ J.\ Theor.\ Phys.\  {\bf 38}, 1113 (1999)]
[arXiv:hep-th/9711200].

\bibitem{Butti:2004pk}
  A.~Butti, M.~Grana, R.~Minasian, M.~Petrini and A.~Zaffaroni,
  ``The baryonic branch of Klebanov-Strassler solution: A supersymmetric family
  of $SU(3)$ structure backgrounds,''
  JHEP {\bf 0503}, 069 (2005)
  [arXiv:hep-th/0412187].

\bibitem{kn}
  I.~R.~Klebanov and N.~A.~Nekrasov,
  ``Gravity duals of fractional branes and logarithmic RG flow,''
  Nucl.\ Phys.\ B {\bf 574}, 263 (2000)
  [arXiv:hep-th/9911096].

\bibitem{kt}
I.~R.~Klebanov and A.~A.~Tseytlin,
``Gravity duals of supersymmetric SU(N) x SU(N+M) gauge theories,''
Nucl.\ Phys.\ B {\bf 578}, 123 (2000)
[arXiv:hep-th/0002159].


\bibitem{ks}
I.~R.~Klebanov and M.~J.~Strassler,
``Supergravity and a confining gauge theory: Duality cascades and  
$\chi$SB-resolution of naked singularities,''
JHEP {\bf 0008}, 052 (2000)
[arXiv:hep-th/0007191].

\bibitem{Herzog:2002ih}
  C.~P.~Herzog, I.~R.~Klebanov and P.~Ouyang,
  ``D-branes on the conifold and N = 1 gauge / gravity dualities,''
  arXiv:hep-th/0205100.

\bibitem{Strassler:2005qs}
  M.~J.~Strassler,
  ``The duality cascade,''
  arXiv:hep-th/0505153.

\bibitem{Seiberg:1994pq}
  N.~Seiberg,
  ``Electric - magnetic duality in supersymmetric nonAbelian gauge theories,''
  Nucl.\ Phys.\ B {\bf 435}, 129 (1995)
  [arXiv:hep-th/9411149].

\bibitem{Hollowood:2004ek}
  T.~J.~Hollowood and S.~Prem Kumar,
  ``An N = 1 duality cascade from a deformation of N = 4 SUSY Yang-Mills,''
  JHEP {\bf 0412}, 034 (2004)
  [arXiv:hep-th/0407029].

\bibitem{Gubser:1998bc}
  S.~S.~Gubser, I.~R.~Klebanov and A.~M.~Polyakov,
  ``Gauge theory correlators from non-critical string theory,''
  Phys.\ Lett.\ B {\bf 428}, 105 (1998)
  [arXiv:hep-th/9802109].

\bibitem{Witten:1998qj}
  E.~Witten,
  ``Anti-de Sitter space and holography,''
  Adv.\ Theor.\ Math.\ Phys.\  {\bf 2}, 253 (1998)
  [arXiv:hep-th/9802150].

\bibitem{Henningson:1998gx}
  M.~Henningson and K.~Skenderis,
  ``The holographic Weyl anomaly,''
  JHEP {\bf 9807}, 023 (1998)
  [arXiv:hep-th/9806087].

\bibitem{bk}
V.~Balasubramanian and P.~Kraus,
``A stress tensor for anti-de Sitter gravity,''
Commun.\ Math.\ Phys.\  {\bf 208}, 413 (1999)
[arXiv:hep-th/9902121].

\bibitem{Emparan:1999pm}
  R.~Emparan, C.~V.~Johnson and R.~C.~Myers,
  ``Surface terms as counterterms in the AdS/CFT correspondence,''
  Phys.\ Rev.\ D {\bf 60}, 104001 (1999)
  [arXiv:hep-th/9903238].

\bibitem{Kraus:1999di}
  P.~Kraus, F.~Larsen and R.~Siebelink,
  ``The gravitational action in asymptotically AdS and flat spacetimes,''
  Nucl.\ Phys.\ B {\bf 563}, 259 (1999)
  [arXiv:hep-th/9906127].

\bibitem{Nojiri:1999jj}
  S.~Nojiri, S.~D.~Odintsov, S.~Ogushi, A.~Sugamoto and M.~Yamamoto,
  ``Axion-dilatonic conformal anomaly from AdS/CFT correspondence,''
  Phys.\ Lett.\ B {\bf 465} (1999) 128
  [arXiv:hep-th/9908066].

\bibitem{Nojiri:2000kh}
  S.~Nojiri, S.~D.~Odintsov and S.~Ogushi,
  ``Finite action in d5 gauged supergravity and dilatonic conformal anomaly
  for dual quantum field theory,''
  Phys.\ Rev.\ D {\bf 62} (2000) 124002
  [arXiv:hep-th/0001122].

\bibitem{deHaro:2000xn}
  S.~de Haro, S.~N.~Solodukhin and K.~Skenderis,
  ``Holographic reconstruction of spacetime and renormalization in the  AdS/CFT
  correspondence,''
  Commun.\ Math.\ Phys.\  {\bf 217}, 595 (2001)
  [arXiv:hep-th/0002230].

\bibitem{Taylor-Robinson:2001pp}
  M.~Taylor-Robinson,
  ``Anomalies, counterterms and the N = 0 Polchinski-Strassler solutions,''
  arXiv:hep-th/0103162.

\bibitem{Muck:2001cy}
  W.~Muck,
  ``Correlation functions in holographic renormalization group flows,''
  Nucl.\ Phys.\ B {\bf 620} (2002) 477
  [arXiv:hep-th/0105270].

\bibitem{bfs1}
M.~Bianchi, D.~Z.~Freedman and K.~Skenderis,
``How to go with an RG flow,''
JHEP {\bf 0108}, 041 (2001)
[arXiv:hep-th/0105276].

\bibitem{bfs2}
M.~Bianchi, D.~Z.~Freedman and K.~Skenderis,
``Holographic renormalization,''
Nucl.\ Phys.\ B {\bf 631}, 159 (2002)
[arXiv:hep-th/0112119].

\bibitem{Skenderis:2002wp}
  K.~Skenderis,
  ``Lecture notes on holographic renormalization,''
  Class.\ Quant.\ Grav.\  {\bf 19}, 5849 (2002)
  [arXiv:hep-th/0209067].

\bibitem{Berg:2002hy}
  M.~Berg and H.~Samtleben,
  ``Holographic correlators in a flow to a fixed point,''
  JHEP {\bf 0212} (2002) 070
  [arXiv:hep-th/0209191].

\bibitem{Papadimitriou:2004rz}
  I.~Papadimitriou and K.~Skenderis,
  ``Correlation functions in holographic RG flows,''
  JHEP {\bf 0410}, 075 (2004)
  [arXiv:hep-th/0407071].

\bibitem{Hollands:2005ya}
  S.~Hollands, A.~Ishibashi and D.~Marolf,
  ``Counter-term charges generate bulk symmetries,''
  arXiv:hep-th/0503105.

\bibitem{Klebanov:2002gr}
  I.~R.~Klebanov, P.~Ouyang and E.~Witten,
  ``A gravity dual of the chiral anomaly,''
  Phys.\ Rev.\ D {\bf 65}, 105007 (2002)
  [arXiv:hep-th/0202056].

\bibitem{bhkt1}
A.~Buchel,
``Finite temperature resolution of the Klebanov-Tseytlin singularity,''
Nucl.\ Phys.\ B {\bf 600}, 219 (2001)
[arXiv:hep-th/0011146].

\bibitem{bt}
 A.~Buchel and A.~A.~Tseytlin,
 ``Curved space resolution of singularity of fractional D3-branes on
 conifold,''
 Phys.\ Rev.\ D {\bf 65}, 085019 (2002)
 [arXiv:hep-th/0111017].

\bibitem{bhkt2}
A.~Buchel, C.~P.~Herzog, I.~R.~Klebanov, L.~A.~Pando Zayas and A.~A.~Tseytlin,
``Non-extremal gravity duals for fractional D3-branes on the conifold,''
JHEP {\bf 0104}, 033 (2001)
[arXiv:hep-th/0102105].

\bibitem{bhkt3}
S.~S.~Gubser, C.~P.~Herzog, I.~R.~Klebanov and A.~A.~Tseytlin,
``Restoration of chiral symmetry: A supergravity perspective,''
JHEP {\bf 0105}, 028 (2001)
[arXiv:hep-th/0102172].

\bibitem{kw}
  I.~R.~Klebanov and E.~Witten,
  ``Superconformal field theory on threebranes at a Calabi-Yau  singularity,''
  Nucl.\ Phys.\ B {\bf 536}, 199 (1998)
  [arXiv:hep-th/9807080].

\bibitem{Deser:1996na}
  S.~Deser,
  ``Conformal anomalies: Recent progress,''
  Helv.\ Phys.\ Acta {\bf 69}, 570 (1996)
  [arXiv:hep-th/9609138].


\bibitem{ne2}
G.~Policastro, D.~T.~Son and A.~O.~Starinets,
``From AdS/CFT correspondence to hydrodynamics,''
JHEP {\bf 0209}, 043 (2002) [arXiv:hep-th/0205052].




\bibitem{ne4}
G.~Policastro, D.~T.~Son and A.~O.~Starinets,
``From AdS/CFT correspondence to hydrodynamics. II: Sound waves,''
JHEP {\bf 0212}, 054 (2002) [arXiv:hep-th/0210220].

\bibitem{ne1}
G.~Policastro, D.~T.~Son and A.~O.~Starinets,
``The shear viscosity of strongly coupled $N = 4$
supersymmetric Yang-Mills  plasma,''
Phys.\ Rev.\ Lett.\  {\bf 87}, 081601 (2001) [arXiv:hep-th/0104066].


\bibitem{bls}
  A.~Buchel, J.~T.~Liu and A.~O.~Starinets,
  ``Coupling constant dependence of the shear viscosity in N = 4 supersymmetric
  Yang-Mills theory,''
  Nucl.\ Phys.\ B {\bf 707}, 56 (2005)
  [arXiv:hep-th/0406264].

\bibitem{r1}
D.~Teaney,
``Effect of shear viscosity on spectra, elliptic flow, and Hanbury Brown-Twiss
radii,''
Phys.\ Rev.\ C {\bf 68}, 034913 (2003).


\bibitem{r2}
E.~Shuryak,
``Why does the quark gluon plasma at RHIC behave as a nearly ideal fluid?,''
Prog.\ Part.\ Nucl.\ Phys.\  {\bf 53}, 273 (2004)
[arXiv:hep-ph/0312227].

\bibitem{r3}
D.~Molnar and M.~Gyulassy,
``Saturation of elliptic flow at RHIC: Results from the covariant elastic
parton cascade model MPC,''
Nucl.\ Phys.\ A {\bf 697}, 495 (2002)
[Erratum-ibid.\ A {\bf 703}, 893 (2002)]
[arXiv:nucl-th/0104073].



\bibitem{bl1}
  A.~Buchel and J.~T.~Liu,
  ``Universality of the shear viscosity in supergravity,''
  Phys.\ Rev.\ Lett.\  {\bf 93}, 090602 (2004)
  [arXiv:hep-th/0311175].




\bibitem{kss1}
  P.~Kovtun, D.~T.~Son and A.~O.~Starinets,
  ``Viscosity in strongly interacting quantum field theories from black hole
  physics,''
  Phys.\ Rev.\ Lett.\  {\bf 94}, 111601 (2005)
  [arXiv:hep-th/0405231].



\bibitem{bh2}
  A.~Buchel,
  ``On universality of stress-energy tensor correlation functions in
  supergravity,''
  Phys.\ Lett.\ B {\bf 609}, 392 (2005)
  [arXiv:hep-th/0408095].




\bibitem{bbs} P.~Benincasa,  A.~Buchel and A.~O.~Starinets,
``Sound waves in strongly coupled non-conformal gauge theory plasma,''
to appear. 

\bibitem{Krasnitz:2000ir}
  M.~Krasnitz,
  ``A two point function in a cascading N = 1 gauge theory from
  supergravity,''
  arXiv:hep-th/0011179.

\bibitem{Krasnitz:2002ct}
  M.~Krasnitz,
  ``Correlation functions in a cascading N = 1 gauge theory from
  supergravity,''
  JHEP {\bf 0212}, 048 (2002)
  [arXiv:hep-th/0209163].

\bibitem{Herzog:2004tr}
  C.~P.~Herzog, Q.~J.~Ejaz and I.~R.~Klebanov,
  ``Cascading RG flows from new Sasaki-Einstein manifolds,''
  JHEP {\bf 0502}, 009 (2005)
  [arXiv:hep-th/0412193].

\bibitem{Witten:1998zw}
  E.~Witten,
  ``Anti-de Sitter space, thermal phase transition, and confinement in  gauge
  theories,''
  Adv.\ Theor.\ Math.\ Phys.\  {\bf 2}, 505 (1998)
  [arXiv:hep-th/9803131].


\bibitem{bpaz}
  A.~Buchel and L.~A.~Pando Zayas,
  ``Hagedorn vs. Hawking-Page transition in string theory,''
  Phys.\ Rev.\ D {\bf 68}, 066012 (2003)
  [arXiv:hep-th/0305179].

\bibitem{bln2}
  A.~Buchel and J.~T.~Liu,
  ``Thermodynamics of the N = 2* flow,''
  JHEP {\bf 0311}, 031 (2003)
  [arXiv:hep-th/0305064].


\bibitem{n2hydro}
  A.~Buchel,
  ``N = 2* hydrodynamics,''
  Nucl.\ Phys.\ B {\bf 708}, 451 (2005)
  [arXiv:hep-th/0406200].

\end{thebibliography}
\end{document}